\documentclass{aa}

\usepackage{graphicx}
\usepackage[dvipsnames]{xcolor}
\usepackage{xfrac}
\usepackage{txfonts}
\usepackage{float}
\usepackage{svg}
\usepackage{multirow}
\usepackage{multicol}
\usepackage{makecell}
\usepackage{colortbl}
\usepackage{natbib}
\usepackage[colorlinks=true,     linkcolor=blue, citecolor=blue, filecolor=blue, urlcolor=blue]{hyperref}

\usepackage{subcaption}
\newcommand{\citespecial}[1]{\citeauthor{#1} \citeyear{#1}}

\definecolor{verde}{RGB}{0, 150, 0}

\begin{document}

    \title{Detecting the edges of galaxies with deep learning}

   %\subtitle{I. Overviewing the $\kappa$-mechanism}

   \author{Jes\'us Fern\'andez \inst{1}
          \and
          Fernando Buitrago\inst{2,3}
          \and
          Benjam\'in Sahelices\inst{1}
          }

    \institute{GIR GCME. Departamento de Inform\'{a}tica. Universidad de Valladolid, 47011 Valladolid, Spain.\\
    \email{jesus.fernandez.iglesias@estudiantes.uva.es} 
    \and 
    Departamento de F\'{i}sica Te\'{o}rica, At\'{o}mica y \'{O}ptica, Universidad de Valladolid, 47011 Valladolid, Spain.
    \and
    Instituto de Astrof\'{\i}sica e Ci\^{e}ncias do Espa\c{c}o, Universidade de Lisboa, OAL, Tapada da Ajuda, PT1349-018 Lisbon, Portugal.\
    }

 \abstract{
Galaxy edges or truncations are low-surface-brightness (LSB) features located in the galaxy outskirts that delimit the distance up to where the gas density enables efficient star formation. As such, they could be interpreted as a non-arbitrary means to determine the galaxy size and this is also reinforced by the smaller scatter in the galaxy mass-size relation when comparing them with other size proxies. However, there are several problems attached to this novel metric, namely, the access to deep imaging and the need to contrast the surface brightness, color, and mass profiles to derive the edge position. While the first hurdle is already overcome by new ultra-deep galaxy observations, we hereby propose the use of machine learning (ML) algorithms  to determine the position of these features for very large datasets. We compare the semantic segmentation by our deep learning (DL) models with the results obtained by humans for HST observations of a sample of 1052 massive (M$_{\rm stellar}$ > 10$^{10}$ M$_{\odot}$) galaxies at $z < 1$. In addition, the concept of astronomic augmentations is introduced to endow the inputs of the networks with a physical meaning. Our findings suggest that similar performances than humans could be routinely achieved, although in the majority of cases, the best results are obtained by combining (with a pixel-by-pixel democratic vote) the output of several neural networks using ensemble learning. Additionally, we find that using edge-aware loss functions allows for the networks  to focus their optimization on the galaxy boundaries and, therefore, to provide estimates that are much more sensitive to the presence of neighboring bodies that may affect the shape of the truncation. The experiments reveal a great similarity between the semantic segmentation performed by the AI compared to the human model. For the best model, an average dice of 0.8969 is achieved, while an average dice of 0.9104 is reached by the best ensemble, where the dice coefficient represents the harmonic mean between the precision and the recall. This methodology will be profusely used in future datasets, such as that of Euclid, to derive scaling relations that are expected to closely follow the galaxy mass assembly. We also offer to the community our DL algorithms in the author's github repository.}


\maketitle
%
%-------------------------------------------------------------------

\section{Introduction}
\label{intro}

% Introduction
Identifying the appropriate method to determine the size of a galaxy is a longstanding problem in astrophysics since galaxies are diffuse objects without sharply defined edges. Many different approaches have been adopted in the last century (see \citespecial{Chamba20_hist} for a thorough review). In recent decades, the effective radius (r$_{e}$, the semi-major axis of the ellipse encompassing half of the light from a galaxy) has been the most widely used proxy for the size. The main reasons behind this decision are twofold: on the one hand, its robustness against different values for signal-to-noise ratio (S/N) and exposure time and, on the other hand, its connection with galaxy surface brightness parametric fitting, namely, the \citet{Sersic1968} functions. However, taking a half-light radius as the proxy for galaxy size is an arbitrary choice and its calculation is tightly connected with the galaxy concentration (see the discussion in the introduction of \citespecial{Trujillo20}).

The aforementioned \citet{Trujillo20} suggested R$_{1}$ (the isomass contour at 1 M$_{\odot}$/pc$^{2}$) as a physically motivated size proxy. 
Considering the two-phase formation scenario for galaxies \citep[see, e.g., ][]{Hilz13,Pulsoni21}, a first period of in situ star formation should be followed by the subsequent accretion of galaxy satellites. Therefore, the cessation of that first stage must be imprinted in the galaxy outskirts, as there is a spatial limit for reaching the gas density threshold enabling star formation \citep{Schaye04}. The gas density translates into a stellar density, assuming a certain (low) efficiency into converting gas to stars\ and this is the reason why R$_{1}$ is connected with a sudden change in the galaxy properties (surface brightness, color, and mass profiles) for Milky Way-like galaxies in the local Universe \citep{Martinez-Lombilla19, Diaz-Garcia22}. However, the 1~M$_{\odot}$/pc$^{2}$ threshold seems to be dependent (at least) on the galaxy stellar mass or (ultimately) on the physical conditions of the star formation episode. To investigate in depth where these features appear, irrespective of their surface mass density value, \citet{Chamba22} and \citet{buitrago2023strong} (with the former study focused in the low-z Universe and the latter at intermediate, z $<$ 1, redshifts) detected these sudden drops or changes in the galaxy outer profiles, studying the evolution on their location with stellar mass and redshift. The series of observational tests conducted in these articles, together with the reduction in the scatter in the mass-size relation by a factor of 2-2.5, solidified this metric as a physically motivated means to define galaxy sizes. Very interestingly, these edges could be identified with the previously known galaxy truncations \citep{vanderKruit1979} and, as such, we utilize these two names interchangeably in the present work.

Galaxy truncations have initially been identified for edge-on galaxy disks whose sizes did not seem to change despite integrating the images for longer exposure times \citep{vanderKruit1981a,vanderKruit1981b}. It is to noteworthy that this is not at variance with the fact of galaxies being fuzzy objects, because there are always stars and light beyond this distance due to the existence of a stellar halo or stellar migration. Nevertheless, they had not been considered useful size proxies in the past because of the low surface brightness (LSB) levels where they appear \citep[$\mu$ > 26-27 mag/arcsec$^{-2}$ in 10$\times$10 arcsec apertures][]{Martin-Navarro12,Martin-Navarro14,Trujillo16}. This situation could be changed with the arrival of future telescope facilities, especially the large-aperture synoptic telescopes that will be able to routinely reach very deep surface brightness levels \citep[$\mu$ > 28-30 mag/arcsec$^{-2}$][]{Borlaff22}. This research has been branded the LSB regime of galaxy formation and evolution \citep{Duc15,Mihos19}. However, if we are to extract the maximum information from these incoming images, our endeavor is not only about building the right telescopes and other instruments, but also to develop the next generation software able to deal with the millions and even billions of objects to be studied, while simultaneously achieving high accuracy in our feature extraction.

As this is a pattern-recognition problem, machine learning (ML) is especially suited for tackling these problems in spite of the diversity of galaxy shapes. Deep learning (DL), and specifically convolutional neural networks (CNN), have shown their power in similar big data projects focused on astrophysical and extragalactic studies \citep[][to name a few]{Huertas-Company15,Dominguez-Sanchez18,Hausen20,Vega-Ferrero21}. 

The DL technique that we present here goes beyond the algorithmic segmentation performed by other codes such as Source-Extractor \citep{Bertin1996} or NoiseChisel \citep{Akhlaghi15,Akhlaghi19} because of the fact that several pieces of evidence (1D or 2D surface brightness and color and mass profiles) must be taken into account in order to derive the truncation position. Generative networks like U-Nets \cite{Ronneberger15} or variational autoencoders (VAE, \citespecial{Kingma13}) are capable of generating pixel level segmentation and separate the background noise from the information of the galaxy. Specific axis and edge detection techniques have been recently proposed \citep{Soria20} along with novel DL models such as EfficientNet \citep{Tan19}, ResNet \citep{He16, He19}, DenseNet \citep{Huang17}, FastFCN \citep{Wu19}, GatedSCNN \citep{Takikawa19} or Mask RCNN \citep{He20} to perform this task. Other proposal \citep{gonzalez2018galaxy} rely on the use of object detection networks such as YOLO \citep{redmon2016look} for automated galaxy detection and classification. Other works \citep{Walmsley20} have suggested leveraging bayesian DL techniques to investigate morphological features of galaxies. Such an approach provides an avenue to incorporate uncertainty into the process of edge estimation.

In the present work, we take a first step on the characterization of truncations and galaxy edges with ML and with the relatively small ($>$1000) albeit meaningful sample from \citet{buitrago2023strong}, which will be the base for the framework that will be developed for the forthcoming Euclid satellite \citep{Laureijs11}. Of course, this approach could also be extended for other future enterprises such as the Vera Rubin Observatory’s Legacy Survey of Space and Time \citep[LSST,][]{2019ApJ...873..111I} or the ARRAKIHS satellite\footnote{https://www.cosmos.esa.int/documents/7423467/7423486/ESA-F2-ARRAKIHS-Phase-2-PUBLIC-v0.9.2.pdf/61b363d7-2a06-1196-5c40-c85aa90c2113?t=1667557422996} among others. The tools to be created will become open-source in the following repository\footnote{https://github.com/jesusferigl} and our aim is that they could be applied to different fits images and galaxy surveys than the ones that we took as a basis for our study.

The paper structure is the following: Section \ref{dataset} summarizes the parent imaging and the adopted methodology in order to infer galaxy edges or truncations, Section \ref{models_training} describes the models we built  to analyze our data, Section \ref{accuracy} details our results and Section \ref{conc} reports our conclusions. Hereafter, our assumed cosmology is $\Omega_m$=0.3, $\Omega_\Lambda$=0.7 and H$_0$=70 km s$^{-1}$ Mpc$^{-1}$. We used the \citet{Chabrier03} initial mass function (IMF), unless otherwise stated. 
Magnitudes are provided in the AB system \citep{Oke1983}.

%%%%%%%%% Dataset
\section{Dataset}
\label{dataset}
% Description of the initial dataset
The parent sample comes from \citet{buitrago2023strong}, where the reader can find all the details from our galaxy sample and associated data. Here, we summarize its most important characteristics, highlighting the necessary changes to adapt our previous datasets to become the inputs for our ML algorithms. 

The images for our study come from the CANDELS fields (the best trade-off between area and depth for HST observations), namely, the CANDELS survey \citep{Grogin11,Koekemoer11} for GOODS-N, UDS, EGS and COSMOS, HLF \citep{Illingworth16} for GOODS-S, while for HUDF we used the dataset in \citet{Beckwith06} for the optical and ABYSS \citep{Borlaff19} in the near-infrared (NIR) bands. We took the galaxy IDs, photometric masses, and spectroscopic redshifts (only those flagged as the ones with top quality) for all galaxies from the CANDELS public catalogs\footnote{https://archive.stsci.edu/prepds/candels/} \citep{Santini15,Stefanon17,Nayyeri17,Barro19}, whereas to increase our sample, we also took galaxies with spectroscopic redshifts from other sources \citep{vanderWel16,Straatman18,Lilly09,Damjanov19,vanderWel21}. Then, we selected galaxies with M$_{\rm stellar}$ $>$ 10$^{10}$ M$_{\odot}$ and z$_{\rm spec}$ $<$ 1.1 (in order to maximize the S/N of our targets). As we are looking for features regarding the limits of gas density enabling star formation, truncations should be more easily visible in disk galaxies \citep{Martin-Navarro12} and, as such, we only selected this type of galaxies according the criteria in \citet{Huertas-Company15}. In this publication, these galaxies are later subdivided as pure disks (DISK), disks with central spheroids (DISKSPH), and irregular disks (DISKIRR). We use these subclasses in the present work. Lastly, any galaxy displaying an observational artifact was also removed from our final sample. In summary, our initial mass-selected sample of 3460 galaxies was reduced to 1052 galaxies after applying all the aforementioned selection criteria with integrated apparent magnitudes of 17.5 $<$ H$_{\rm AB}$ $<$ 22.5.

The observed images for the galaxies in our sample were cut with a fixed stamp size of 12$\times$12 arcsec, ergo 400$\times$400 pixels for the optical ACS images, for the F606W (V) and F814W (I) filters, while 200$\times$200 pixels for the near-infrared WFC3 images, for the F125W (J) and F160W (H) filters; however,  these latter ones were later resampled to the near-infrared resolution to match the same pixel scale. A representative sky background value was subtracted from each image (details in \citespecial{buitrago2023strong}). From these postage stamps we derived $g$, $r$, $i,$ and $z$ Sloan Digital Sky Survey (SDSS)-restframe equivalent images in the same manner as the profiles in \citet{Buitrago17} or \citet{buitrago2023strong}, but this time on a   pixel-by-pixel basis. Basically,  we interpolated linearly between contiguous observed bands converted to surface brightness using the formula:

$$\mu = -2.5\log_{10}(\text{flux}) + zeropoint + 5\log_{10}(\rm \text{pixel~scale}),$$

while taking as a reference the mean wavelength of each filter\footnote{See http://svo2.cab.inta-csic.es/svo/theory/fps3/index.php}. These SDSS-restframe images were also corrected by galaxy inclination according to the model in \citet{Trujillo20}, along with the extinction from our galaxy\footnote{Values taken from https://ned.ipac.caltech.edu/extinction\_calculator} and cosmological dimming. To avoid numerical errors, in the case of negative pixel values in the images to be interpolated, we substituted their values with the noise level of each image. This limiting magnitude was calculated utilizing 10$^4$ 1$\times$1 arcsec$^{2}$ boxes to match the characteristic spatial scales of the galaxy in our sample, and then deriving the 3$\sigma$ clipped standard deviations of those.

We obtained color images (g-i, g-r, g-z, i-z, r-i, r-z) from these SDSS-restframe equivalent images by subtracting one from another. Overly high ($>$ 5) or low values ($<$ -1) were trimmed out setting not-a-number (NaN) values instead. Finally, these SDSS-restframe images and their color images were combined pixel-by-pixel using the recipes in \citet{Roediger15} for retrieving mass images. We note that by using these methods, we obtained 24 mass images (one per combination of base filter and color) and thus our final mass image is the median of all of them in units of M$_{\odot}$ pc$^{-2}$. Again, pixels displaying too high ($>$ 10$^{5}$) or too low ($<$ 10$^{-3}$) values were discarded. 

Summarizing, our final inputs to our neural networks are all the previously enumerated 200$\times$200 pixel images, namely, the HST observations, the SDSS-restframe equivalent images, and the color and mass images. Our labels are also 200$\times$200 pixel images set to 1 for all the pixels belonging to the central galaxy and 0 otherwise. Any neighbor overlapping the galaxy has also been set to 0 after a thorough masking. The borders between the central galaxy and the rest of the pixels are our galaxy edges or truncations (see Figure \ref{Fig: Imagen-Mascara-Truncacion}). They are defined as ellipses whose semi-major axis is the truncation position or edge radius found in \citet{buitrago2023strong}, with the axis ratio and position angle being representative of those found for the galaxy outer parts in the H-band. Mind this is approximate labelling, i.e. the truncation might be found at slightly different radial distances azimuthally or they might be erased in certain parts due to previous galaxy merging. These aspects will be expanded in future publications (Raji et al. in prep., Fern\'andez et al. in prep.). Finally, it is to note that our sample (both galaxies and their associated truncation values) differs somewhat from the one in \citet{buitrago2023strong} in terms of the number of objects (four more in our analysis) and truncation values (96 reassessed different values, with a median variation of 1.25 kpc). These changes are not important for our objectives, since our purpose is to create a neural network that is able to reproduce the decisions of human astronomers based on certain data and inputs; thus, it is to be expected (and this is indicated by the preliminary results) that with the new data, the DL models will also be able to adapt with a high level of precision.

\subsection{Astronomic augmentations (AAs)}
\label{AA}

The process of labeling galaxy edges in LSB areas is a complex task that  requires time for an experienced researcher to analyze all the input data. For these reasons, we did not have access to a large number of galaxies available for our study. It is well known that one of the main problems derived from the scarcity of data is overfitting in the training phase. That is why the application of data augmentation techniques is very relevant to achieve the best possible learning., namely, performing slight modifications of the input images to be fed into the networks. In this work, we have used some of the most standard ones, such as random 90-degree-multiple rotations ($0^\circ$, $90^\circ$, $180^\circ$, or $270^\circ$), flipping the images (horizontally, vertically, or not at all) and adding Gaussian noise (standard deviations of 0.01, 0.05, or 0.1).

Nevertheless, we wanted to increase the number of augmentations with the goal of not only gaining access to more examples for learning features, but also to change how the network learns as well. We combined the information that researchers use when detecting truncations and we utilized this data in the training phase, thus providing extra sets of physically meaningful inputs to the networks. The use of multiple combinations of all the available data maximizes the variability of the information used in the training phase, always referring to the same physical reality, which greatly improves the robustness of the training. We enumerate these new augmentations here, that we call ``astronomic" augmentations (given their origin), with the corresponding names to each one of their types: 1) visual: creating RGB images by using all possible range of observational bands sorted by wavelength. Consequently, the different possibilities are IJH, VIJ, VIH, and VJH; 2) color: creating all possible color images: g-i, g-r, g-z, i-z, r-i, and r-z; 3) sloan: creating b/w and color images with the SDSS-restframe equivalent images: g, r, i, z, gri, giz, riz, grz; 4) and mass: creating the mass per pixel images, as detailed in Section \ref{dataset}.

%%%%%%%%% Modelos
\section{Models and training}
\label{models_training}
\begin{figure}[t]
        \centering
        \includegraphics[width=\linewidth]{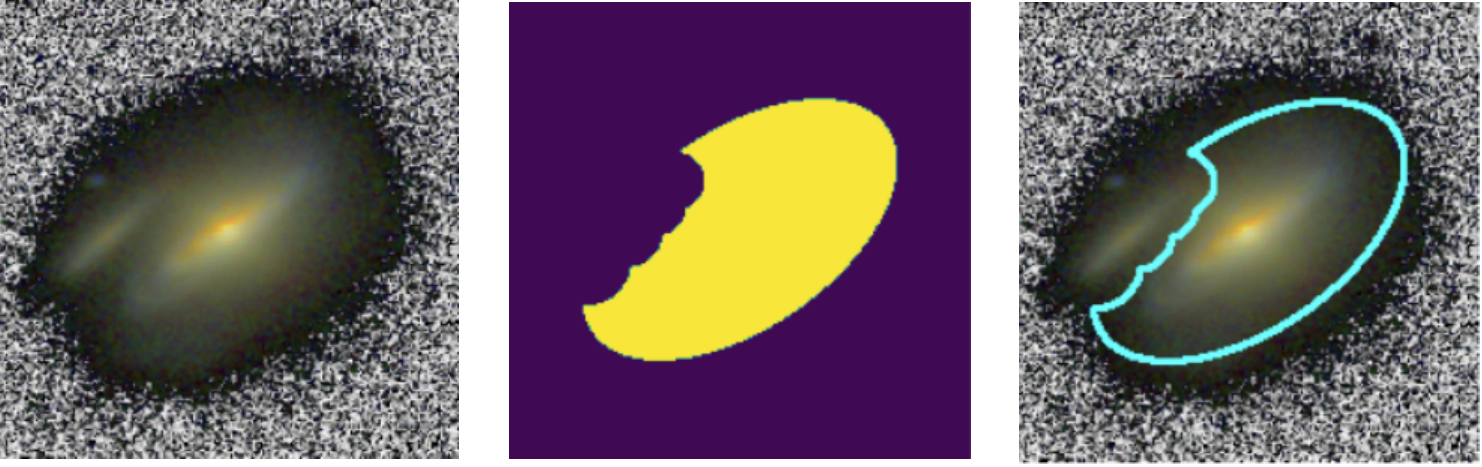}
        \caption{Image-mask-truncation equivalence. Left: Image created combining the V, I, and H bands. Center: associated segmentation mask, where the only active pixels are those belonging to the main galaxy. Right: Determination of the truncation from the perimeter of the segmentation mask. Note: the jagged profile is due to the superposition of a neighboring galaxy.}
        \label{Fig: Imagen-Mascara-Truncacion}
\end{figure}

To detect the galaxy edges, we used a DL-based semantic segmentation approach. Semantic segmentation is aimed at classifying all the pixels in an input, grouping together those pixels that share the same characteristics or belong to the same type of object. The aim of the DL methodology developed in this work is to determine and isolate the bodies of the central galaxies from the noise and other secondary galaxies. Therefore, each of the different algorithms attempts to classify the pixels of the images into two categories, main galaxy, and the rest of the pixels (where this latter  category includes both background noise and secondary galaxies) as seen in Figure \ref{Fig: Imagen-Mascara-Truncacion}. Using the masks generated by the semantic segmentation model, the radius of the associated truncations can be easily determined. Hence, it is crucial that the forthcoming DL models exhibit a good performance in the pixels situated near to the galaxy edges. The current section presents the architecture configurations of our models along with the training specifications.

\subsection{Architectural configurations}
\label{ArchitecturalConfigurations}

There are several CNN architectures that have been proposed over the years to solve the problem of semantic segmentation. The best known examples are U-Nets \citep{Ronneberger15}, nested U-Nets \citep{Zhou18}, or the DeepLab \citep{Chen18} family of architectures. In this work, we decided to use the U-Nets as the base scheme from which to derive the models for estimating galactic truncations. This type of network is able to achieve reasonable quality segmentation with a relatively small training sample size, which is a very relevant restriction in this work. They are composed of two symmetrical stages. The first one, the contraction stage, is carried out by an encoder and tries to extract the most relevant features from the inputs (downsampling). The second one, the expansion stage, is performed by a decoder and reconstructs the segmentation mask using the previously obtained features. In order to avoid a vanishing gradient problem (due to the concatenation of many layers) and to enhance the retrieval of spatial information (that may be lost during downsampling) skip connections are utilized. These links allow the cross flow of information between the symmetrical layers of the encoder and decoder.

Due to their morphology, U-Nets allow for different types of convolutional neural networks to be used as encoders. In this work, three families of CNNs are tested: ResNet \citep{He16}, EfficientNet \citep{Tan19} and DenseNet \citep{Huang17}. These families correspond to well-known models that have obtained state-of-the-art results in a wide range of computer vision tasks. Specifically, from the ResNet family, we used the 18-hidden layers and the 50-hidden layers models. From the EfficientNet family, variants B1, B2, and B6 have been used. Lastly, from the DenseNet family, the DenseNet-161 and DenseNet-201 architectures have been used as encoders. These choices reflect the requirement to include encoders of different sizes, considering the size as the number of training parameters of the model.  All the encoders have been initialized with pretrained weights from the ImageNet dataset \citep{Deng2009imagenet}. The use of pretrained deep neural networks usually improves the performance and generalization capacity. The main reason is that they use previously acquired knowledge related to high and low-level features (edges, textures) that are useful for many computer vision tasks, such as the identification of galaxy truncations.

\begin{figure*}[ht]
        \centering
        \includegraphics[width=\linewidth]{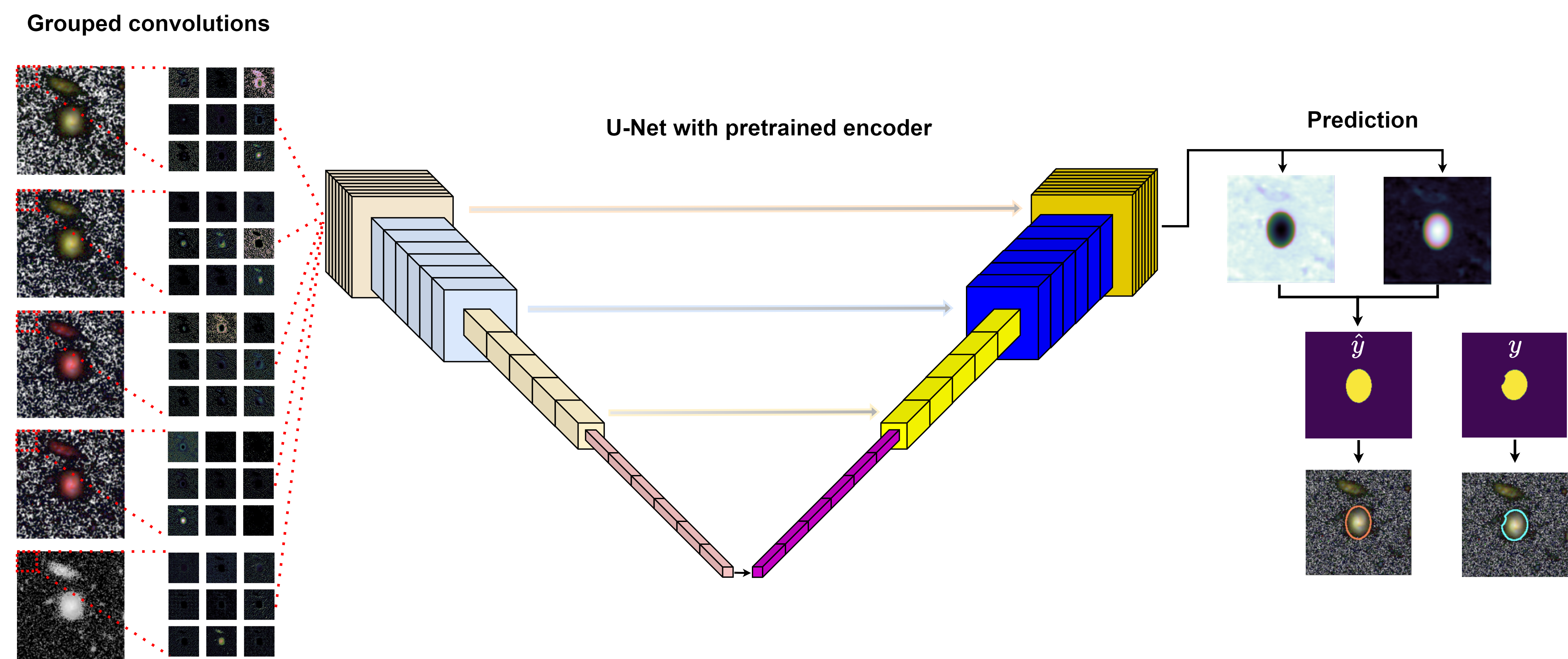}
        \caption{Example of the architecture designed for experiments with Mass-type AAs. The architecture consists of two stages. In the first stage, grouped convolutions process each image in isolation and then a U-Net with a pre-trained encoder uses all the information from the different AAs to infer the truncations.}
        \label{Fig: Architecture}
\end{figure*}

AAs are represented as different images generated for each galaxy, as explained in the Section \ref{AA}. The simplest solution to combine the AAs in the training phase is to concatenate the generated images in some dimension of the input tensors. However, in this work, we decide to exploit the concept of grouped convolutions as a way to solve the problem. Grouped convolutions were primarily introduced in \cite{Krizhevsky17}, as part of the well-known AlexNet architecture, with the main objective of reducing memory needs during training. However, further studies showed that grouped convolutions have a clear advantage in improving the representation of the training data than classical convolutions, as for instance in \cite{Xie2016AggregatedRT}. Therefore, in this work, grouped convolutions on the different inputs are used to extract independently the most relevant spatial information from each image and, subsequently, a second network (U-Net with pre-trained encoder) is used to estimate the galactic truncations from all this information (Figure \ref{Fig: Architecture}).

\subsection{Hyperparameters and training specifications}
\label{Hyperparameters}

In order to estimate the true error rate of the DL methods, we split the original dataset into two subsets, training and test, using the hold-out technique. The number of observations making up the training set is 952, while the number of instances reserved for the test set is 100. Furthermore, the training set is divided into two subsets, training and validation, based on a proportion of 80~\% and 20~\%, respectively. The validation set is used at the end of each epoch to assess learning and make decisions about training hyperparameters for the next epoch.

We train our DL models using a weighted categorical cross-entropy (CCE) loss function. Weights are necessary in order to counterbalance the strong imbalance of categories present in the images. Otherwise, there is a high risk that models only predict the noise category for each pixel, due to the high dominance of these pixels with respect to the ones representing the central galaxy. Weights for the loss function have been estimated empirically. Their values are $\omega_n$ = 0.075, for the class representing noise, and $\omega_g$ = 0.925, for the class representing the main galaxy. This way, hits and misses in pixels belonging to galaxies are weighted more heavily than in pixels belonging to noise. The batch size was set to 32, learning rate was fixed to $10^{-3}$, and the maximum number of training epochs is 1000. The Adam optimizer, from \cite{Kingma2015AdamAM}, was used to update the networks' weights.

In order to mitigate the possible overfitting, which commonly occurs in the domain of deep convolutional neural networks, three regularization techniques were used: data augmentation, early stopping, and ridge penalty. For data augmentation each training image, before being introduced into the network, is randomly rotated $0^\circ$, $90^\circ$, $180^\circ$, or $270^\circ$ and, once this transformation is applied, the image is randomly kept without further changes, mirrored around the abscissa axis or mirrored with respect to the ordinate axis. Thus, there are 11 possible combinations that induce images different from the original one, making it much more difficult for the network to memorize the training examples. After this process, random noise from a normal distribution is added to each training example. The noise added to each image comes from a normal distribution with $\mu = 0$ and $\sigma$ chosen randomly (0.01, 0.05, or 0.1) for each electromagnetic band of each image.

The second regularization technique is the selection of the model that maximizes the performance over the validation set, which is very similar in interpretation to early stopping. Finally, ridge regularization, also known as $l_2$ or the Tikhonov regularization, is used. This technique, originally presented in \cite{Hoerl70} for a linear regression problem, imposes a penalty on the estimated model coefficients. This way, when applied to a DL optimization problem, a penalty is added to the loss function. Equation \ref{Ridge-regularization-equation} shows the application of the technique, where $J(y, h_{\theta}(x))$ represents the traditional loss function between the labels ($y$) and the model outputs ($h_{\theta}(x)$) and $\lambda \sum_{i = 1}^p w_i^2$ represents the penalty (sum of the squared model weights driven by a $\lambda$ penalty factor). The effect of applying this technique is to reduce proportionally the value of all the model's weights, without reducing them to zero. The result is that the flexibility of the relationship between the model predictions and the target variable is reduced, acting as a powerful technique for avoiding overfitting.

\begin{equation}
    L(y, h_{\theta}(x)) = J(y, h_{\theta}(x)) + \lambda \sum_{i = 1}^p w_i^2
    \label{Ridge-regularization-equation}
\end{equation}

For the performance evaluation of the model we used three metrics: precision, recall, and Sørensen-dice or dice coefficient (Equation \ref{Equation:Dice}), also known as $\mathcal{F}1$ score in some contexts. We did not use a pixel-level hit rate because it is not sensitive to the cost of misclassification. The dice coefficient takes into account the cost of misclassification and adapts to the imbalance present in the images. It represents the harmonic mean between the precision, which measures the influence of false positives, and the recall, which measures the influence of false negatives. A perfect precision could be achieved by correctly predicting a single pixel as a galaxy, predicting all other pixels as noise. A perfect recall could be achieved by predicting all pixels in the image as a galaxy. Therefore, it is necessary to perform a weighting of both metrics. The harmonic mean, being  more demanding and pessimistic than its arithmetic and geometric counterparts, guarantees that it accurately reflects the quality of the results obtained by our DL models.

\begin{equation}
    \text{Dice} = \frac{2 \times \text{precision} \times \text{recall}}{\text{precision} + \text{recall}} = \frac{2 \times TP}{2 \times TP + FP + FN}
    \label{Equation:Dice}
\end{equation}

%%%%%%%%% Resultados
\section{Segmentation results}
\label{accuracy}

This section shows the results obtained with the different DL models trained according to the specifications described in the Section \ref{models_training}. All the results shown in this section have been obtained on the test set of galaxies. This set consists of 100 randomly chosen galaxies representative of the whole dataset and which have not been used during the training phase. First of all, the different encoders without AAs are evaluated in order to check the feasibility of our proposal. Next, the impact of the AAs is evaluated and a complete analysis of the best performing model is carried out. Finally, the results of a complete ensemble analysis are shown.

\subsection{Base experiment}
\label{Base}

The base experiment uses the seven models and the training specifications described in Sections \ref{ArchitecturalConfigurations} and \ref{Hyperparameters} without applying AAs, that is, the training is done with the images of galaxies formed by the H, J, and I bands. Table \ref{tab:ResultadosBase} shows dice, precision, and recall for each of the seven encoders embedded in the U-Net's backbone. The general trend reveals that recall is higher than precision for all experiments. This behavior indicates that networks tend to give slightly optimistic estimates of galaxy size, placing the truncations further away from the center of the galaxy than they actually are. The high value for recall shows that the vast majority of the image regions belonging to the galaxies of interest are correctly inferred, however, the slightly lower value of precision shows that there are areas outside the galaxy that have been incorrectly labeled. The dice index shows reasonably good segmentation qualities (> 0.88 for all encoders), oscillating narrowly between the different schemes. The worst value is reached by the Efficient-B2 encoder, and the best one is reached by DenseNet-161.

\begin{table}[t]
    \centering
    \begin{tabular}{ c | c | c | c }
        \hline\hline
        \textbf{Encoder} &  \textbf{Dice} & \textbf{Precision} & \textbf{Recall}\\\hline\hline
        \tiny ResNet18 &  \cellcolor[RGB]{135,  29,  90}\textcolor{white}{0.8849} &  \cellcolor[RGB]{89, 162, 207}\textcolor{white}{0.8531} &  \cellcolor[RGB]{92, 184, 106}\textcolor{white}{0.9438}\\\hline
        \tiny ResNet50 &  \cellcolor[RGB]{74,  28,  73}\textcolor{white}{0.8832} &  \cellcolor[RGB]{8,  48, 107}\textcolor{white}{0.8340} &  \cellcolor[RGB]{247, 252, 245} 0.9591\\\hline
        \tiny EfficientNetB1 &  \cellcolor[RGB]{200,  25,  80}\textcolor{white}{0.8867} &  \cellcolor[RGB]{98, 168, 210}\textcolor{white}{0.8543} &  \cellcolor[RGB]{125, 200, 125}\textcolor{white}{0.9461}\\\hline
        \tiny EfficientNetB2 &  \cellcolor[RGB]{2,   4,  25}\textcolor{white}{0.8809} &  \cellcolor[RGB]{28, 107, 176}\textcolor{white}{0.8437} &  \cellcolor[RGB]{144, 209, 141}\textcolor{white}{0.9475}\\\hline
        \tiny EfficientNetB6 &  \cellcolor[RGB]{84,  29,  78}\textcolor{white}{0.8835} &  \cellcolor[RGB]{154, 199, 224}\textcolor{white}{0.8604} &  \cellcolor[RGB]{0, 109,  44}\textcolor{white}{0.9351}\\\hline
        \tiny DenseNet161 &  \cellcolor[RGB]{250, 234, 220} 0.8926 &  \cellcolor[RGB]{247, 251, 255} 0.8768 &  \cellcolor[RGB]{0,  68,  27}\textcolor{white}{0.9316}\\\hline
        \tiny DenseNet201 &  \cellcolor[RGB]{200,  25,  80}\textcolor{white}{0.8867} &  \cellcolor[RGB]{143, 193, 221}\textcolor{white}{0.8592} &  \cellcolor[RGB]{45, 150,  77}\textcolor{white}{0.9397}\\\hline
        \tiny Average value &  \cellcolor[RGB]{142,  28,  91}\textcolor{white}{0.8851} &  \cellcolor[RGB]{99, 169, 211}\textcolor{white}{0.8545} & \cellcolor[RGB]{84, 180, 102}\textcolor{white}{0.9433}\\\hline
    \end{tabular}
    \caption{Results for the base experiment. Last row shows the average values for all encoders. Dice, precision, and recall values are represented in reddish, bluish, and greenish tones respectively, where lighter colors represent higher values and darker colors represent lower values.}
    \label{tab:ResultadosBase}
\end{table}

\subsection{Other experiments: Effects of AAs}
\label{EffectsAA}

In this section, we analyze the variability induced in the learning process by the introduction of the different AAs which are classified into four groups: 1) the visual group is made up of RGB files created with the four possible combinations of filters V, I, J, and H, always assigning the bluer channels to shorter wavelength bands (IJH, VIJ, VIH, and VJH). In other words, the visual group is formed by the base experiment plus the images formed by the VIJ, VIH and VJH combinations; 2) the color group is composed of the visual group to which six images are added, each one created by each of the six possible colors (g-i, g-r, g-z, i-z, r-i, and r-z); 3) the sloan group is formed by the visual group plus eight images formed both by the combination of the SDSS filters ordered by wavelength (gri, giz, riz, and grz) and by each of the filters individually (g, r, i, and z); 4) finally, the mass group is formed by the visual group together with the mass image.

Table \ref{tab:ResultadosAA}, analogous to Table \ref{tab:ResultadosBase}, shows results over the test set in terms of dice, precision, and recall indicators per encoder embedded in the U-Net's backbone and AA (visual, mass, color, and sloan). All average values of the dice coefficient are higher than in the base experiment case, showing that the introduction of the AAs leads to an improvement in the performance of the DL models. The average recall takes lower values with AAs compared with the base experiment. However, having obtained higher mean dice values means that the precision values have to rise considerably. This behavior could suggest that the estimates of the truncations with AAs are somewhat more conservative, placing them closer to the centre of the galaxy than in the case of the base experiment. In other words, AAs cause a substantial decrease in the number of image regions misdetected as galaxy, maintaining a very good proportion of pixels correctly categorized as a galaxy, leading to a marked improvement in the overall quality of the inferences. 

%%%%%%%%%%%%%%%%%%%%%%%%%
%%EXAMPLE OF LOSS PLOTS%%
%%%%%%%%%%%%%%%%%%%%%%%%%

\begin{figure*}[ht]
    \centering
    \begin{subfigure}{0.32\textwidth}
        \centering
        \includegraphics[width=\linewidth]{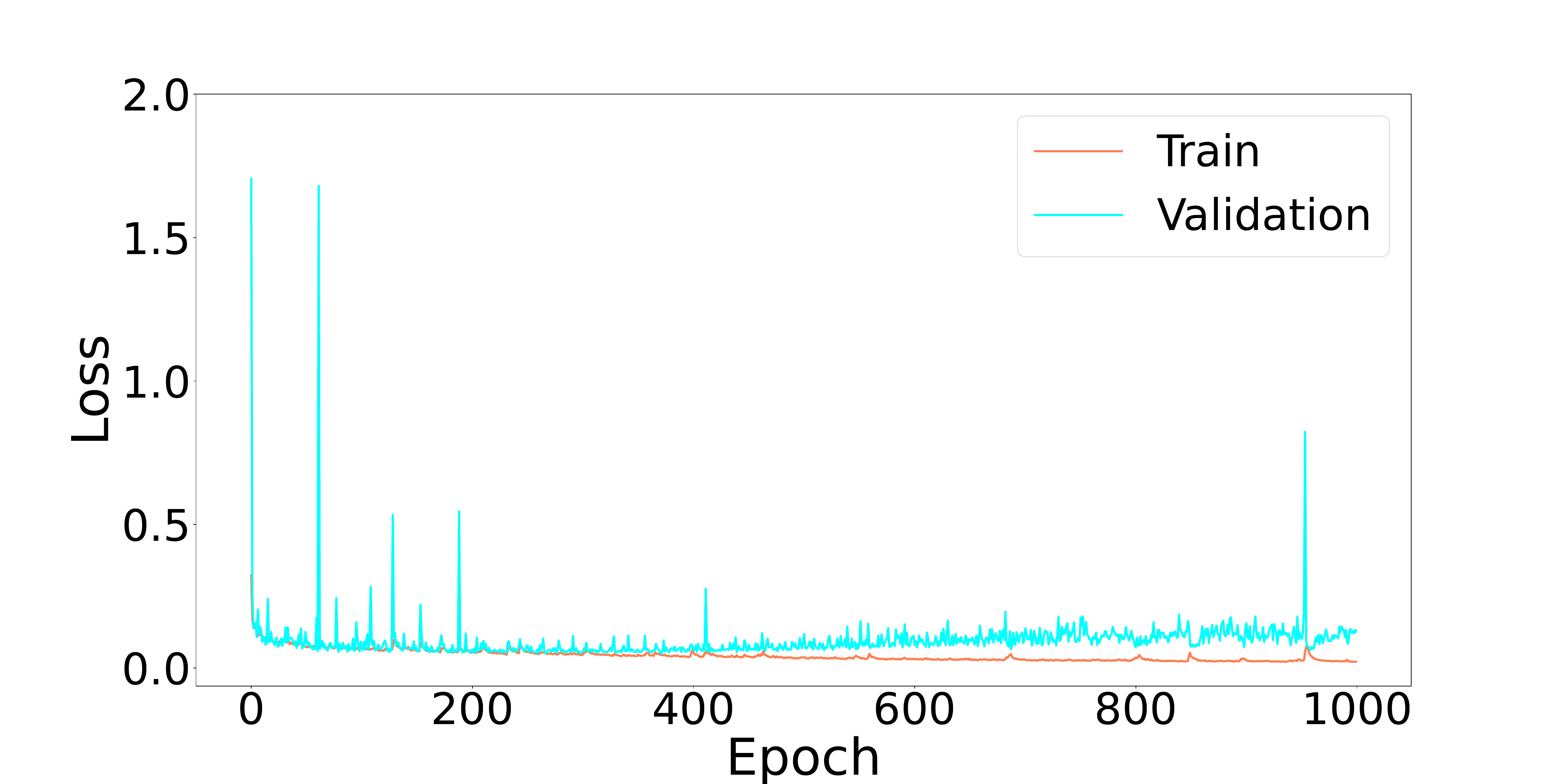}
        \caption{ResNet50 with  base scheme.}
        \label{ResNet-50-Base}
    \end{subfigure}
    \begin{subfigure}{0.32\textwidth}
        \centering
        \includegraphics[width=\linewidth]{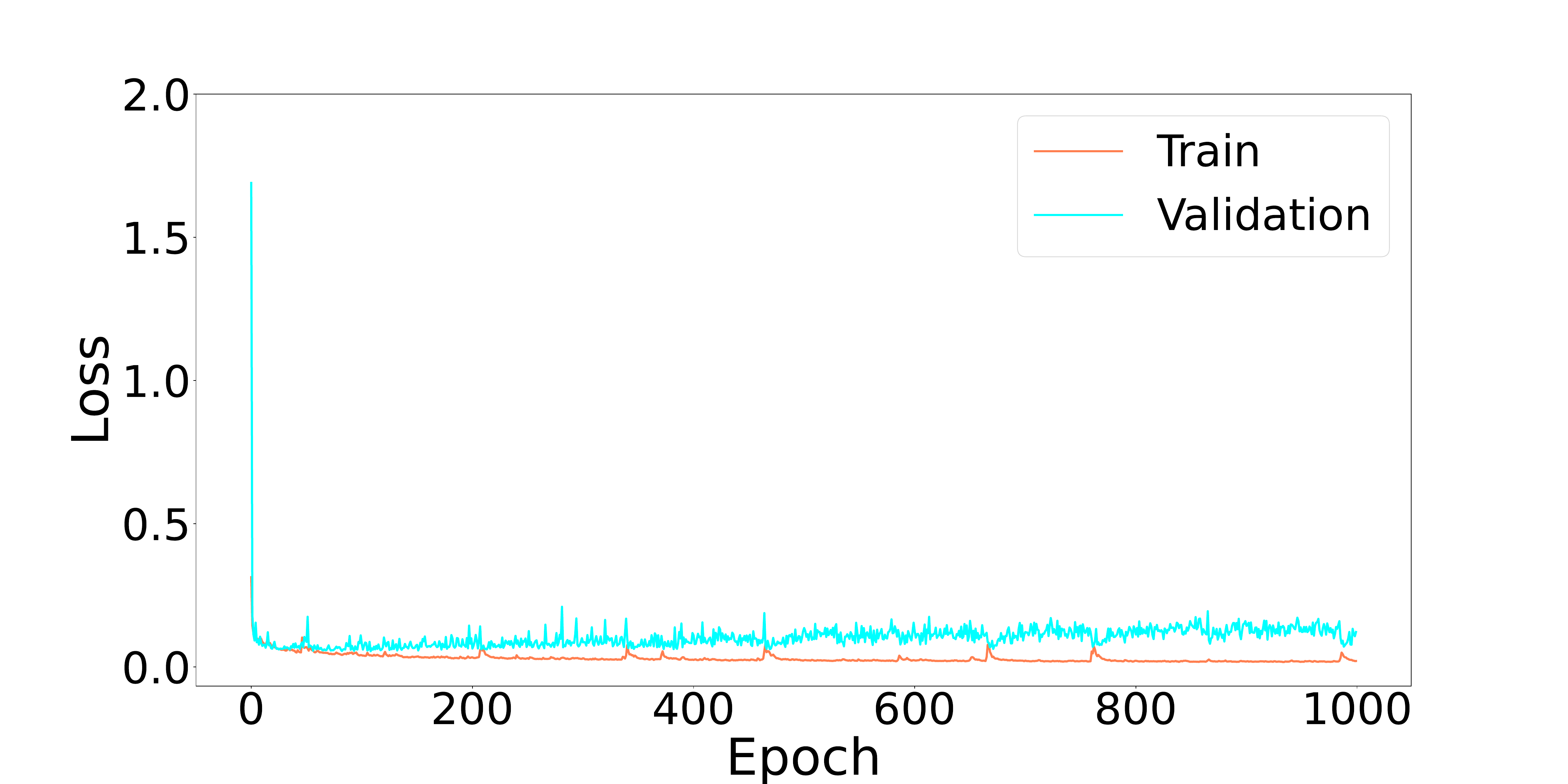}
        \caption{EfficientNetB6 with visual AA.}
        \label{EffB6-Visual}
    \end{subfigure}
    \begin{subfigure}{0.32\textwidth}
        \centering
        \includegraphics[width=\linewidth]{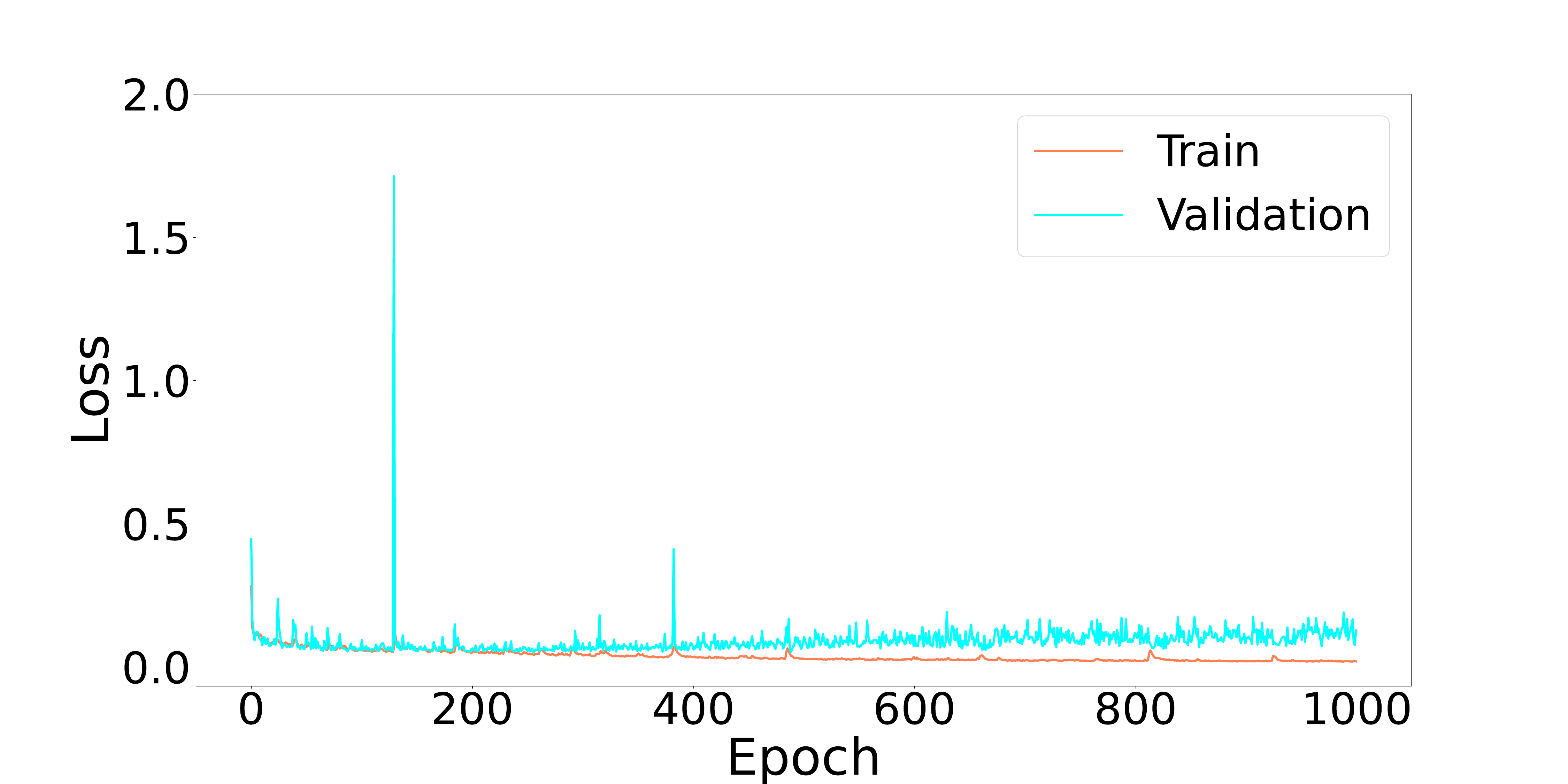}
        \caption{DenseNet161 with sloan AA.}
        \label{DenseNet161-Visual+Sloan}
    \end{subfigure}

    \begin{subfigure}{0.32\textwidth}
        \centering
        \includegraphics[width=\linewidth]{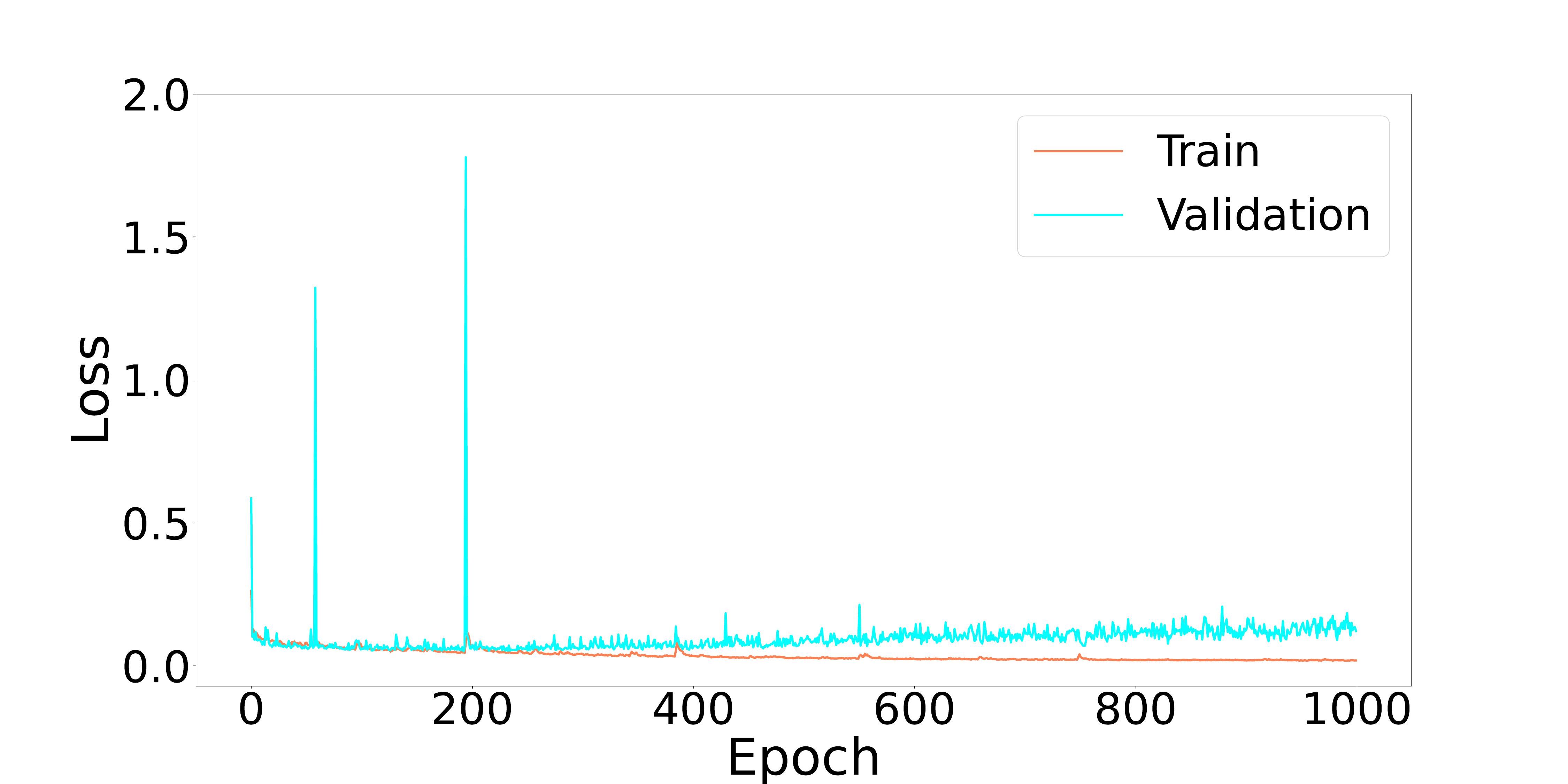}
        \caption{ResNet18 with color AA.}
        \label{ResNet18-Visual+Color}
    \end{subfigure}
    \begin{subfigure}{0.32\textwidth}
        \centering
        \includegraphics[width=\linewidth]{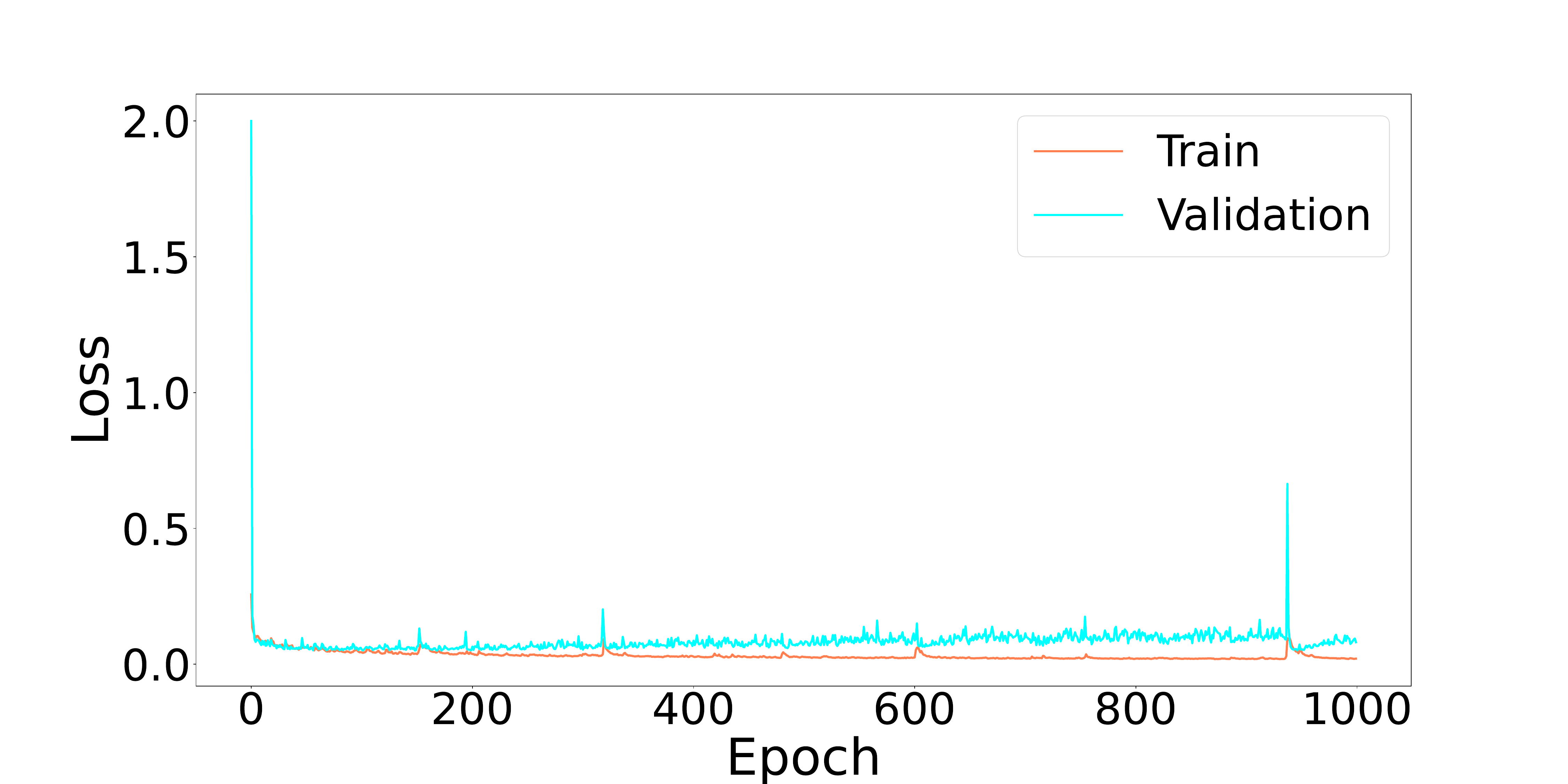}
        \caption{EfficientNetB2 with mass AA.}
        \label{EffB2-Visual+Masa}
    \end{subfigure}
    \caption{Learning curves showing the loss evolution over train and validation datasets of some DL configurations using the base scheme (a) and different AAs (b, c, d, e).}
    \label{Example-of-losses}
\end{figure*}

%%%%%%%%%%%%%%%%%%%%%%%%%
%%%%%%%%%%%%%%%%%%%%%%%%%
%%%%%%%%%%%%%%%%%%%%%%%%%

According to the results shown in Table \ref{tab:ResultadosAA}, mass (visual+mass) displays the best results, visual ranks second, while color and sloan have the worst results. Given that both color and sloan include visual augmentations, it may be assumed that these two sets of images do not improve the accurate prediction of truncations by DL models. In any case, the performance offered by all configurations, regardless of the inputs or architectures used, is remarkably uniform. None of the experiments obtained notably worse results than the rest. All configurations produce quality results, which are product of a stable training. This is illustrated in Figure \ref{Example-of-losses}, which shows the evolution of the loss over the epochs on the training and validation datasets. Five combinations of DL models and AAs representative of the general behavior have been selected. It can be seen that the learning patterns are consistent and do not show large oscillations. Moreover, there is no obvious overfitting, which is evidence of a good convergence towards optimal models. This provides evidence of stable training processes and, therefore, quality of results across all AAs and architectures.

\begin{table*}[t]
    \centering
    \begin{tabular}{ c | c | c | c | c | c | c | c | c | c | c | c | c }
    \hline\hline
    \textbf{Encoder} & \multicolumn{3}{c|}{\textbf{Visual}} & \multicolumn{3}{c|}{\textbf{Sloan}} & \multicolumn{3}{c|}{\textbf{Color}} & \multicolumn{3}{c}{\textbf{Mass}}\\\hline\hline
     & \tiny Dice & \tiny Precision & \tiny Recall & \tiny Dice & \tiny Precision & \tiny Recall & \tiny Dice & \tiny Precision & \tiny Recall & \tiny Dice & \tiny Precision & \tiny Recall\\\hline
    \tiny  ResNet18 & \tiny \cellcolor[RGB]{216,  38,  71}\textcolor{white}{0.8894} & \tiny \cellcolor[RGB]{130, 186, 219} \textcolor{white}{0.8769} & \tiny \cellcolor[RGB]{130, 202, 130} \textcolor{white}{0.9269} & \tiny \cellcolor[RGB]{174,  23,  89} \textcolor{white}{0.8876} & \tiny \cellcolor[RGB]{97, 167, 210} \textcolor{white}{0.8718} & \tiny \cellcolor[RGB]{144, 209, 141} \textcolor{white}{0.9293} & \tiny \cellcolor[RGB]{204,  27,  78}\textcolor{white}{0.8888} & \tiny \cellcolor[RGB]{14,  89, 162} \textcolor{white}{0.8530} & \tiny \cellcolor[RGB]{235, 247, 231} 0.9502 & \tiny \cellcolor[RGB]{246, 188, 153} 0.8949 & \tiny \cellcolor[RGB]{247, 251, 255} 0.9036 & \tiny \cellcolor[RGB]{29, 134,  65} \textcolor{white}{0.9080}\\\hline
    \tiny  ResNet50 & \tiny \cellcolor[RGB]{238,  85,  63}\textcolor{white}{0.8911} & \tiny \cellcolor[RGB]{152, 199, 223} \textcolor{white}{0.8803} & \tiny \cellcolor[RGB]{125, 200, 125} \textcolor{white}{0.9260} & \tiny \cellcolor[RGB]{126,  30,  89}\textcolor{white}{0.8858} & \tiny \cellcolor[RGB]{34, 114, 181} \textcolor{white}{0.8589} & \tiny \cellcolor[RGB]{200, 233, 193} 0.9401 & \tiny \cellcolor[RGB]{245, 174, 135} 0.8943 & \tiny \cellcolor[RGB]{202, 221, 240} 0.8898 & \tiny \cellcolor[RGB]{91, 183, 105} \textcolor{white}{0.9206} & \tiny \cellcolor[RGB]{222,  45,  68}\textcolor{white}{0.8897} & \tiny \cellcolor[RGB]{136, 190, 220} \textcolor{white}{0.8779} & \tiny \cellcolor[RGB]{125, 200, 125} \textcolor{white}{0.9258}\\\hline
    \tiny  EfficientNetB1 & \tiny \cellcolor[RGB]{195,  23,  82}\textcolor{white}{0.8884} & \tiny \cellcolor[RGB]{15,  91, 163} \textcolor{white}{0.8534} & \tiny \cellcolor[RGB]{232, 246, 228} 0.9490 & \tiny \cellcolor[RGB]{116,  30,  88}\textcolor{white}{0.8854} & \tiny \cellcolor[RGB]{149, 197, 223} \textcolor{white}{0.8797} & \tiny \cellcolor[RGB]{75, 176,  97} \textcolor{white}{0.9182} & \tiny \cellcolor[RGB]{31,  18,  46}\textcolor{white}{0.8819} & \tiny \cellcolor[RGB]{8,  55, 117} \textcolor{white}{0.8452} & \tiny \cellcolor[RGB]{220, 241, 214} 0.9451 & \tiny \cellcolor[RGB]{243, 123,  85}\textcolor{white}{0.8924} & \tiny \cellcolor[RGB]{46, 126, 188} \textcolor{white}{0.8616} & \tiny \cellcolor[RGB]{227, 244, 222} 0.9470\\\hline
    \tiny  EfficientNetB2 & \tiny \cellcolor[RGB]{2,   4,  25} \textcolor{white}{0.8805} & \tiny \cellcolor[RGB]{20,  97, 168} \textcolor{white}{0.8549} & \tiny \cellcolor[RGB]{187, 228, 181} \textcolor{white}{0.9373} & \tiny \cellcolor[RGB]{71,  28,  72} \textcolor{white}{0.8836} & \tiny \cellcolor[RGB]{15,  91, 163} \textcolor{white}{0.8533} & \tiny \cellcolor[RGB]{208, 236, 201} 0.9420 & \tiny \cellcolor[RGB]{200,  25,  80} \textcolor{white}{0.8886} & \tiny \cellcolor[RGB]{50, 130, 190} \textcolor{white}{0.8625} & \tiny \cellcolor[RGB]{201, 234, 194} 0.9403 & \tiny \cellcolor[RGB]{208,  30,  76} \textcolor{white}{0.8890} & \tiny \cellcolor[RGB]{152, 199, 223} \textcolor{white}{0.8802} & \tiny \cellcolor[RGB]{107, 191, 113} \textcolor{white}{0.9230}\\\hline
    \tiny  EfficientNetB6 & \tiny \cellcolor[RGB]{121,  30,  89}\textcolor{white}{0.8856} & \tiny \cellcolor[RGB]{64, 144, 197} \textcolor{white}{0.8657} & \tiny \cellcolor[RGB]{162, 217, 156} \textcolor{white}{0.9324} & \tiny \cellcolor[RGB]{240,  95,  67}\textcolor{white}{0.8914} & \tiny \cellcolor[RGB]{35, 115, 182} \textcolor{white}{0.8590} & \tiny \cellcolor[RGB]{230, 245, 225} 0.9478 & \tiny \cellcolor[RGB]{93,  30,  81}\textcolor{white}{0.8845} & \tiny \cellcolor[RGB]{29, 108, 177} \textcolor{white}{0.8574} & \tiny \cellcolor[RGB]{199, 233, 192} \textcolor{white}{0.9397} & \tiny \cellcolor[RGB]{245, 152, 111}\textcolor{white}{0.8935} & \tiny \cellcolor[RGB]{188, 215, 235} \textcolor{white}{0.8867} & \tiny \cellcolor[RGB]{112, 194, 116} \textcolor{white}{0.9237}\\\hline
    \tiny  DenseNet161 & \tiny \cellcolor[RGB]{215,  37,  72}\textcolor{white}{0.8893} & \tiny \cellcolor[RGB]{220, 233, 246} 0.8955 & \tiny \cellcolor[RGB]{24, 129,  61} \textcolor{white}{0.9067} & \tiny \cellcolor[RGB]{150,  27,  91}\textcolor{white}{0.8867} & \tiny \cellcolor[RGB]{102, 170, 212} \textcolor{white}{0.8727} & \tiny \cellcolor[RGB]{128, 201, 127} \textcolor{white}{0.9264} & \tiny \cellcolor[RGB]{155,  26,  91}\textcolor{white}{0.8869} & \tiny \cellcolor[RGB]{171, 207, 229} \textcolor{white}{0.8834} & \tiny \cellcolor[RGB]{68, 172,  94} \textcolor{white}{0.9173} & \tiny \cellcolor[RGB]{247, 212, 187} 0.8959 & \tiny \cellcolor[RGB]{163, 204, 227} \textcolor{white}{0.8820} & \tiny \cellcolor[RGB]{152, 212, 147} \textcolor{white}{0.9304}\\\hline
    \tiny  DenseNet201 & \tiny \cellcolor[RGB]{250, 234, 220} 0.8969 & \tiny \cellcolor[RGB]{186, 214, 234} \textcolor{white}{0.8863} & \tiny \cellcolor[RGB]{133, 204, 132} \textcolor{white}{0.9274} & \tiny \cellcolor[RGB]{137,  29,  90}\textcolor{white}{0.8862} & \tiny \cellcolor[RGB]{8,  48, 107} \textcolor{white}{0.8434} & \tiny \cellcolor[RGB]{247, 252, 245} 0.9550 & \tiny \cellcolor[RGB]{57,  25,  64}\textcolor{white}{0.8830} & \tiny \cellcolor[RGB]{232, 241, 250} 0.8991 & \tiny \cellcolor[RGB]{0,  68,  27} \textcolor{white}{0.8937} & \tiny \cellcolor[RGB]{244, 135,  95}\textcolor{white}{0.8928} & \tiny \cellcolor[RGB]{88, 161, 206} \textcolor{white}{0.8701} & \tiny \cellcolor[RGB]{190, 229, 183} \textcolor{white}{0.9379}\\\hline
    \tiny  {\bf Average Value} & \tiny \cellcolor[RGB]{203,  26,  79}\textcolor{white}{0.8887} & \tiny \cellcolor[RGB]{106, 173, 213} \textcolor{white}{0.8733} & \tiny \cellcolor[RGB]{146, 210, 142} \textcolor{white}{0.9294} & \tiny \cellcolor[RGB]{150,  27,  91}\textcolor{white}{0.8867} & \tiny \cellcolor[RGB]{51, 131, 190} \textcolor{white}{0.8627} & \tiny \cellcolor[RGB]{185, 227, 178} \textcolor{white}{0.9370} & \tiny \cellcolor[RGB]{155,  26,  91}\textcolor{white}{0.8869} & \tiny \cellcolor[RGB]{88, 161, 206} \textcolor{white}{0.8701} & \tiny \cellcolor[RGB]{146, 210, 142} \textcolor{white}{0.9296} & \tiny \cellcolor[RGB]{244, 128,  89}\textcolor{white}{0.8926} & \tiny \cellcolor[RGB]{152, 199, 223} \textcolor{white}{0.8803} & \tiny \cellcolor[RGB]{137, 206, 135} \textcolor{white}{0.9280}\\\hline
\end{tabular}
    \caption{Results with AAs over the test sample. Last row shows the average values for all encoders. Color coding corresponds to the one used in Table \ref{tab:ResultadosBase}, i. e., lighter colors represent better results.}
    \label{tab:ResultadosAA}
\end{table*}

Nevertheless, beyond the fact that some AAs induce better or worse results, the variability and singularities they introduce in the training process can be used a posteriori to build a democratic system that takes advantage of the majorities. It would therefore be interesting to demonstrate statistically that the different AAs influence the learning process (and, thus, the performance) of the different U-Nets. In order to demonstrate this assertion, we carried out a Friedman two-way analysis of variance by ranks non-parametric test, as used in \cite{Sheskin2000}. To this end, we transformed the dice results obtained over the test set in Table \ref{tab:ResultadosAA} into a rank-order format for each model, as can be seen in Table \ref{tab:RankingModelos}, having $n = 7$ experiments and $k = 5$ experimental conditions in which every experiment is assessed.
\begin{table}[H]
    \centering
    \begin{tabular}{| c | c | c | c | c | c |}
        \hline
        \tiny {\bf Encoder} & \tiny {\bf Base} & \tiny {\bf Visual} & \tiny {\bf Sloan} & \tiny {\bf Color} & \tiny {\bf Mass} \\\hline
        \tiny ResNet18 & \tiny 5 & \tiny 2 & \tiny 4 & \tiny 3 & \tiny 1\\\hline
        \tiny ResNet50 & \tiny 5 & \tiny 2 & \tiny 4 & \tiny 1 & \tiny 3\\\hline
        \tiny EfficientNetB1 & \tiny 3  & \tiny 2 & \tiny 4 & \tiny 5 & \tiny 1\\\hline
        \tiny EfficientNetB2 & \tiny 4  & \tiny 5 & \tiny 3 & \tiny 2 & \tiny 1\\\hline
        \tiny EfficientNetB6 & \tiny 5  & \tiny 3 & \tiny 2 & \tiny 4 & \tiny 1\\\hline
        \tiny DenseNet161 & \tiny 2     & \tiny 3 & \tiny 5 & \tiny 4 & \tiny 1\\\hline
        \tiny DenseNet201 & \tiny 3 & \tiny 1 & \tiny 4 & \tiny 5 & \tiny 2\\\hline
        \tiny $\sum_i R_i$ & \tiny 27 & \tiny 18 & \tiny 26 & \tiny 24 & \tiny 10\\\hline
        \tiny {\bf $\overline{R}$} & \tiny \textbf{3.8571} & \tiny \textbf{2.5714} & \tiny \textbf{3.7143} & \tiny \textbf{3.4286} & \tiny \textbf{1.4286}\\\hline
    \end{tabular}
    \caption{Rank-ordered results for AAs. Penultimate row shows the sum of all the rankings obtained for each AA. Last row shows average rank for each AA.}
    \label{tab:RankingModelos}
\end{table}

\begin{equation}
\chi_r^2 = \frac{12}{nk\left( k+1 \right)}  \left[ \sum_{j=1}^{k}\left( \sum_{i = 1}^{n} R_{i,j} \right)^2 \right] - 3n\left( k+1 \right) = 11.4286.
\label{EcuacionTest-I}
\end{equation}

The null hypothesis represents the absence of a significant difference between any of the $k = 5$ population medians. To reject the null hypothesis the computed value $\chi_r^2$ (see Equation \ref{EcuacionTest-I}) must be equal to or greater than the tabled critical chi-square value at the prespecified level of significance, 5~\% in this work, with $k-1$ degrees of freedom ($df$). For $df = 4$, the reported critical 5~\% chi-square value is 9.4877. Since the computed value $\chi_r^2$ is greater than the chi-square value, the alternative hypothesis is supported at the prespecified level, showing that training with different AAs influence the U-Nets performance for at least two of the five experimental conditions. 

\begin{table}[H]
    \centering
    \begin{tabular}{| c | c |}
        \hline
        \bf Differences & \bf Result\\\hline\hline
        $|\sum R_{\text{Base}} - \sum R_{\text{Visual}}|$ & |27 - 18| = 9\\\hline\hline
        $|\sum R_{\text{Base}} - \sum R_{\text{Sloan}}|$ & |27 - 26| = 1\\\hline\hline
        $|\sum R_{\text{Base}} - \sum R_{\text{Color}}|$ & |27 - 24| = 3\\\hline\hline
        \cellcolor[RGB]{75, 176,  97}  \textcolor{white}{$|\sum R_{\text{Base}} - \sum R_{\text{Mass}}|$} & \cellcolor[RGB]{75, 176,  97} \textcolor{white}{|27 - 10| = 17}\\\hline\hline
        $|\sum R_{\text{Visual}} - \sum R_{\text{Sloan}}|$ & |18 - 26| = 8\\\hline\hline
        $|\sum R_{\text{Visual}} - \sum R_{\text{Color}}|$ & |18 - 24| = 6\\\hline\hline
        $|\sum R_{\text{Visual}} - \sum R_{\text{Mass}}|$ & |18 - 10| = 8\\\hline\hline
        $|\sum R_{\text{Sloan}} - \sum R_{\text{Color}}|$ & |26 - 24| = 2\\\hline\hline
        \cellcolor[RGB]{75, 176,  97}  \textcolor{white}{$|\sum R_{\text{Sloan}} - \sum R_{\text{Mass}}|$} & \cellcolor[RGB]{75, 176,  97} \textcolor{white}{|26 - 10| = 16}\\\hline\hline
        $|\sum R_{\text{Color}} - \sum R_{\text{Mass}}|$ & |24 - 10| = 14\\\hline
    \end{tabular}
    \caption{Difference scores between the pairs of sums of ranks of Table \ref{tab:RankingModelos}. Green rows show the significantly different AAs.}
    \label{tab:Bonferroni}
\end{table}

To find out which of the AAs are significantly different from each other, it is necessary to conduct comparisons contrasting specific AAs between them. For this purpose, several sources \citep{daniel1990applied, siegel1988nonparametric} describe a comparison procedure based on the application of the Bonferroni-Dunn method. Using Equation \ref{EcuacionTest-II}, we can compute the minimum required difference (designated as $CD_F$) between the sums of the ranks of any two AAs at the prespecified level of significance (5~\% in this work): 

\begin{equation}
CD_F = z_{\text{adj}} \sqrt{\frac{nk(k+1)}{6}} = 2.58 \sqrt{35} = 15.2635.
\label{EcuacionTest-II}
\end{equation}

When comparisons are not planned beforehand, the  $z_{\text{adj}}$ value of Equation \ref{EcuacionTest-II} can be established by dividing the maximum tolerated familywise Type I error by the total number of comparisons conducted \citep[see][]{Sheskin2000}. In this case, the error is 5~\% and the total number of comparisons to be performed is 10, giving the quotient of both 0.005 (directional hypothesis). In the table of the normal distribution, the $z$ value that corresponds to that quotient is 2.58. Therefore, the significant comparisons are those whose ranks differences are equal to or greater than the value computed in Equation \ref{EcuacionTest-II}. The significant comparisons found are highlighted in green in Table \ref{tab:Bonferroni}. Thus, we can state that there is a significant difference between the results of the U-Nets when using base and mass AAs and when using sloan and mass AAs. Once it can be concluded that learning processes differ due to the introduction of the AAs, we can take advantage of this variability to develop the concept of democratic ensemble learning in Section \ref{ensemble}.

\subsection{Best model performance}
\label{BestModel}

\begin{figure*}[ht]
    \centering
    \begin{subfigure}{0.49\textwidth}
        \centering
        \includegraphics[width=\linewidth]{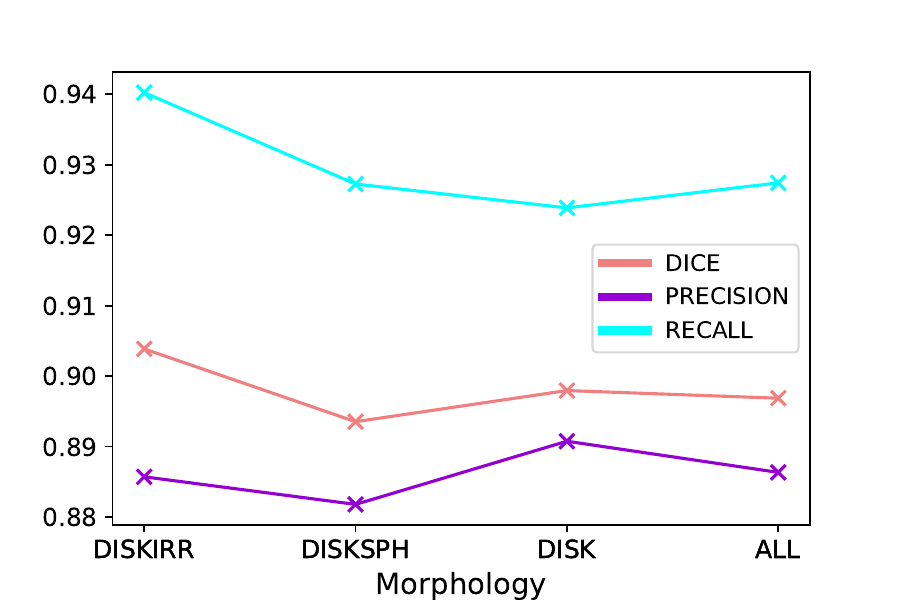}
        \caption{}
        \label{BestModel1-1}
    \end{subfigure}
    \begin{subfigure}{0.49\textwidth}
        \centering
        \includegraphics[width=\linewidth]{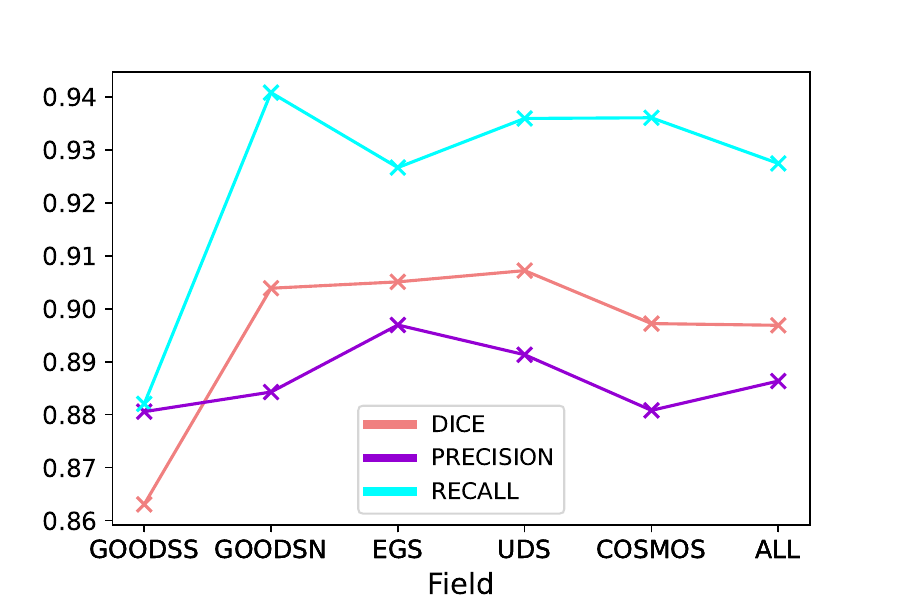}
        \caption{}
        \label{BestModel1-2}
    \end{subfigure}

    \begin{subfigure}{0.49\textwidth}
        \centering
        \includegraphics[width=\linewidth]{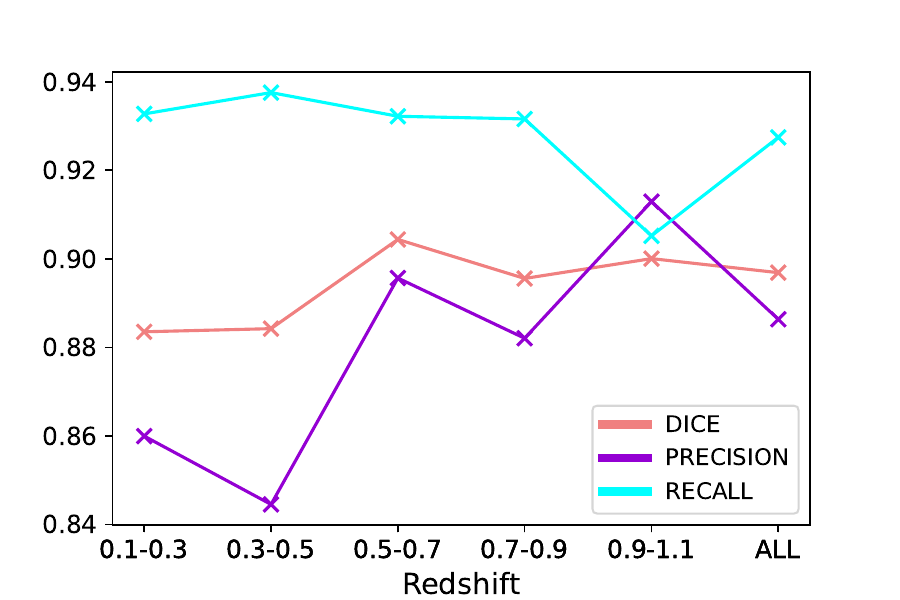}
        \caption{}
        \label{BestModel1-3}
    \end{subfigure}
    \begin{subfigure}{0.49\textwidth}
        \centering
        \includegraphics[width=\linewidth]{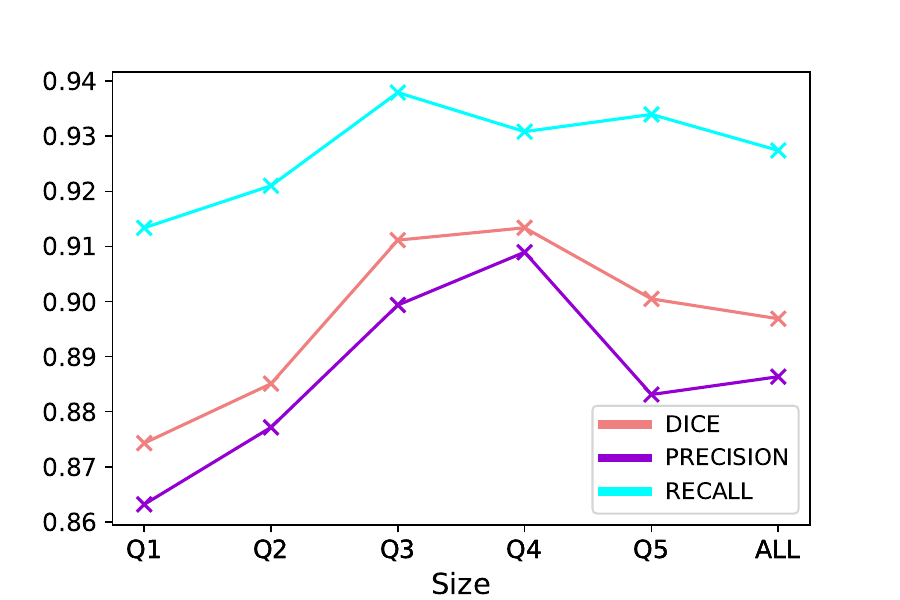}
        \caption{}
        \label{BestModel1-4}
    \end{subfigure}
    \caption{Metrics over the test set by galaxy morphology (a), field of view (b), redshift (c), and size (d) for our best performing model. The first plot (a) shows mean dice, precision, and recall for {\sc diskirr}, {\sc disksph}, {\sc disk,} and all morphologies. The second plot (b) displays the same metrics for all our galaxy fields, namely {\sc egs}, {\sc goodsn}, {\sc cosmos}, {\sc uds}, and {\sc goodss}. The third and fourth plots shows the same metrics for the five redshift bins and the five size quintiles the galaxy sample is split into.}
    \label{BestModel1}
\end{figure*}

\begin{figure*}[t]
    \centering
    \begin{subfigure}{.49\textwidth}
        \centering
        \includegraphics[width=\linewidth]{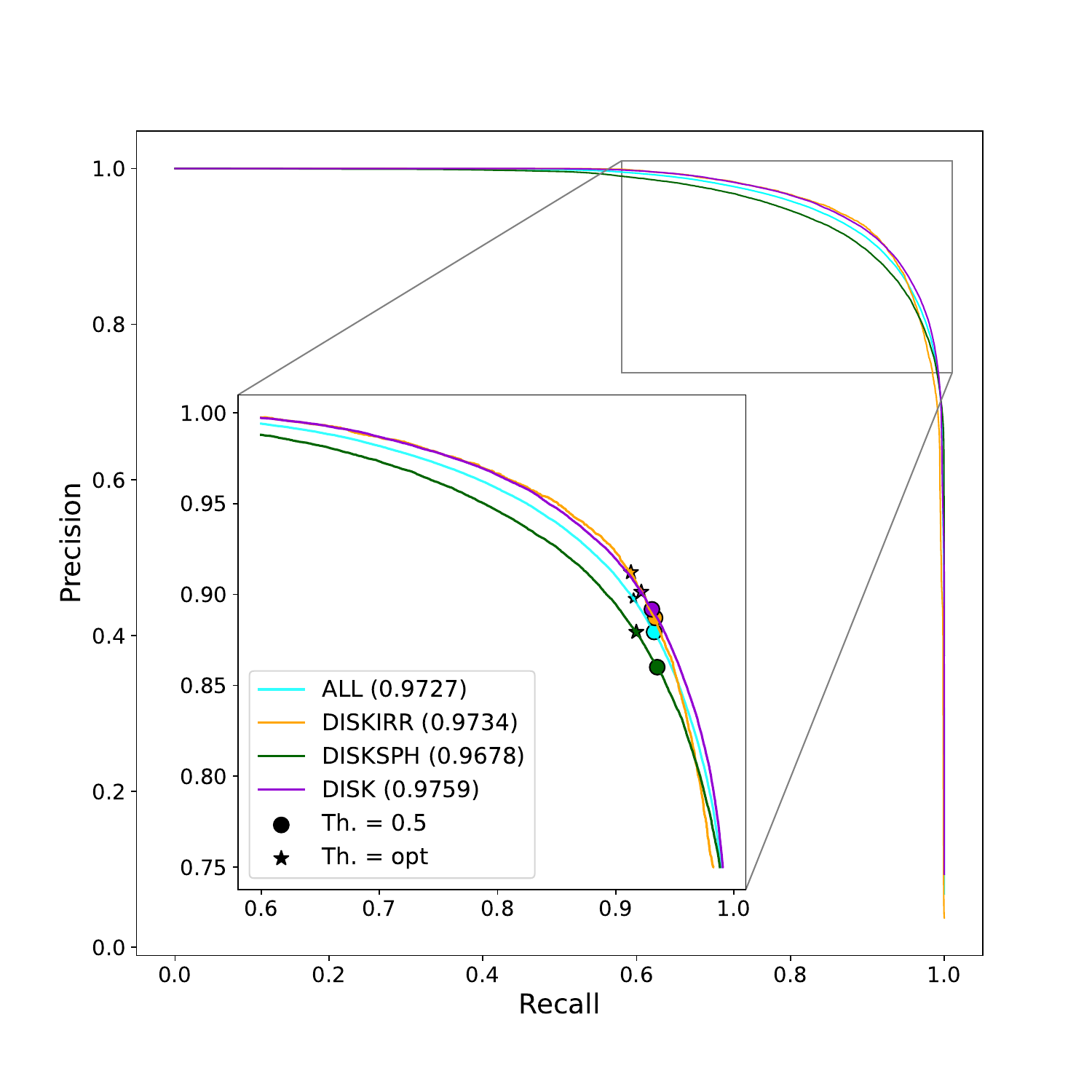}
        \caption{}
        \label{BestModel2-1}
    \end{subfigure}%
    \begin{subfigure}{0.49\textwidth}
        \centering
        \includegraphics[width=\linewidth]{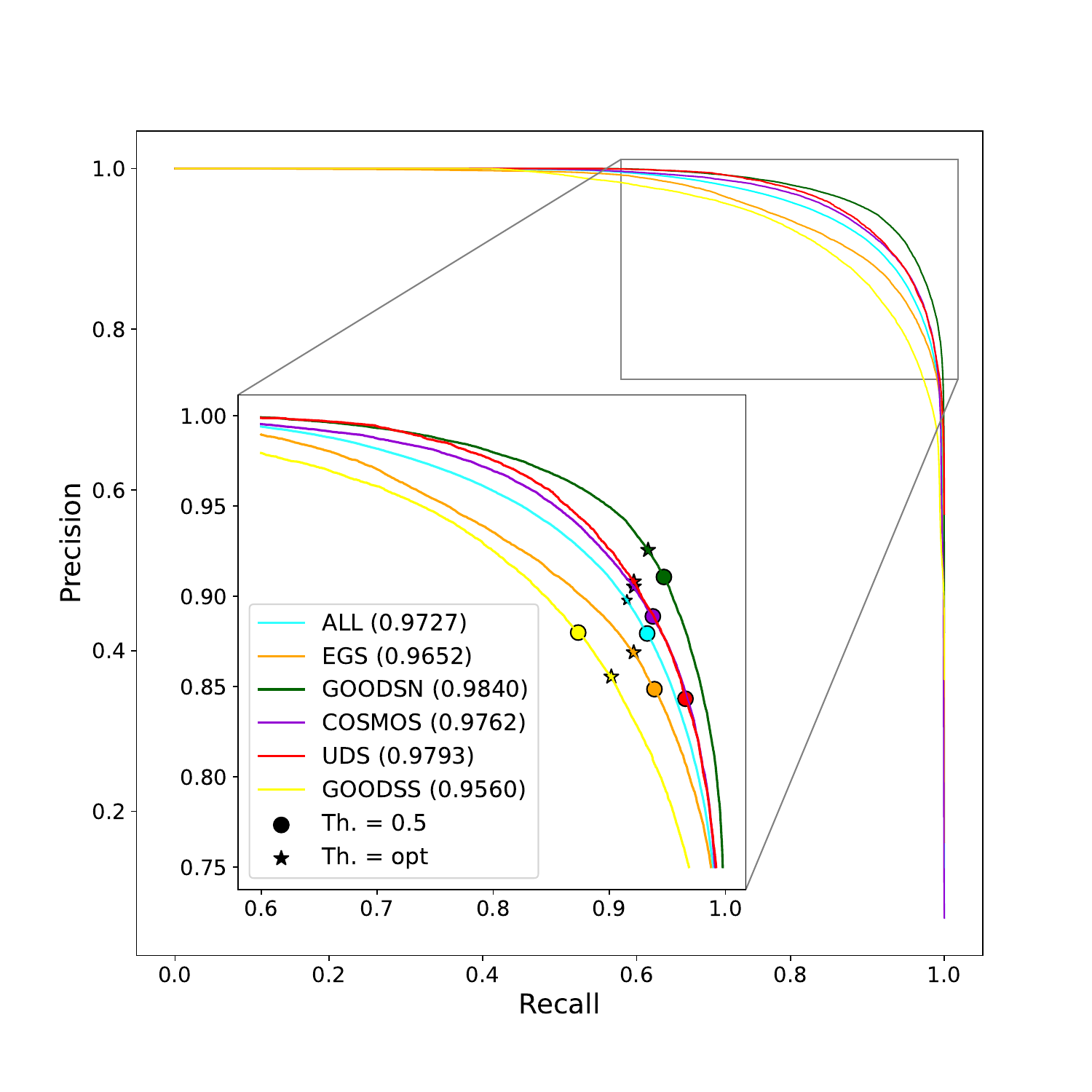}
        \caption{}
        \label{BestModel2-2}
    \end{subfigure}
    \begin{subfigure}{0.49\textwidth}
        \centering
        \includegraphics[width=\linewidth]{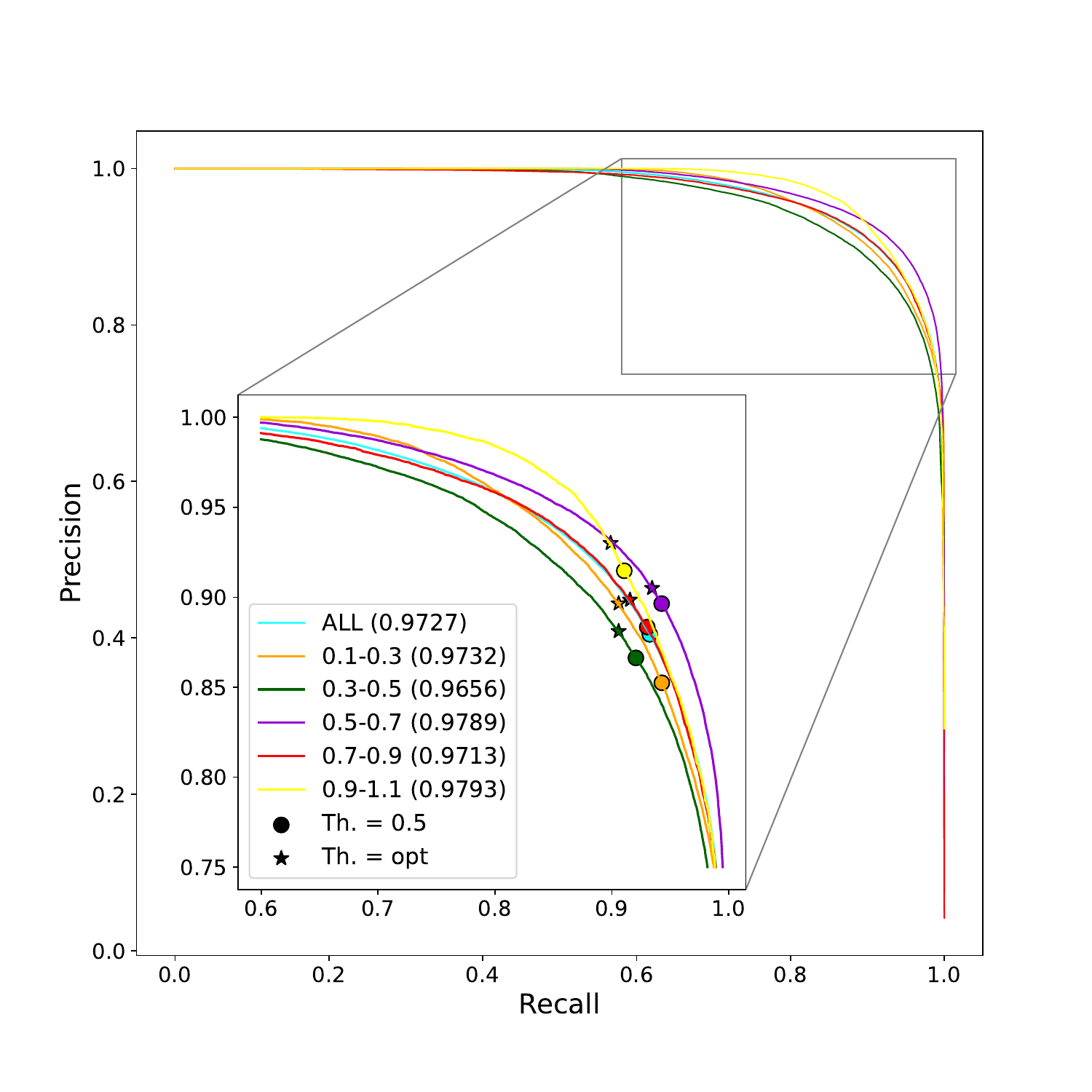}
        \caption{}
        \label{BestModel2-3}
    \end{subfigure}
    \begin{subfigure}{0.49\textwidth}
        \centering
        \includegraphics[width=\linewidth]{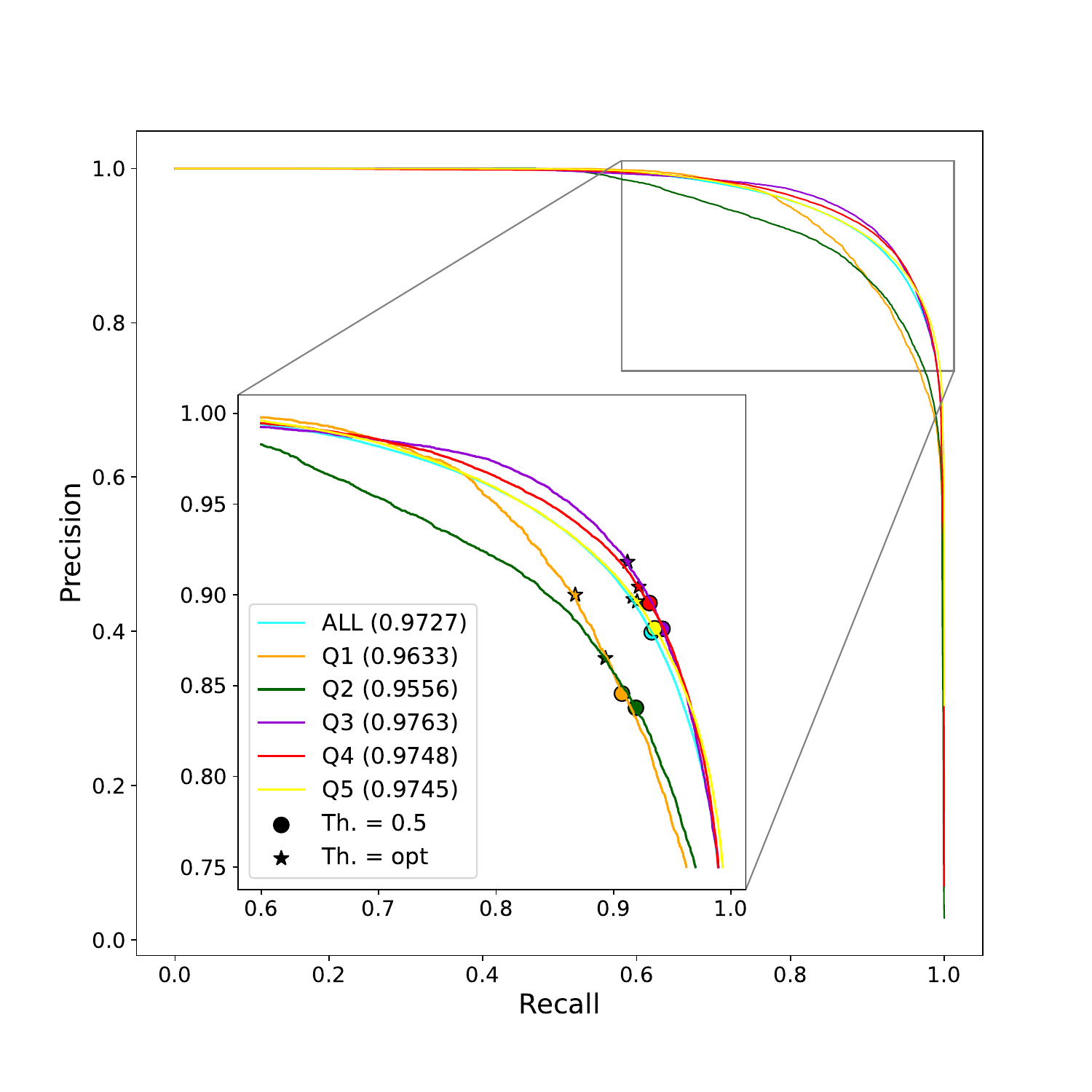}
        \caption{}
        \label{BestModel2-4}
    \end{subfigure}
    \caption{Precision-recall curves, along with the area under the curves (AUC), over the test set by galaxy morphology (a), galaxy field (b), redshift (c), and size (d). The categories for each plot are the same as in Figure \ref{BestModel1}. The dot symbol represents the location of the threshold 0.5 (the one used for binary classification). The star symbol shows the optimal threshold that maximizes the dice index.}
\label{BestModel2}
\end{figure*}

This section shows the results obtained by the best performing model, that is, the one using a DenseNet-201 neural network as encoder and trained with  visual astronomical augmentations. We isolated the performance of the model in terms of different characteristics into which the test sample can be grouped. Specifically, these characteristics are the galaxy morphology ({\sc disk}, {\sc diskirr,} and {\sc disksph}), the exposure field ({\sc egs}, {\sc goodsn}, {\sc cosmos}, {\sc uds}, and {\sc goodss}), the redshift and the size. Figure \ref{BestModel1} shows the results in terms of dice, precision and recall estimators. Figure \ref{BestModel2} shows the precision-recall (P-R) curves. We did not use ROC curves due to the imbalance of our data. Each P-R curve includes both the point where the 0.5 threshold is located (the one used to give a binary classification between galaxy and noise) and the point where the optimal dice index is reached. The results show that the morphological type for which the worst performance is obtained is {\sc disksph}, as expected given of the fuzzier nature of these objects (Figure \ref{BestModel1-1}). Ultimately, the {\sc disk} and {\sc diskirr} morphologies perform best. Differences between precision and recall are much more marked for the {\sc diskirr} morphology than for the {\sc disk}. This indicates that the estimated truncations are much more balanced in the latter case, leading to a higher number of false positives for {\sc diskirr} galaxies. In any case, the results obtained on {\sc diskirr} galaxies are highly valuable. The reason is that this morphology is characterized by the presence of asymmetric features \citep{Huertas_Company_2015}, which makes the precise estimation of its edges more challenging than for {\sc disk}-type galaxies. Although the dice index for the 0.5 threshold is higher for the {\sc diskirr} morphology, the quality over the whole spectrum of thresholds is higher for inferences of {\sc disk} galaxies, as can be seen in Figure \ref{BestModel2-1}. For all morphologies, all thresholds that induce an optimal dice index are greater than 0.5. This may indicate that the model is overconfident, and it is necessary for the network to have a high certainty that a pixel belongs to the galaxy class.

\begin{figure*}[t]
    \centering
    \begin{subfigure}{.49\textwidth}
        \centering
        \includegraphics[width=\linewidth]{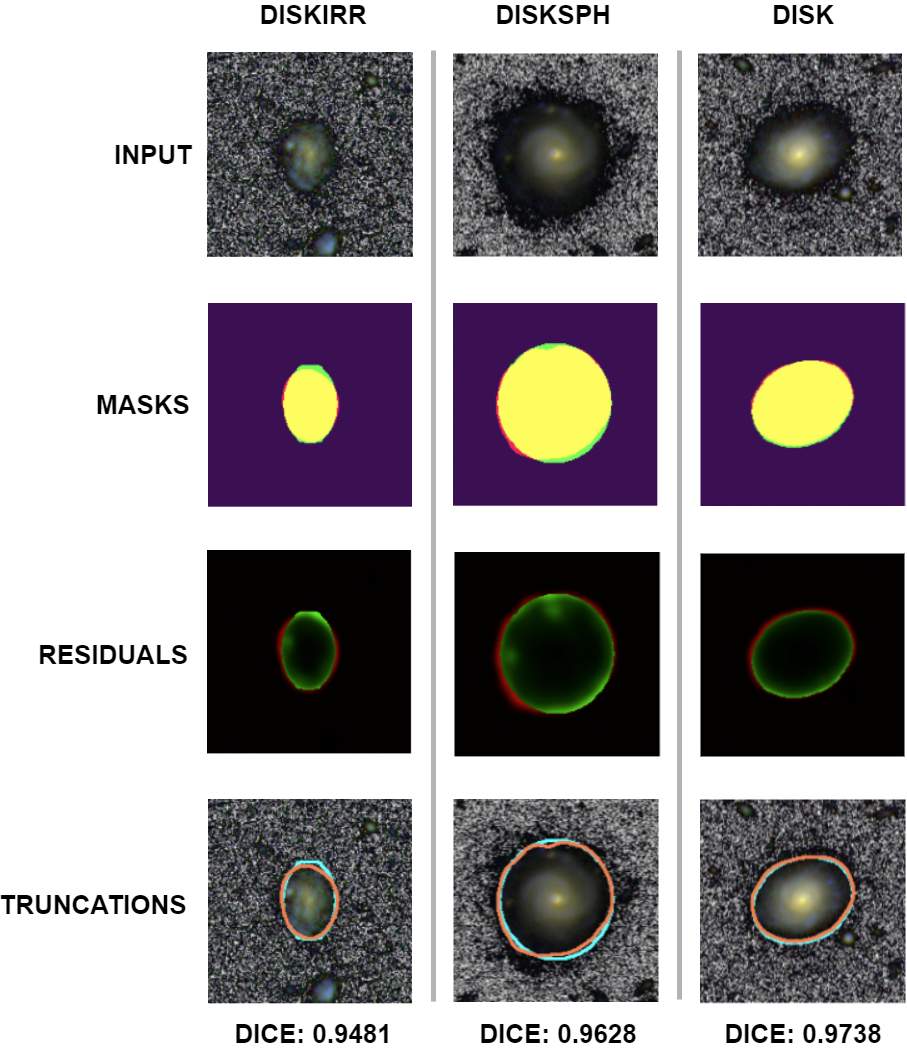}
        \caption{}
        \label{Fig: BestModelBestAndWorstPredictions-best}
    \end{subfigure}%
    \begin{subfigure}{0.49\textwidth}
        \centering
        \includegraphics[width=\linewidth]{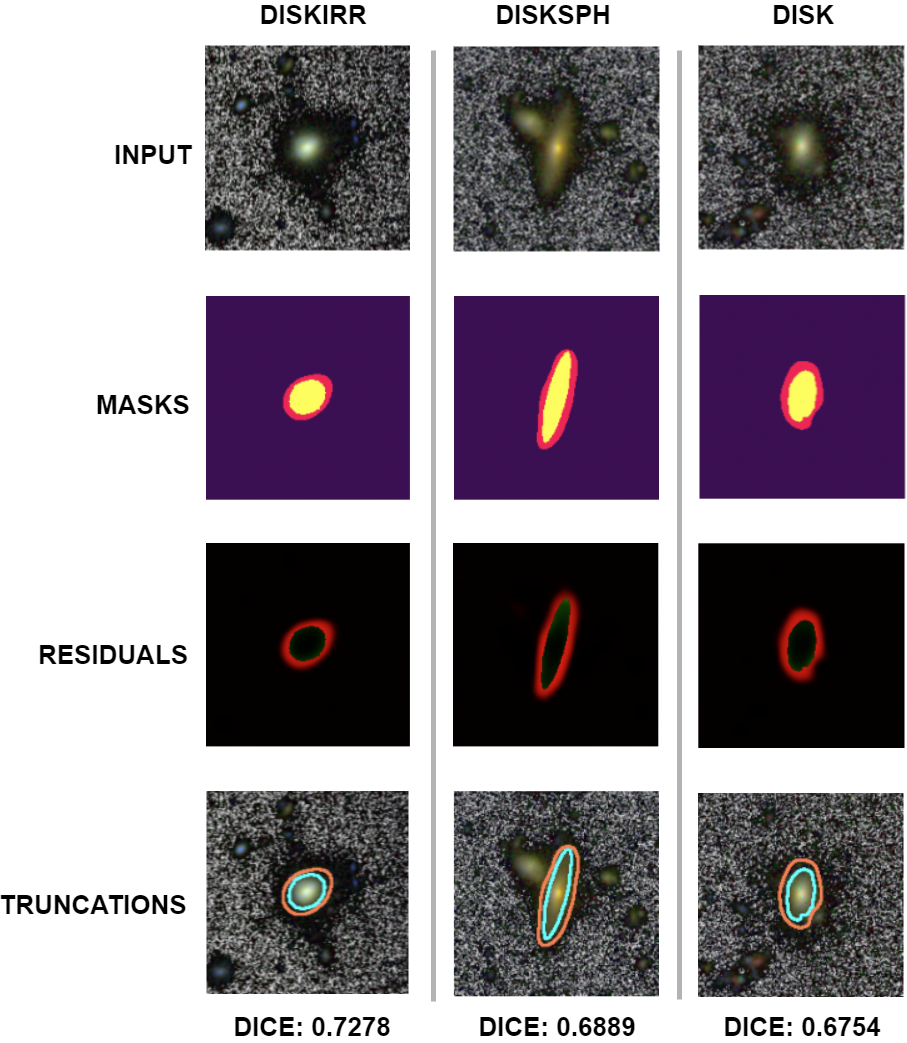}
        \caption{}
        \label{Fig: BestModelBestAndWorstPredictions-worse}
    \end{subfigure}
    \caption{Greatest (a) and least (b) accurate inferences by morphology for the best-performing model on the test dataset. First row shows the images of the galaxies. Second row displays the overlap of the predicted and real masks, where yellow, red and green zones represent true positives, false positives, and false negatives, respectively. The third row shows the residuals, where reddish and greenish tones represent negative and positive residuals, and the value (from light to dark) indicates their magnitude (from higher to lower). The last row shows the predicted and real truncations superimposed on the images.}
\label{Fig: BestModelBestAndWorstPredictions}
\end{figure*}

Assessing the metrics by field of exposure (Figure \ref{BestModel1-2}), the performance is very similar for all of them except for the {\sc goodss} field where we retrieve the worst values. The images in this field are, on average, the deepest (i.e., with the least noise) that we are analyzing. Therefore, our network better discerns the central galaxy from any other neighbor, and so our network becomes more conservative in detecting pixels belonging to the central body. This leads to a smaller recall, which, in turn, produces a reduction in the dice index. Looking at the P-R curves (Figure \ref{BestModel2-2}), it has been observed that the threshold inducing an optimal dice coefficient is higher than 0.5 for {\sc egs}, {\sc goodsn}, {\sc cosmos,} and {\sc uds}. However, in the case of {\sc goodss}, this threshold is below the threshold used for binary classification. This result suggests that while the model exhibits a high level of confidence, with respect to the first fields it is cautious when categorizing pixels as galaxies in the {\sc goodss} field. This may be another reason that explains the worse behavior of the model on the {\sc goodss} field than on the rest of the fields.
 
There is no marked variation in the quality of the estimated truncations for the different redshift values (Figure \ref{BestModel1-3}). The dice index does not decrease uniformly as galaxies become more distant, which is reasonable considering the high quality of the whole data sample for all redshifts. Nevertheless, it can be seen that the trends in precision and recall are opposite. While the former increases along with the redshift, the latter gradually decreases. This means that as galaxies become more distant, the DL model infers masks with fewer false positives and more false negatives, namely, the truncations are tighter, but some areas that are actually galaxies are not detected as such. To explain this, it should be noted that the higher the redshift, the higher the effect of the cosmological dimming decreasing the contrast between galaxy and background noise pixels, thereby making the 2D galaxy surface brightness profiles appear fainter. Consequently, the number of false positive pixels is progressively smaller and thus increasing the precision. The P-R curves and their inner area also reflect the same results for the different redshift levels (Figure \ref{BestModel2-3}). Similar to what was observed for the different morphologies, all thresholds inducing an optimal dice index are greater than 0.5, suggesting some overconfidence.

Finally, when the results are analyzed by galaxy size, it can be found that the performance varies considerably (Figure \ref{BestModel1-4}). The pattern seems to indicate that, on average, the dice index improves with increasing galaxy size. Furthermore, while the recall evolution appears to be slightly more stable, the main improvement lies in the precision. One of the possible reasons may be that as a galaxy gets larger, it is easier for the DL model not to make false positive errors, which leads to an increase in the precision. For very large galaxies (Q5) the results worsen. This may be because in such galaxies, the amount of noise that appears in the stamp is very small, with most of the image occupied by the galaxy and the galaxy-noise transition. As a result, it is more difficult for the model to locate the truncation and the performance worsens. The information provided by the P-R curves aligns with this pattern, as galaxies of larger sizes consistently yield to better results when using different thresholds (Figure \ref{BestModel2-4}). Again, all thresholds inducing an optimal dice index are greater than 0.5, reflecting the ease of the model to be very confident in its predictions about the galaxy class.

Figure \ref{Fig: BestModelBestAndWorstPredictions} shows the most and least accurate inferences by galaxy morphology. From the first row to the last one, it can be appreciated the images, the superimposed real and predicted masks, the residuals and the superimposed real and predicted truncations. For the masks, true positives, false positives and false negatives are represented by yellowish, reddish, and greenish tones, respectively. The information provided by the residuals is richer. Reddish tones represent those areas where the real mask is 0, greenish tones represent those areas where the real mask is 1 and black tones indicate residuals very close to 0, that is,\ there are almost identical behaviors seen between the predictions and labels. Analyzing the best cases (Figure \ref{Fig: BestModelBestAndWorstPredictions-best}) it can be seen that for the three types of morphologies, the predictions reveal a high degree of similarity with respect to the ground truths (dice $> 0.94$). Regarding the masks, reddish and greenish tones are barely visible, leading to very high values of both precision and recall. Concerning the residuals, it can be seen that the largest ones (less dark colors) appear in the boundary regions of the galaxies, that is, the zones close to the truncation. This event reflects that the model has more doubts about the category of these pixels, which is a desirable behavior. For the worst cases (Figure \ref{Fig: BestModelBestAndWorstPredictions-worse}), the neural network infers the shape of the truncation remarkably well. However, the penalty suffered in the dice index is caused by an optimistic estimation of the limits. This can be easily seen by analyzing both the overlap of the masks and the residuals. For the former, reddish tones predominate around the yellowish areas, causing an increase in the number of false positives and penalizing the precision. For the latter, residuals in bright red clearly indicate an overconfidence of the model about the presence of galaxy, estimating much larger truncations than those reflected in the masks.

Lastly, we trained this best performance model with the final masks in \citet{buitrago2023strong} as explained toward the end of Section \ref{AA}. The numbers are very similar to those already quoted in the paper: the mean dice is 0.8933, the mean precision is 0.8685, and the mean recall is 0.9411. We also conducted this experiment with other architectures and AAs. The outcome of all these tests is that there is not a substantial change by using or not these new labels, reassuring us of the stability of our results.

\subsection{Prioritizing boundary proximity: A close look at edge-aware loss functions}
\label{LossBundaryProximity}

The DL models presented so far have been optimized using a loss function that equally weights all pixels belonging to the same class ($\omega_g = 0.925$ for galaxy class and $\omega_n = 0.075$ for noise class). Thus, a pixel located at the center of the galaxy will have the same importance as a pixel placed just inside the truncation. Similarly, a noise pixel that is far away from the central body will be worth the same as a noise pixel just outside the truncation. This situation might potentially induce the network to minimize the loss by focusing its attention on those areas that are easy to learn, so that precise behavior is not necessary in the most doubtful areas to categorize (those close to the truncations). In order to overcome this situation, in this section we explore the performance of some loss functions that aim to ensure that DL models give more importance to those regions that are close to the boundaries of the galaxies.

\begin{figure*}[ht]
        \centering
        \includegraphics[width=1\linewidth]{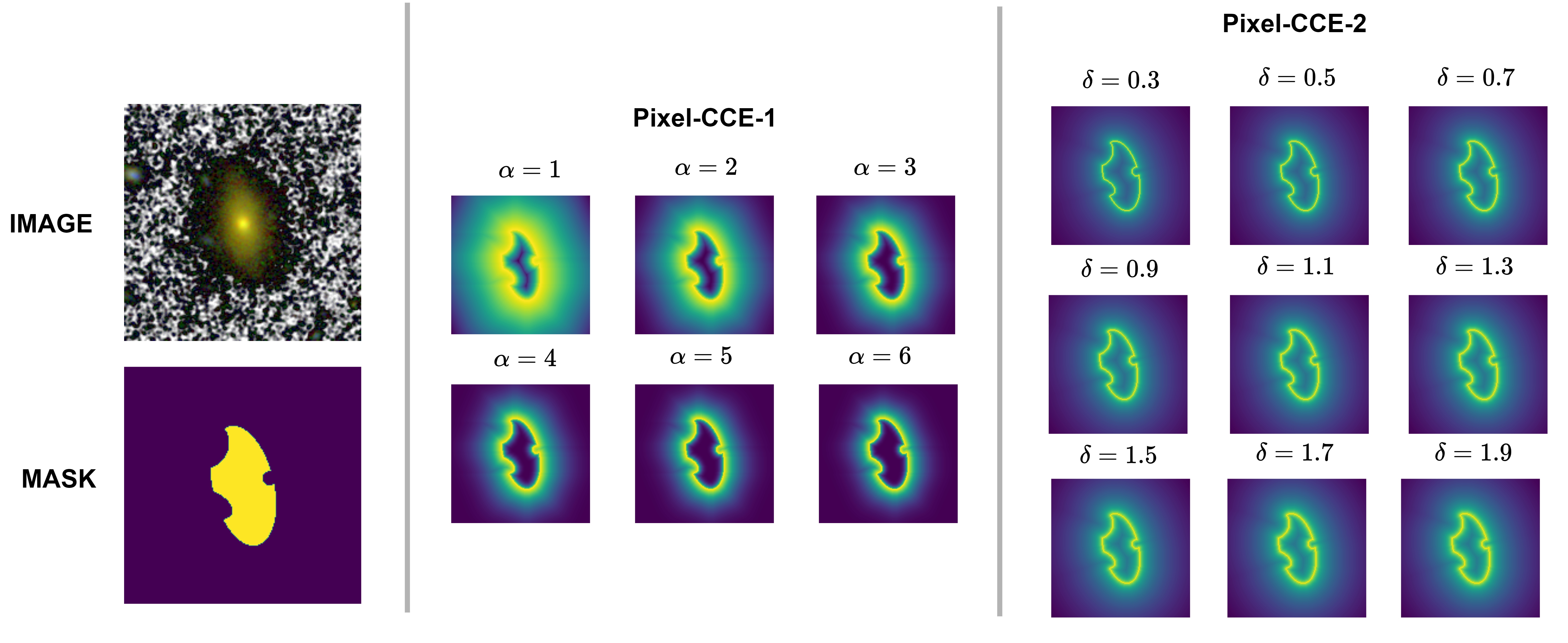}
        \caption{IJH image with its associated mask (left) and different weight matrices using the $\text{pixel-CCE}_1$ (center) and $\text{pixel-CCE}_2$ (right) losses.}
        \label{Fig: WeightsByDeltaAndAlpha}
\end{figure*}

We propose to use a variant of the CCE loss, where the weighting is performed per pixel rather than per class. Two different formulas are suggested for assigning the weight of each pixel. The first one ($\text{Pixel-CCE}_1$ from now on) is in Equation \ref{EcuacionWeightsCCE-1}, where $d((i,j), c)$ represents the Euclidean distance from the pixel $(i,j)$ to the point $c$, while $\mathcal{C}$ represents the set of all pixels that make up the truncation contour and $\alpha$ acts as a weighting factor. This factor marks the weight intensity in the neighboring areas of the truncation compared to the more distant zones, as can be seen in (Figure \ref{Fig: WeightsByDeltaAndAlpha}). Weights belonging to the internal and external parts of the truncation are scaled independently.

\begin{equation}
W_{i,j} = 1 - D(i,j)^\alpha \text{, where } D(i,j) = \min_{\forall c \in \mathcal{C}} d((i,j), c)
\label{EcuacionWeightsCCE-1}
\end{equation}

The second one ($\text{Pixel-CCE}_2$ hereafter) is in Equation \ref{EcuacionWeightsCCE-2}. The terms $d((i,j), c)$ and $\mathcal{C}$ have the same meaning as in the previous equation, while $\delta$ acts as an offset. That offset determines how broad the truncation-surrounding area that is highly weighted is (Figure \ref{Fig: WeightsByDeltaAndAlpha}).

\begin{equation}
W_{i,j} = \log{\frac{1}{\min_{\forall c \in \mathcal{C}} d((i,j), c) + \delta}}
\label{EcuacionWeightsCCE-2}
\end{equation}

\begin{figure*}[ht]
\centering
\begin{subfigure}[b]{1\textwidth}
   \includegraphics[width=1\linewidth]{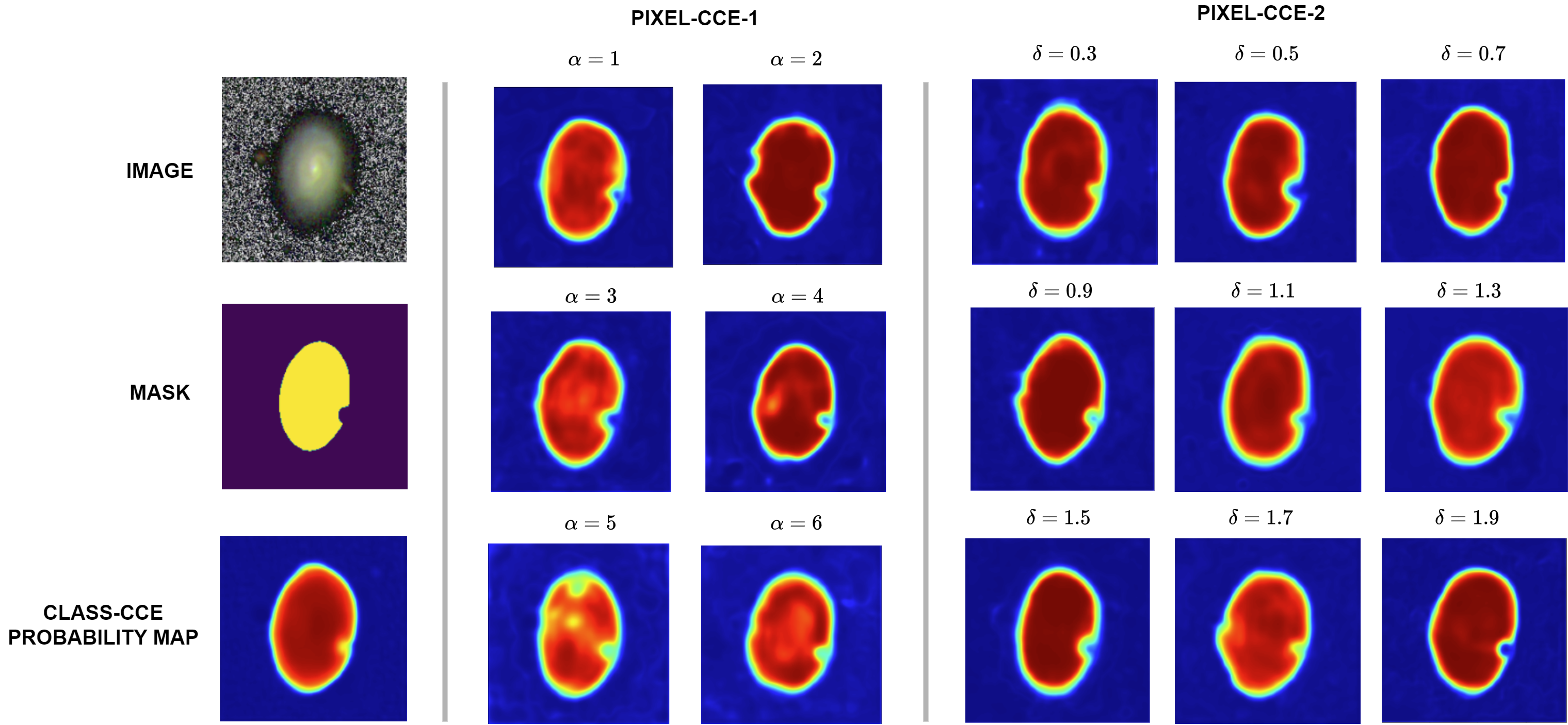}
   \caption{}
   \label{Fig: ProbabilityMaps-a} 
\end{subfigure}

\begin{subfigure}[b]{1\textwidth}
   \includegraphics[width=1\linewidth]{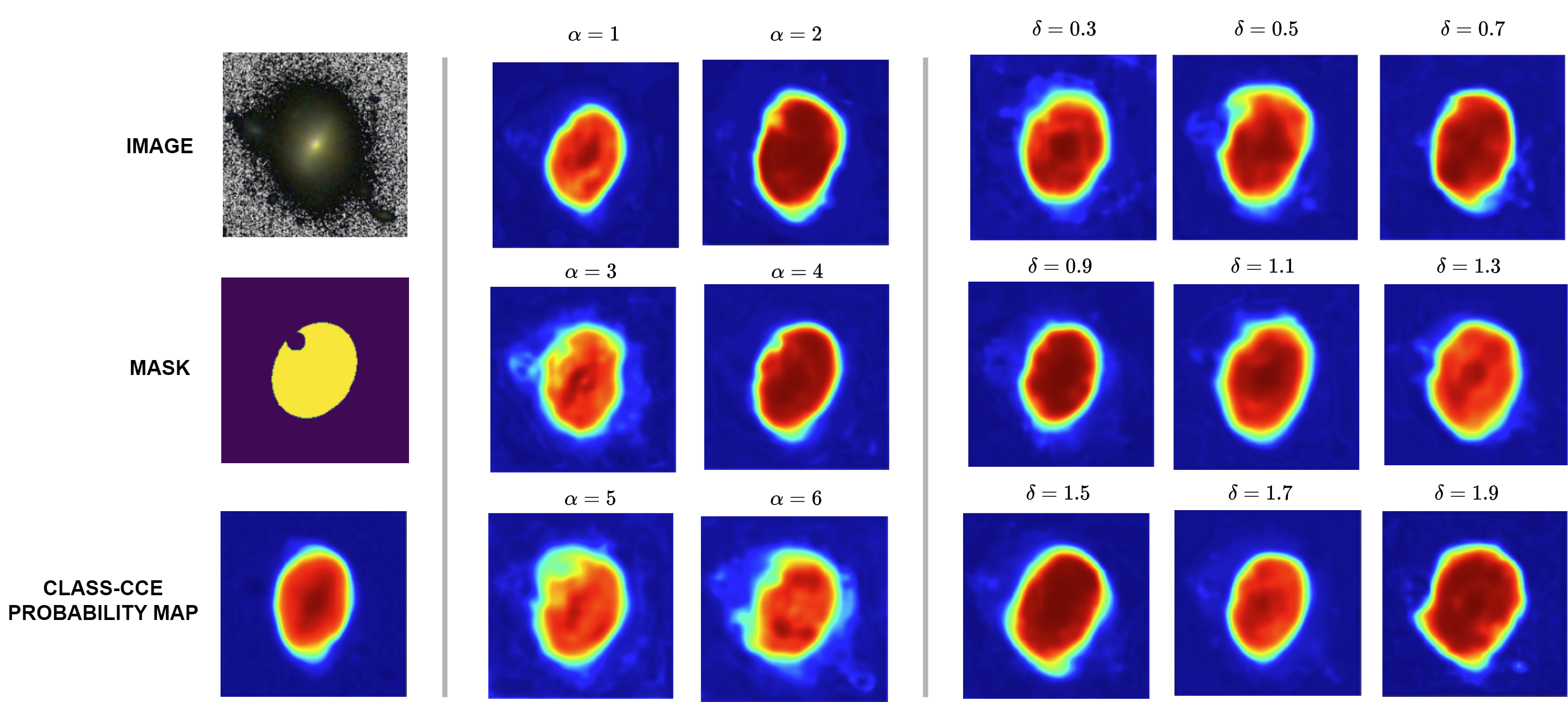}
   \caption{}
   \label{Fig: ProbabilityMaps-b}
\end{subfigure}

\caption{Probability maps obtained by class-CCE, $\text{pixel-CCE}_1$ and $\text{pixel-CCE}_2$ losses for two galaxies (a, b) of the test set. Dark red tones show confidence values close to 1 (high probability of galaxy), while dark blue tones reflect values close to 0 (high probability of noise).}
\label{Fig: ProbabilityMaps}
\end{figure*}

In order to assess the performance of the presented loss functions, we used the configuration that yielded the best-performing model (DenseNet-201 encoder with visual AAs). We trained the model using different configurations for both pixel CCE approaches. In the first case, we launched six experiments, varying the values of the hyperparameter $\alpha$. In the second case, we ran nine experiments, using different values for the hyperparameter $\delta$. The results obtained on the test set can be seen in Table \ref{Pixel-CCE-Hausdorff-Results}. In the case of $\text{pixel-CCE}_1$, none of the six configurations obtained a better dice index than the obtained using class-CCE. For $\text{pixel-CCE}_2$, two of the nine configurations were able to improve the results. In both cases, and for all configurations tested, no clear pattern is observed indicating that these functions are a better alternative to class-CCE. When the $\text{pixel-CCE}_1$ $\alpha$ hyperparameter increases, the results seem to worsen markedly. This suggests that excessive weighting of the zones adjacent to the truncation, in relation to the rest of the parts, leads to a decrease in performance. On the other side, when the $\text{pixel-CCE}_2$ $\delta$ hyperparameter is varied, the results do not appear to follow a linear trend. The only two configurations that lead to better results than class-CCE are achieved with $\delta = 0.9$ and $\delta = 1.7$. However, the intermediate values between these deltas show an oscillating and irregular behavior.

Although the average performance of both losses does not seem to improve the behavior of class-CCL, sometimes their use may be worthwhile. In Figure \ref{Fig: ProbabilityMaps}, we illustrate a significant example of the difference between the different loss functions. We chose two representative galaxies from the test dataset in which differences between the approaches can be seen. It can be observed the probability maps obtained by class-CCE and by all the tested $\text{pixel-CCE}_1$ and $\text{pixel-CCE}_2$ variants. In the first case (Figure \ref{Fig: ProbabilityMaps-a}), we can notice the presence of a foreign body on the right side of the galaxy. In the respective mask, it can be seen the impact this object has on the truncation. The class-CCE configuration reflects some doubt about whether that zone should be categorized as a main galaxy or as noise (yellowish tones). However, several $\text{pixel-CCE}_1$ ($\alpha = 1, 2, 3, 4$) and $\text{pixel-CCE}_2$ ($\delta = 0.5, 0.7, 1.1, 1.3, 1.5, 1.9$) configurations reflect more clearly the notion that this zone is not a central galaxy and it corresponds to noise (blue-greenish and bluish colors). Furthermore, although it does not appear to affect the truncation since it is more distant, the galaxy also has a neighboring object on its left side. The class-CCE configuration interprets this object as an obvious area of noise (very dark blue tone), which does not impact the shape of the truncation. However, certain configurations (primarily $\text{pixel-CCE}_1$ with $\alpha=2$, and to a lesser extent $\text{pixel-CCE}_1$ with $\alpha=3,5$ and $\text{pixel-CCE}_2$ with $\delta=0.7, 0.9, 1.7$) seem to be more sensitive to this object, reflecting lower probabilities of belonging to the central galaxy and thus affecting the shape of the truncation. This behavior, which, in terms of the dice coefficient, leads to a decrease in performance (due to the absence of this object in the mask), may be more interesting and reliable when estimating truncations in images with numerous clustered bodies. In the second case (Figure \ref{Fig: ProbabilityMaps-b}), the galaxy also has a neighboring object, this time in the upper-left part.  This object affects its boundaries and, consequently, the shape of its truncation. Once again, the probability map provided by Class-CCE is uniform in that area and is not affected by the presence of the object. However, several configurations of the edge-aware alternatives (primarily $\text{pixel-CCE}_1$ with $\alpha=2,3,4,5,6$ and $\text{pixel-CCE}_2$ with $\delta=0.5$) detect the presence of this object. Consequently, they give much lower probabilities of galaxy membership (greenish tones) to the pixels in that area. In a general sense (for both galaxies), it can be observed that for the majority of configurations, edge-aware alternatives produce probability maps that are less pronounced. This can be verified by observing fewer sharp transitions between intense reddish areas and dark blue areas. More uncertainty may be beneficial in situations where a range of predictions needs to be considered, ranging from scenarios where high confidence in the pixels being galaxies is required (resulting in smaller truncations) to scenarios where greater flexibility in dealing with model false positives is allowed (resulting in more lenient truncations).

\begin{table}[t]
    \centering
      \begin{tabular}{| c | c | c | c | c |}
        \hline\hline
        \textbf{Loss} & \textbf{Config} & \textbf{Dice} & \textbf{Precision} & \textbf{Recall} \\
        \hline\hline
        Class-CCE & & \cellcolor[RGB]{244, 137,  97} \textcolor{white}{0.8969} & \cellcolor[RGB]{107, 174, 214} \textcolor{white}{0.8863} & \cellcolor[RGB]{232, 246, 227} 0.9274 \\
        \Xhline{4\arrayrulewidth}
        \multirow{6}{*}{$\text{Pixel-CCE}_1$} & $\alpha = 1$ & \cellcolor[RGB]{25,  16,  42} \textcolor{white}{0.8819} & \cellcolor[RGB]{168, 206, 228} \textcolor{white}{0.8939} & \cellcolor[RGB]{0,  74,  29} \textcolor{white}{0.8980} \\
        \cline{2-5}
        & $\alpha = 2$ & \cellcolor[RGB]{226,  52,  65} \textcolor{white}{0.8930} & \cellcolor[RGB]{214, 229, 244} 0.9023 & \cellcolor[RGB]{27, 132,  63} \textcolor{white}{0.9048}\\
        \cline{2-5}
        & $\alpha = 3$ & \cellcolor[RGB]{198,  24,  81} \textcolor{white}{0.8911} & \cellcolor[RGB]{214, 229, 244} 0.9023 & \cellcolor[RGB]{10, 117,  51} \textcolor{white}{0.9028} \\
        \cline{2-5}
        & $\alpha = 4$ & \cellcolor[RGB]{152,  27,  91} \textcolor{white}{0.8887} & \cellcolor[RGB]{199, 219, 239} 0.8986 & \cellcolor[RGB]{15, 122,  55} \textcolor{white}{0.9034} \\
        \cline{2-5}
        & $\alpha = 5$ & \cellcolor[RGB]{81,  29,  76} \textcolor{white}{0.8850} & \cellcolor[RGB]{8,  67, 136} \textcolor{white}{0.8659} & \cellcolor[RGB]{240, 249, 237} 0.9293 \\
        \cline{2-5}
        & $\alpha = 6$ & \cellcolor[RGB]{2,   4,  25} \textcolor{white}{0.8804} & \cellcolor[RGB]{8,  48, 107} \textcolor{white}{0.8622} & \cellcolor[RGB]{221, 241, 215} 0.9256 \\
        \Xhline{4\arrayrulewidth}
        \multirow{9}{*}{$\text{Pixel-CCE}_2$} & $\delta = 0.3$ & \cellcolor[RGB]{34,  19,  48} \textcolor{white}{0.8824} & \cellcolor[RGB]{62, 142, 196} \textcolor{white}{0.8796} & \cellcolor[RGB]{135, 204, 133} \textcolor{white}{0.9159} \\
        \cline{2-5}
        & $\delta = 0.5$ & \cellcolor[RGB]{243, 120,  82} \textcolor{white}{0.8960} & \cellcolor[RGB]{181, 211, 233} \textcolor{white}{0.8957} & \cellcolor[RGB]{153, 213, 148} \textcolor{white}{0.9176} \\
        \cline{2-5}
        & $\delta = 0.7$ & \cellcolor[RGB]{244, 134,  94} \textcolor{white}{0.8967} & \cellcolor[RGB]{88, 161, 206} \textcolor{white}{0.8836} & \cellcolor[RGB]{247, 252, 245} 0.9309 \\
        \cline{2-5}
        & $\delta = 0.9$ & \cellcolor[RGB]{246, 175, 137} \textcolor{white}{0.8988} & \cellcolor[RGB]{131, 187, 219} \textcolor{white}{0.8891} & \cellcolor[RGB]{239, 249, 236} 0.9291 \\
        \cline{2-5}
        & $\delta = 1.1$ & \cellcolor[RGB]{179,  22,  87} \textcolor{white}{0.8901} & \cellcolor[RGB]{133, 188, 219} \textcolor{white}{0.8894} & \cellcolor[RGB]{123, 199, 124} \textcolor{white}{0.9148} \\
        \cline{2-5}
        & $\delta = 1.3$ & \cellcolor[RGB]{228,  55,  64} \textcolor{white}{0.8931} & \cellcolor[RGB]{247, 251, 255} 0.9103 & \cellcolor[RGB]{0,  68,  27} \textcolor{white}{0.8973} \\
        \cline{2-5}
        & $\delta = 1.5$ & \cellcolor[RGB]{222,  45,  68} \textcolor{white}{0.8926} & \cellcolor[RGB]{81, 156, 204} \textcolor{white}{0.8826} & \cellcolor[RGB]{222, 242, 216} 0.9257 \\
        \cline{2-5}
        & $\delta = 1.7$ & \cellcolor[RGB]{250, 234, 220} 0.9022 & \cellcolor[RGB]{239, 246, 252} 0.9086 & \cellcolor[RGB]{110, 193, 115} \textcolor{white}{0.9137} \\
        \cline{2-5}
        & $\delta = 1.9$ & \cellcolor[RGB]{138,  29,  91} \textcolor{white}{0.8880} & \cellcolor[RGB]{84, 158, 205} \textcolor{white}{0.8830} & \cellcolor[RGB]{149, 211, 145} \textcolor{white}{0.9172} \\
        \hline
      \end{tabular}
    \caption{Results over the test set using the edge-aware loss functions and the encoder DenseNet-201 with visual AAs. Color-coding corresponds to the one used in Table \ref{tab:ResultadosBase}.}
  \label{Pixel-CCE-Hausdorff-Results}
\end{table}

\subsection{Approaching human expert behavior via democratic ensemble learning}
\label{ensemble}

\begin{figure*}
        \centering
        \includegraphics[width=1\linewidth]{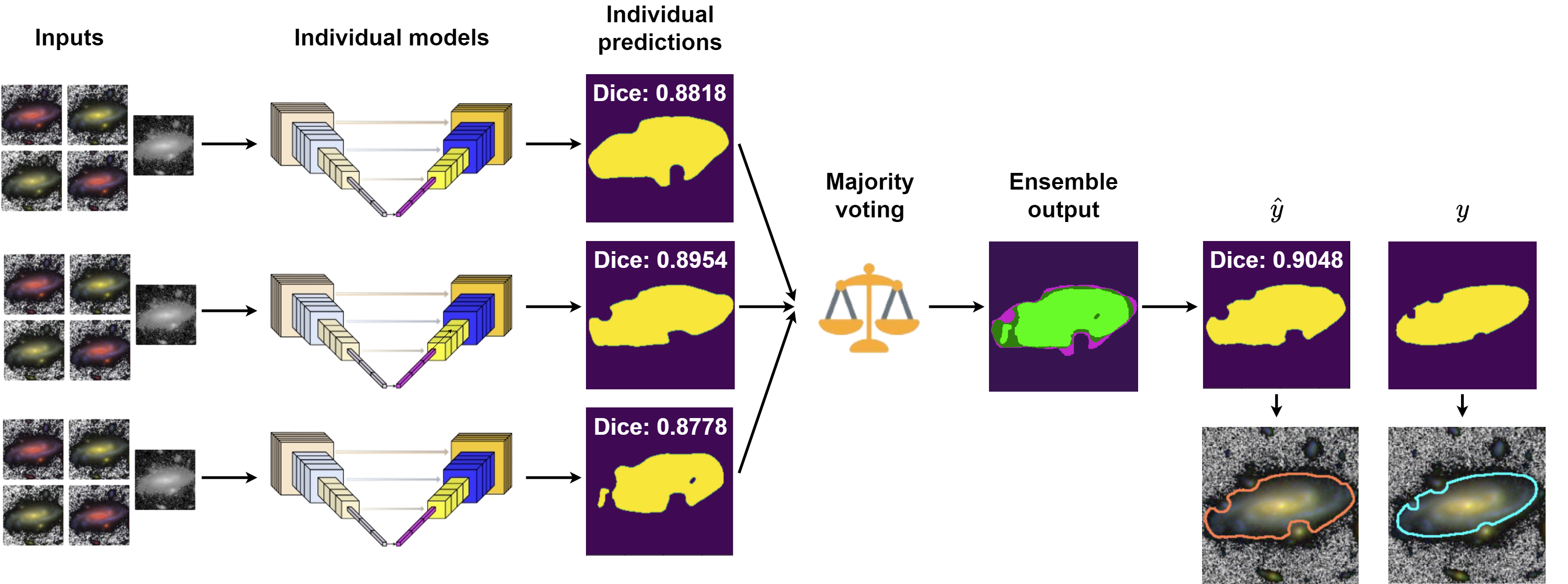}
        \caption{Designed ensemble architecture. Predictions from the individual models are combined using a majority vote criterion to provide a final output.}
        \label{Fig: EnsembleCombination}
\end{figure*}

In this section, the use of ensemble learning is proposed in order to improve the quality of the segmentation performed by DL models. Ensemble learning combines more than one base algorithm to obtain, through an aggregation mechanism, a unitary prediction. Intuitively, decision-making improves when diverse opinions are contrasted. Thus, the algorithms to be combined need to be diverse, that is, they need to make errors on different instances. The experimental grid, in which different AAs are combined with different encoders, results in the derivation of highly diverse models. Therefore, the 35 segmentation networks that have been presented throughout the work are used as the base algorithms. To keep the computation within reasonable limits, it was decided to combine the base models in groups of three. This gives rise to 6545 unique ensembles, which are combined by the democratic criterion of the majority vote (mode), as can be seen in Figure \ref{Fig: EnsembleCombination} We note that this way of aggregating predictions forces to take an odd number of base models to form the ensemble.

\begin{figure*}
    \centering
    \begin{subfigure}{0.49\textwidth}
        \centering
        \includegraphics[width=\linewidth]{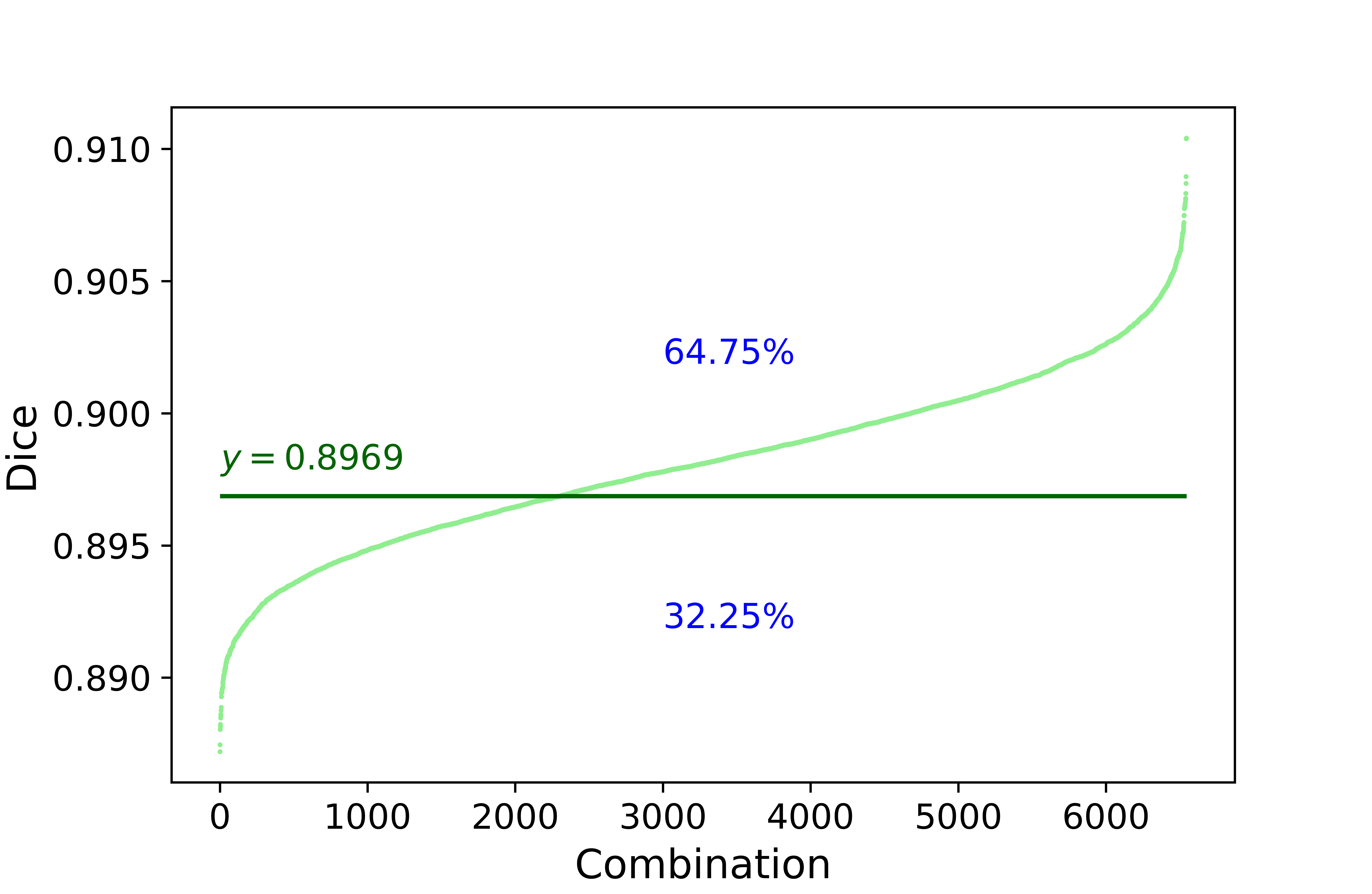}
        \caption{}
        \label{Fig:Ensemble-1}
    \end{subfigure}
    \begin{subfigure}{0.49\textwidth}
        \centering
        \includegraphics[width=\linewidth]{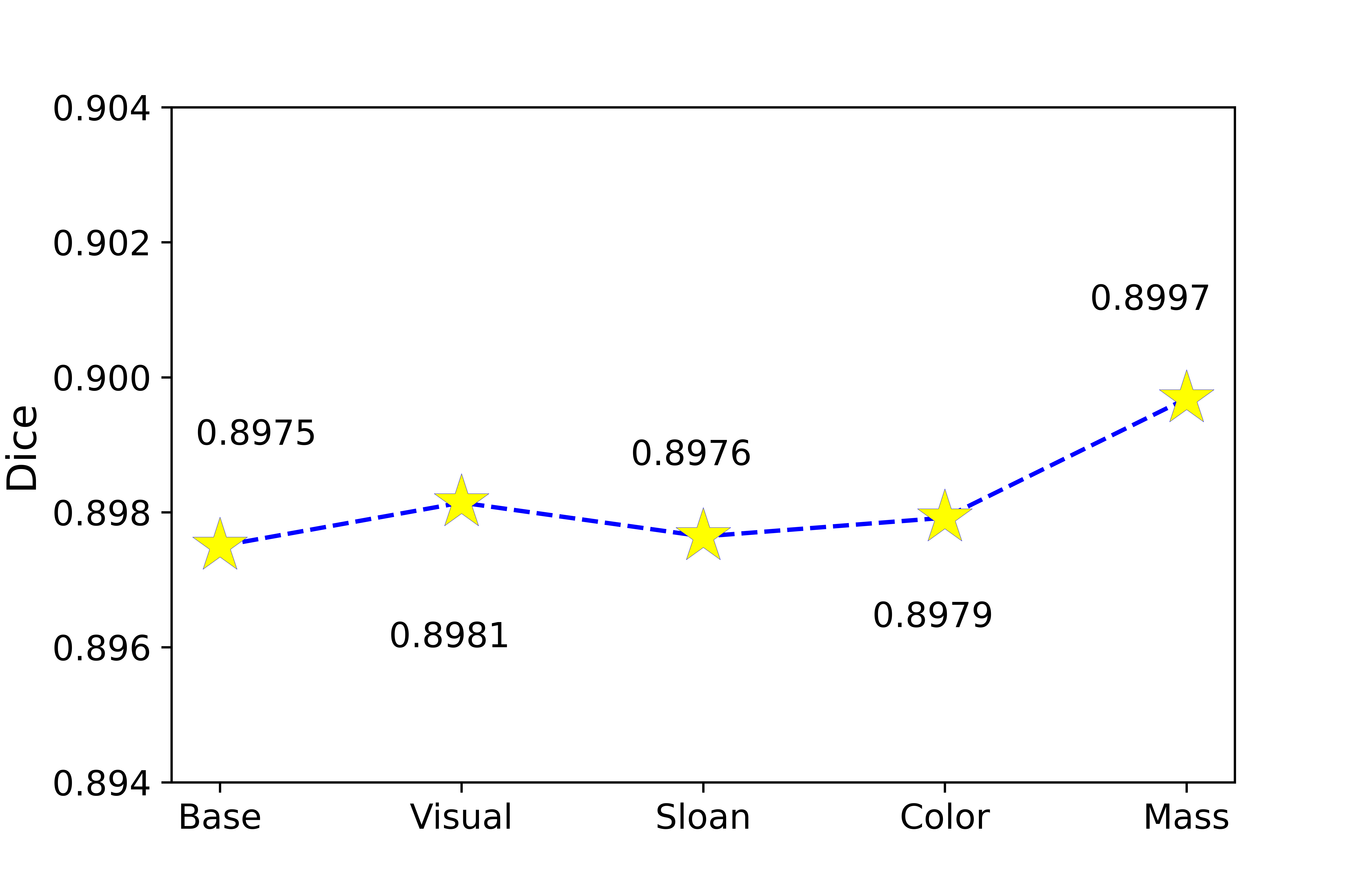}
        \caption{}
        \label{Fig:Ensemble-2}
    \end{subfigure}
    \caption{Metrics summarizing the behavior of the ensembles. The left plot (a) shows the sorted dice index values for all built ensembles. The right plot (b) displays, for each AA, the average dice index of all assemblies in which that AA participates at least one time.}
\label{Fig:Ensemble}
\end{figure*}

In the Figure \ref{Fig:Ensemble-1}, it can be seen how 64.75~\% of the constructed ensembles exceed the highest dice index obtained with any of the individual models (namely DenseNet-201 trained with visual AAs). These data show the effectiveness of the ensemble learning approach to approximate human behavior in finding the galaxy truncations. In fact, the approach can be linked to the creation of a committee of expert astronomers in which the opinion of the majority prevails. The best configuration found is made up of the models EfficinetNet-B6, ResNet-18, and DenseNet-161, all of them trained with mass-type AAs, reaching a dice index of 0.9104, a precision of 0.9016, and a recall of 0.9353. On the other hand, Figure \ref{Fig:Ensemble-2} shows that the AA that, on average, induces ensembles with a higher dice index is mass. The visual, color, and sloan type AAs (in that order) are placed later. At the end,  we have the ensembles consisting of at least one base model. In order to quantify the improvement caused by the participation of each AA in the ensemble, we order the results from lowest to highest (base $<$ sloan $<$ color $<$ visual $<$ mass) and compute the improvement ratio caused by each AA. This way, the sloan improvement over base is 1.1 per ten thousand, the color improvement over sloan is 3.3 per ten thousand, the visual improvement over color is 2.2 per ten thousand, and the mass improvement over visual is 1.8 per thousand. The improvement induced by the mass-type AA is one order of magnitude greater than any of the other improvements, which indicates that this type of AA is clearly the one that induces the best performance. These results are coherent with those shown in the Sections \ref{Base} and \ref{EffectsAA}.

\begin{figure*}[ht]
    \centering
    \begin{subfigure}{0.48\textwidth}
        \centering
        \includegraphics[width=\linewidth]{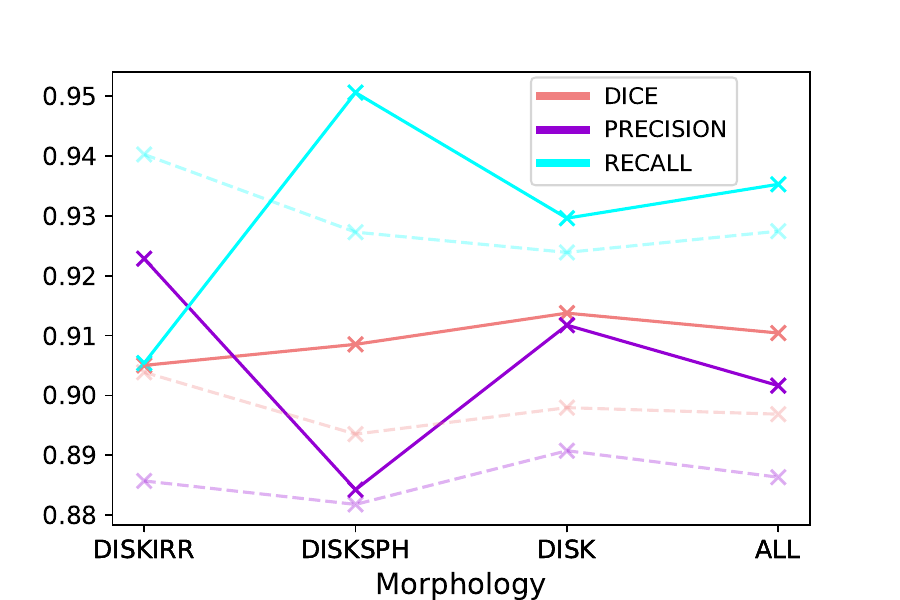}
        \caption{}
        \label{EnsembleByMorphology-1}
    \end{subfigure}
    \begin{subfigure}{0.48\textwidth}
        \centering
        \includegraphics[width=\linewidth]{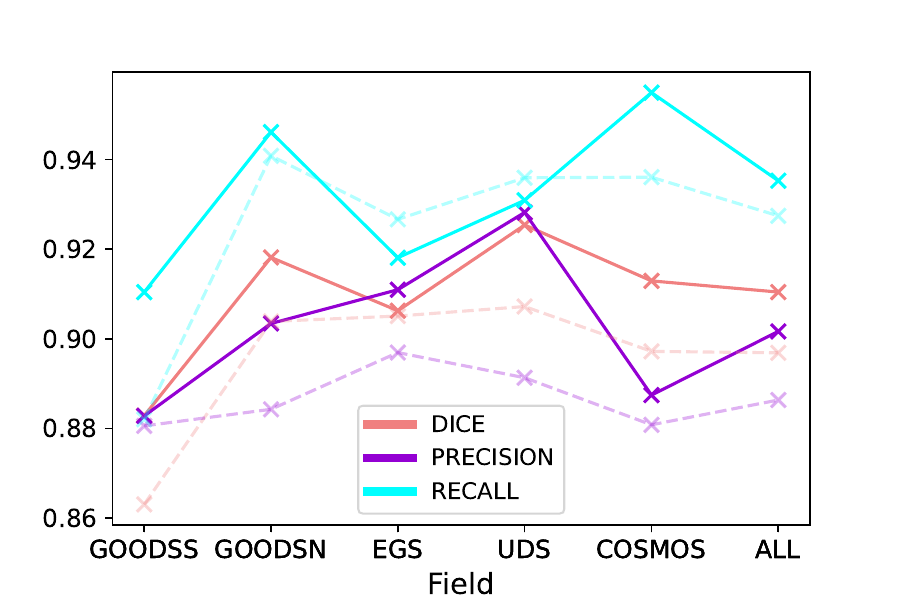}
        \caption{}
        \label{EnsembleByMorphology-2}
    \end{subfigure}

    \begin{subfigure}{0.48\textwidth}
        \centering
        \includegraphics[width=\linewidth]{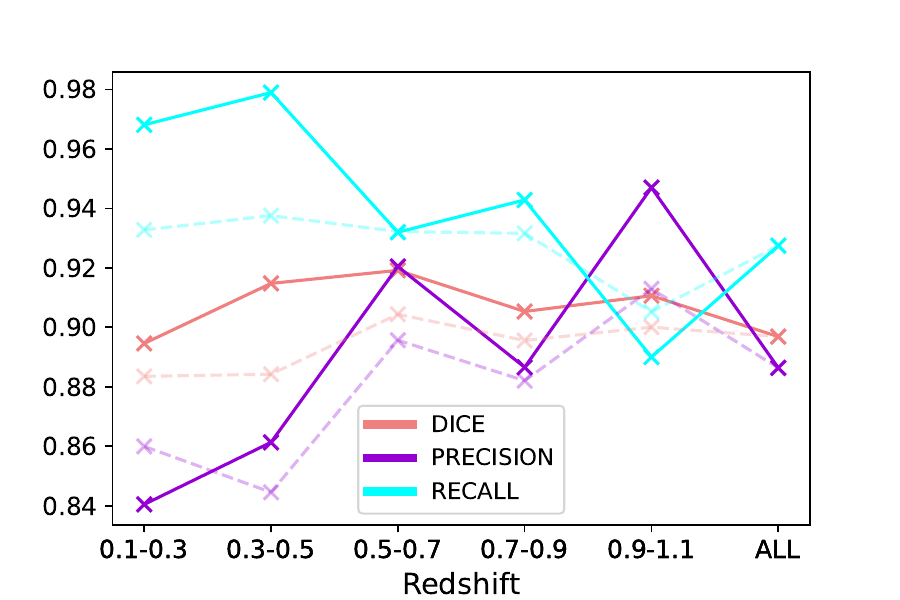}
        \caption{}
        \label{EnsembleByMorphology-3}
    \end{subfigure}
    \begin{subfigure}{0.48\textwidth}
        \centering
        \includegraphics[width=\linewidth]{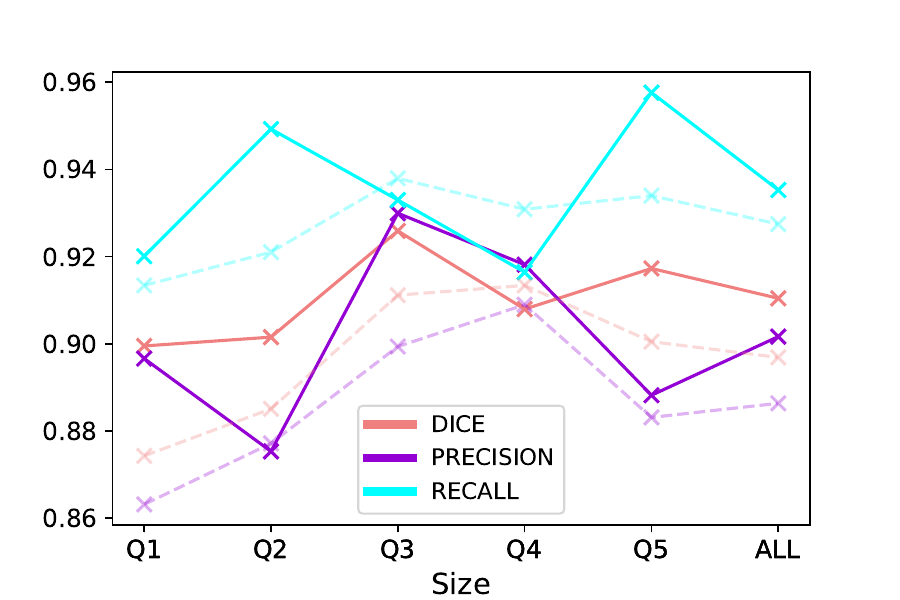}
        \caption{}
        \label{EnsembleByMorphology-4}
    \end{subfigure}
    \caption{Metrics over the test set by galaxy morphology (a), field of view (c), redshift (c), and size (d) for the best ensemble configuration. The categories for each plot are the same as in Figure \ref{BestModel1}. Dashed (more transparent) lines show the results for the best individual model.}
    \label{Fig:EnsembleByMorphology}
\end{figure*}

Analogously to the best individual model, Figure \ref{Fig:EnsembleByMorphology} shows the dice, precision, and recall estimators by galaxy morphology, field of view, redshift, and size for the best ensemble configuration. In order to simplify comparisons with the best individual model, its results have also been plotted in Figure \ref{Fig:EnsembleByMorphology} with dashed more-transparent lines. Analyzing the morphology, in terms of the dice index, {\sc disk} performs the best, followed by {\sc disksph} and {\sc diskirr,} respectively. It is worthwhile to analyze the oscillation of precision and recall for the different morphologies. In the case of {\sc disk} and {\sc disksph}, the recall is higher than the precision, however, for {\sc diskirr} the opposite happens. The inferred masks for {\sc diskirr} type galaxies show far fewer false positives than for the other morphologies, while the inferred masks for {\sc disk} and {\sc disksph} type galaxies show fewer false negatives. Comparing the results with those of the best single model, the main change occurs over {\sc diskirr} galaxies. While the individual model had a high recall and a lower precision, the ensemble model greatly increases the recall and decreases the precision. This means that the truncations offered by the ensemble for these types of galaxies are much more optimistic than those offered by the individual alternative. For {\sc disksph} and {\sc disk} type galaxies, the ensemble improves the results proportionally, improving both precision and recall. Regarding the field of view, {\sc uds} achieves the best results, as already found for the best individual model in Section \ref{BestModel}. Similarly, the worst performance is also obtained for the {\sc goodss} type galaxies. {\sc goodsn}, {\sc cosmos,} and {\sc egs} are in the mid-range, with a dice index that goes from best to worst, respectively. While for the {\sc goodsn}, {\sc cosmos,} and {\sc goodss} fields, the recall is much higher than the precision, for the {\sc egs} and {\sc uds} fields, the predictions are much more balanced in terms of false positives and negatives. Performance, as in the case of the best single model, shows signs of invariance to redshift. Again, the precision seems to increase as the galaxies become more distant, while the recall decreases. In terms of size, the conclusions are similar to the case of the best individual model. The best results are achieved for medium-size galaxies (in the third and fourth quintiles). For the smallest and largest galaxies, lower quality inferences are obtained.

In order to explore the performance of ensemble learning when combining more than three models, we have selected some relevant cases. The results can be seen in Table \ref{MoreEnsembles}. First, we built the ensembles consisting of all the models that have been trained with the same astronomic augmentation (Table \ref{MoreEnsembles-1}). Thus, for each AA, the results of the seven encoders were combined. The results are consistent with those obtained in Tables \ref{tab:ResultadosBase} and \ref{tab:ResultadosAA}. The best ensemble is the one that uses all seven encoders trained with mass-type AAs. The worst results were still obtained when no AA is used (base case). Between them, ranked from best to worst, we have:\ visual, color, and sloan AAs. Nevertheless, all five ensembles improve the results of the best individual model, showing the ability of this technique to enhance and complement the results of models that make diverse errors. We have also calculated the ensemble consisting of all available individual models (35). Although the result is excellent, achieving a dice of 0.9054 and surpassing the best individual model, it does not overcome the best ensemble formed by three models (dice of 0.9104). In order to explore more possibilities, we also constructed the ensembles formed by all the models using the same encoder (Table \ref{MoreEnsembles-2}). Thus, for each encoder, the five versions that have been trained using different AAs are combined. All ensembles, except the one formed by the EfficientNetB2 encoder, outperform the best single model. The best results are obtained by DenseNet encoder ensembles (DenseNet201 and DenseNet161), followed by the ResNet ones (ResNet18 and ResNet50) and, finally, by the EfficientNet ones (EfficientNetB1, EfficientNetB6, and EfficientNetB2).

Appendix \ref{app:appendix} shows the truncations predicted by the best combination of ensemble learning for the 100 galaxies in the test set.

\begin{table*}[t]
    \centering
    \begin{subfigure}[c]{.49\textwidth}
        \centering
        \begin{tabular}{| c | c | c | c |}
        \hline\hline
        \textbf{Combination} & \textbf{Dice} & \textbf{Precision} & \textbf{Recall} \\
        \hline\hline
        All base & \cellcolor[RGB]{2,   4,  25} \textcolor{white}{0.8973} & \cellcolor[RGB]{8,  48, 107} \textcolor{white}{0.8631} & \cellcolor[RGB]{247, 252, 245} 0.9537 \\
        \hline
        All visual & \cellcolor[RGB]{172,  23,  89} \textcolor{white}{0.9026} & \cellcolor[RGB]{191, 216, 236} 0.8857 & \cellcolor[RGB]{0,  68,  27} \textcolor{white}{0.9401} \\
        \hline
        All sloan & \cellcolor[RGB]{73,  28,  73} \textcolor{white}{0.8997} & \cellcolor[RGB]{57, 137, 193} \textcolor{white}{0.8737} & \cellcolor[RGB]{142, 208, 139} \textcolor{white}{0.9479} \\
        \hline
        All color & \cellcolor[RGB]{103,  30,  84} \textcolor{white}{0.9006} & \cellcolor[RGB]{112, 177, 215} \textcolor{white}{0.8790} & \cellcolor[RGB]{39, 144,  72} \textcolor{white}{0.9438} \\
        \hline
        All mass & \cellcolor[RGB]{250, 234, 220} 0.9097 & \cellcolor[RGB]{247, 251, 255} 0.8941 & \cellcolor[RGB]{12, 119,  52} \textcolor{white}{0.9424} \\
        \hline
        All & \cellcolor[RGB]{239,  89,  64} \textcolor{white}{0.9054} & \cellcolor[RGB]{155, 200, 224} \textcolor{white}{0.8823} & \cellcolor[RGB]{156, 214, 151} \textcolor{white}{0.9484} \\
        \hline
      \end{tabular}
        \caption{}
        \label{MoreEnsembles-1}
    \end{subfigure}%
    \begin{subfigure}[c]{.49\textwidth}
        \centering
      \begin{tabular}{| c | c | c | c |}
        \hline\hline
        \textbf{Combination} & \textbf{Dice} & \textbf{Precision} & \textbf{Recall} \\
        \hline\hline
        All ResNet18 & \cellcolor[RGB]{245, 154, 113} \textcolor{white}{0.9029} & \cellcolor[RGB]{143, 193, 221} \textcolor{white}{0.8823} & \cellcolor[RGB]{122, 198, 123} \textcolor{white}{0.9437} \\
        \hline
        All ResNet50 & \cellcolor[RGB]{227,  53,  65} \textcolor{white}{0.9008} & \cellcolor[RGB]{81, 156, 204} \textcolor{white}{0.8781} & \cellcolor[RGB]{132, 203, 131} \textcolor{white}{0.9441} \\
        \hline
        All EfficientNetB1 & \cellcolor[RGB]{137,  29,  90} \textcolor{white}{0.8985} & \cellcolor[RGB]{8,  48, 107} \textcolor{white}{0.8674} & \cellcolor[RGB]{247, 252, 245} 0.9515 \\
        \hline
        All EfficientNetB2 & \cellcolor[RGB]{2,   4,  25} \textcolor{white}{0.8951} & \cellcolor[RGB]{8,  51, 111} \textcolor{white}{0.8677} & \cellcolor[RGB]{185, 227, 178} \textcolor{white}{0.9467} \\
        \hline
        All EfficientNetB6 & \cellcolor[RGB]{81,  29,  76} \textcolor{white}{0.8972} & \cellcolor[RGB]{23, 100, 171} \textcolor{white}{0.8725} & \cellcolor[RGB]{140, 207, 138} \textcolor{white}{0.9445} \\
        \hline
        All DenseNet161 & \cellcolor[RGB]{250, 234, 220} 0.9049 & \cellcolor[RGB]{247, 251, 255} 0.8927 & \cellcolor[RGB]{0,  68,  27} \textcolor{white}{0.9353} \\
        \hline
        All DenseNet201 & \cellcolor[RGB]{246, 180, 142} \textcolor{white}{0.9035} & \cellcolor[RGB]{152, 199, 223} \textcolor{white}{0.8829} & \cellcolor[RGB]{110, 193, 115} \textcolor{white}{0.9432} \\
        \hline
      \end{tabular}
        \caption{}
        \label{MoreEnsembles-2}
    \end{subfigure}%
    \caption{Metrics over the test set obtained by the ensembles, grouping the individual models by AA (a) and by encoder (b). In Table \ref{MoreEnsembles-1}, each row collects the results of the ensemble formed by the seven models (one for each encoder) that have been trained with the corresponding AA. Last row shows the ensemble made up of the 35 trained models. In Table \ref{MoreEnsembles-2}, each row shows the results of the ensemble formed by the five models (one for each AA) that have been trained with the same encoder. Color coding corresponds to the one used in Table \ref{tab:ResultadosBase}.}
  \label{MoreEnsembles}
\end{table*}

\section{Conclusions}
\label{conc}
% Conclusions
Galaxies are thought to be diffuse objects without clear borders and, as a result, their nature impedes measurements of physically motivated galaxy sizes. However, LSB observations have shown that galaxy outskirts display sudden drops in their mass density profiles, which also display a counterpart in their surface brightness profiles. These features, which trace a perimeter much closer to our intuitive idea of the end of a galaxy, are linked with the radial location of the gas density threshold enabling star formation and actually produce size-related scaling relations with a much smaller scatter \citep{Trujillo20}. \citet{Chamba22} and \citet{buitrago2023strong} also identified them with the galaxy truncations detected from the seminal works by Piet van der Kruit \citep{vanderKruit1979}. Current ultra-deep observations and next-generation synoptic telescopes will reveal these features for millions of galaxies. Because of the numerous pieces of evidence needed to derive the positions of these edges, and considering the inherent pattern-recognition nature of this problem, using DL (in general) and CNNs (in particular) appears to be the only way forward to address these measurements for the swarm of data to come in the near future.

Based on the detections of these edges or truncations for a sample of 1052 massive (M$_{\rm stellar}$ $>$ 10$^{10}$ M$_{\odot}$) disk galaxies at z$_{\rm spec}$ $<$ 1.1 in the CANDELS fields \citep{buitrago2023strong}, we built a series of U-Nets that make use of convolutional neural networks as encoders. Our final objective is both to construct an algorithm able to detect efficiently these LSB features as well as understanding the best inputs to achieve our aims. This last point has to do with the fact that our neural networks do not only analyze the HST parent sample images, but they are also feed with ancillary data, namely, SDSS-restframe equivalent images, as well as color and mass images (see Section \ref{dataset}). In this way, we are providing the same information as the one given to humans when inferring the positions of these edges or truncations. We called these transformations of the original images ``astronomic augmentations," as they are similar to standard machine learning augmentations, but they are based on our astronomical knowledge. All average values of the dice coefficient improve in comparison with those of the base experiment when making use of such ``astronomic augmentations."

The several experiments that we carried out, namely, visual (HST RGB images), sloan (visual + SDSS bands and SDSS RGB images), color (visual + SDSS colors), and  mass (visual + mass images), display an inherent variability in their outcome metrics for the various U-Net architectures we utilize (see Section \ref{accuracy} and Table \ref{tab:ResultadosBase} and Table \ref{tab:ResultadosAA}). This ensures that our neural networks learn in a different manner based on different inputs (see Section \ref{ArchitecturalConfigurations}). Among all the AAs, the best results are recurrently obtained when using the mass-type ones, which actually makes physical sense as it is the channel that directly refers to the drop in the galaxies' stellar mass density profiles. Even though we highlight the best model performance (see Section \ref{BestModel}), which we obtained for the encoder DenseNet-201 trained with visual AAs (dice: 0.8969, precision: 0.8863, recall: 0.9274; see Table \ref{tab:ResultadosAA}), we want to emphasize that the best results are retrieved for ensembles where different neural networks democratically vote the best solution in a pixel-by-pixel mode (see Figure \ref{Fig:Ensemble}). Using this method, 64.75~\% of the ensembles improve the highest dice value in comparison with our best individual model.

This best ensemble is the combination of EfficientNet-B6, ResNet-18, and DenseNet-161, all of them (very remarkably, as it is similar to what human astronomers need) trained with mass AAs. The metrics in this case are dice: 0.9104, precision: 0.9016, and recall: 0.9353. Finally, we also performed a series of tests checking our results versus morphological disk type, depth (i.e., galaxy field), redshift, and size, in order to ensure their robustness (see Section \ref{BestModel}).

Galaxy edges or truncations will permit us to obtain physically motivated galaxy sizes for images fulfilling the adequate depth and data reduction requirements. Our study shows that ML is able to reach a similar performance to that of humans and it will be mandatory to apply this technique for the billions of galaxies and tens of thousands of square degrees to be imaged in the future. In addition, the extraction of these features could be beneficial to other studies of galaxy morphology, as they will help identify any excess of light in the galaxy outer parts, thus indicating the presence of stellar haloes or minor merging. In summary, analyses with state-of-the-art tools as the one we are presenting here, will pave the way for solidifying galaxy edges or truncations as a viable physically motivated proxy for galaxy sizes.

%__________________________________________________________________

\begin{acknowledgements}
We acknowledge the referee for the insightful comments and feedback on the paper.
We have used extensively the following software packages: TOPCAT \citep{Taylor05}, ALADIN \citep{Bonnarel2000}, SEGMENTATION MODELS \citep{Iakubovskii:2019}, MATPLOTLIB \citep{matplotlib_ref} and PYTORCH \citep{NEURIPS2019_9015}.
JF and FB acknowledge the support from the grant PID2020-116188GA-I00 by the Spanish Ministry of Science and Innovation, while FB also acknowledges PID2019-107427GB-C32.
This work has made use of the Rainbow Cosmological Surveys Database, which is operated by the Centro de Astrobiolog\'ia (CAB/INTA), partnered with the University of California Observatories at Santa Cruz (UCO/Lick,UCSC).
Based on zCOSMOS observations carried out using the Very Large Telescope at the ESO Paranal Observatory under Programme ID: LP175.A0839. We also thank Jes\'us Vega-Ferrero, Ignacio Trujillo, Nushkia Chamba and Helena Dom\'inguez-S\'anchez for their advice.

\end{acknowledgements}

\bibliographystyle{aa.bst}
\bibliography{refs.bib}

\begin{appendix}
\section{Test set inferences}
\label{app:appendix}

Here, we present the inferences for all the galaxies in the test set (one row per galaxy) using the best ensemble configuration. The first column shows the RGB color image created from the concatenation of the I, J and H bands. The second one represents the overlap between the label and the output of the neural networks (true positives in yellow, true negatives in dark blue, false positives in red, and false negatives in green). The third column depicts the RGB image with the human-created (cyan) and the model-created (coral) truncations superimposed. The fourth column shows the output of the ensemble in a probabilistic way. Thus, the pixels represented in dark blue indicate that the three models that make up the ensemble have categorized them as noise. The violet pixels indicate that while two of the models have categorized them as noise, one has categorized them as a galaxy. Pixels represented in dark green indicate that while one model has categorized them as noise, two have categorized them as galaxies. Finally, the pixels represented in light green show total agreement for the categorization as a galaxy. Therefore, we note that the mask predicted by the ensemble corresponds with the area colored in light green plus the area colored in dark green.

\begin{figure}[!ht]
        \centering
        \includegraphics[width=1\linewidth]{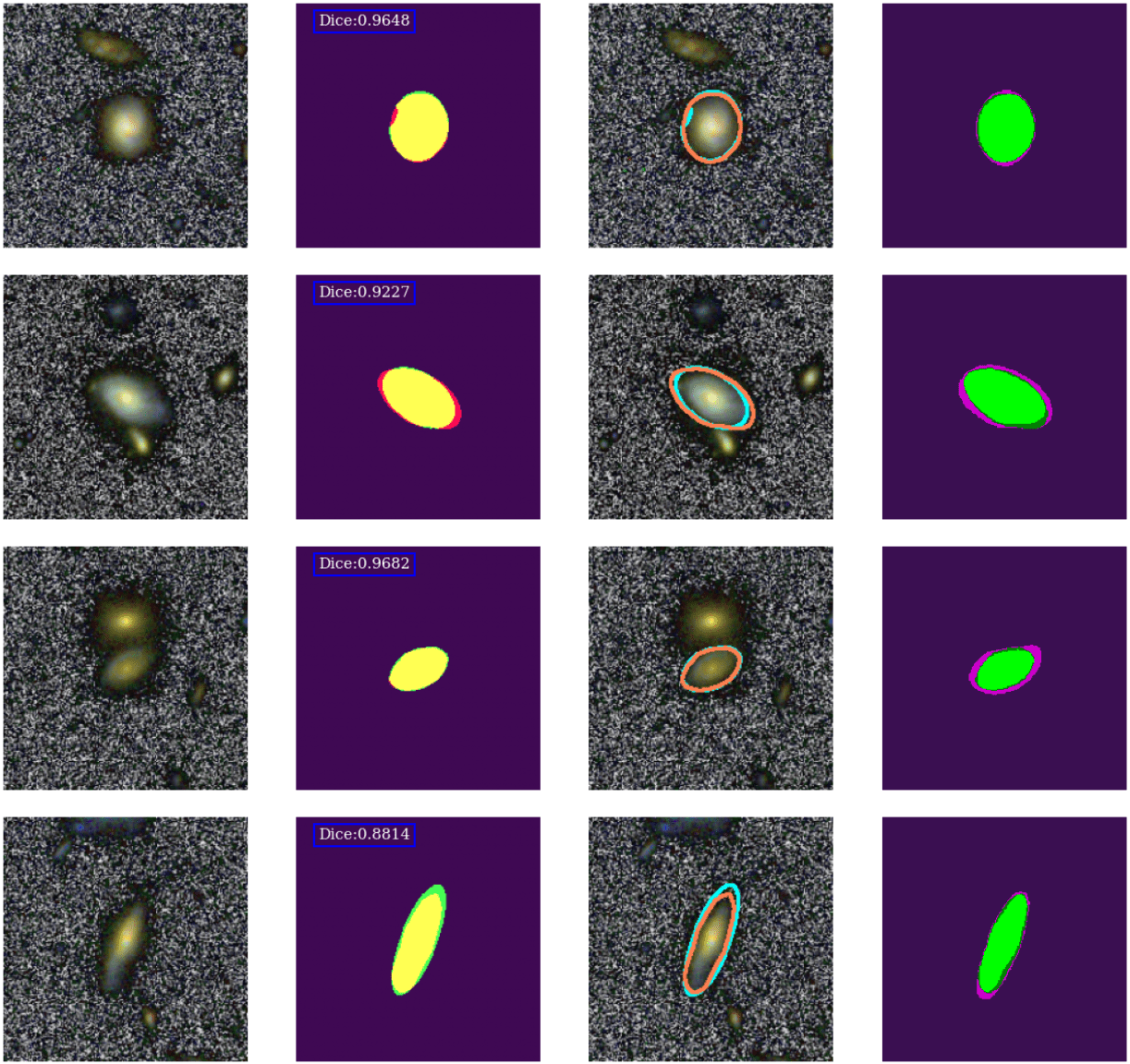}
        \caption{RGB images (first column), superimposed predicted and real masks (second column), superimposed predicted and real truncations (third column), and probabilistic ensemble outputs for galaxies 1 to 4 in the test set.}
        \label{Appendix-Figure-1}
\end{figure}

\begin{figure}[!ht]
        \centering
        \includegraphics[width=1\linewidth]{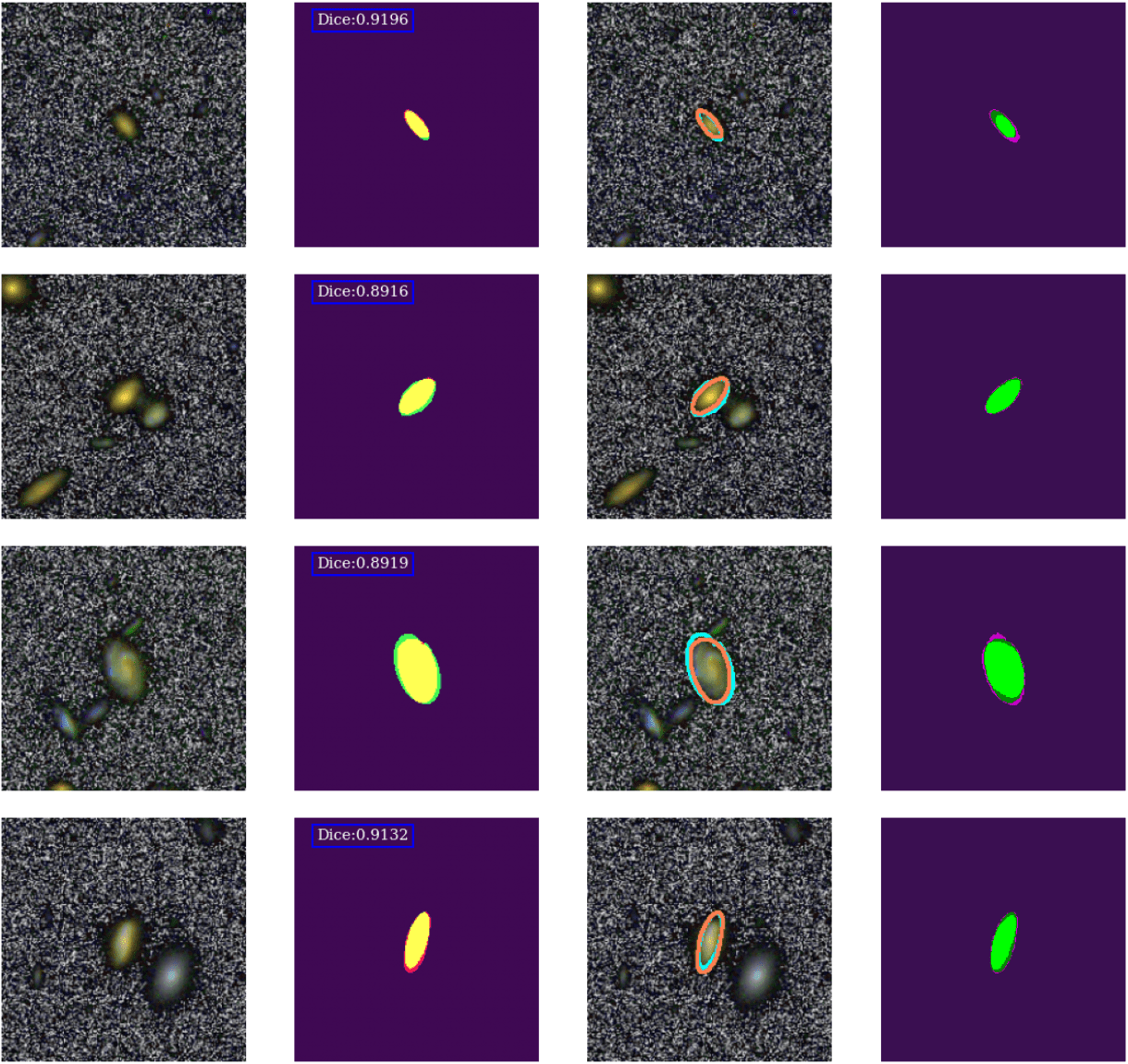}
         \caption{RGB images (first column), superimposed predicted and real masks (second column), superimposed predicted and real truncations (third column), and probabilistic ensemble outputs for galaxies 5 to 8 in the test set.}
        \label{Appendix-Figure-2}
\end{figure}

\begin{figure}[!ht]
        \centering
        \includegraphics[width=1\linewidth]{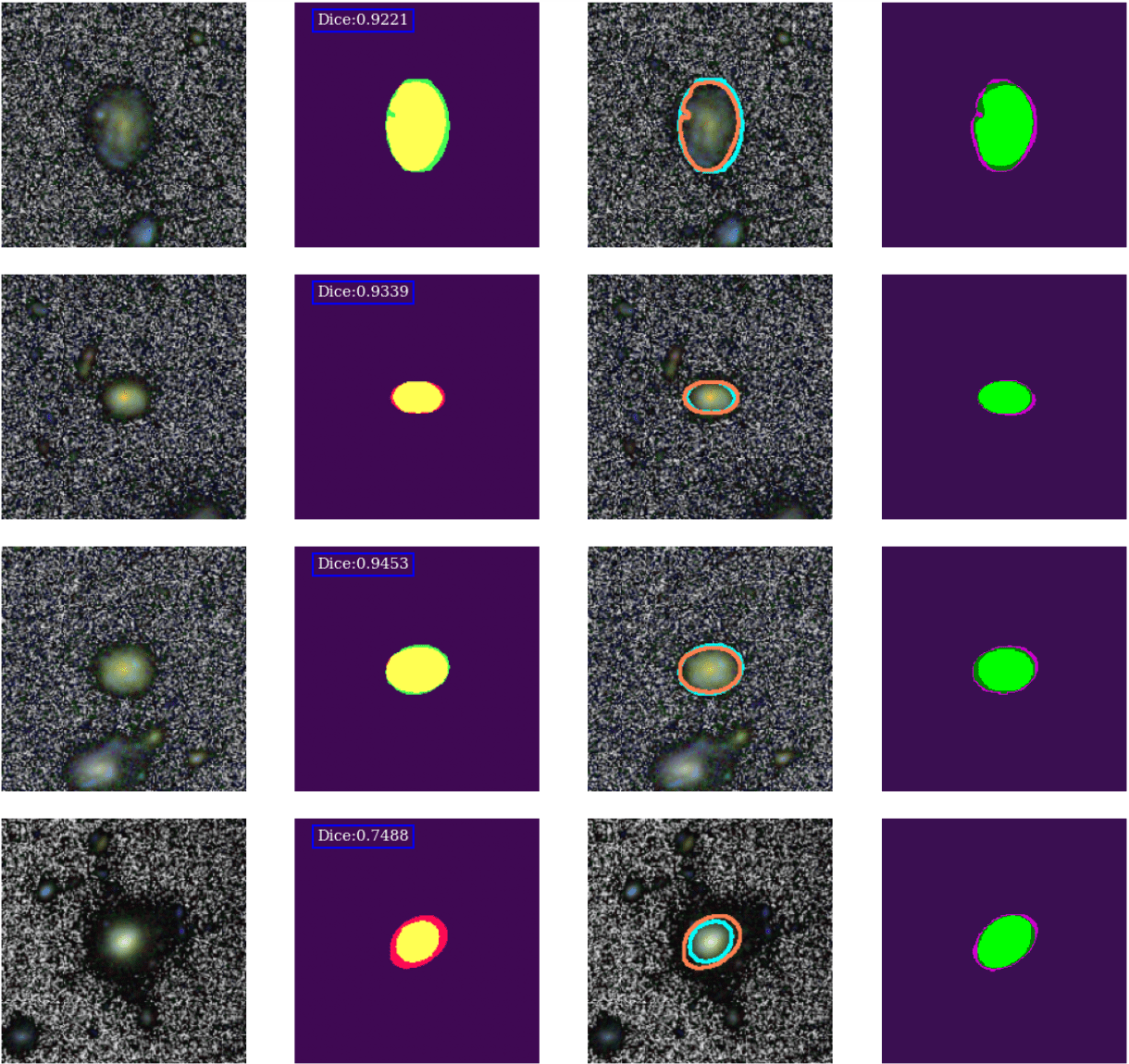}
         \caption{RGB images (first column), superimposed predicted and real masks (second column), superimposed predicted and real truncations (third column), and probabilistic ensemble outputs for galaxies 9 to 12 in the test set.}
        \label{Appendix-Figure-3}
\end{figure}

\begin{figure}[!ht]
        \centering
        \includegraphics[width=1\linewidth]{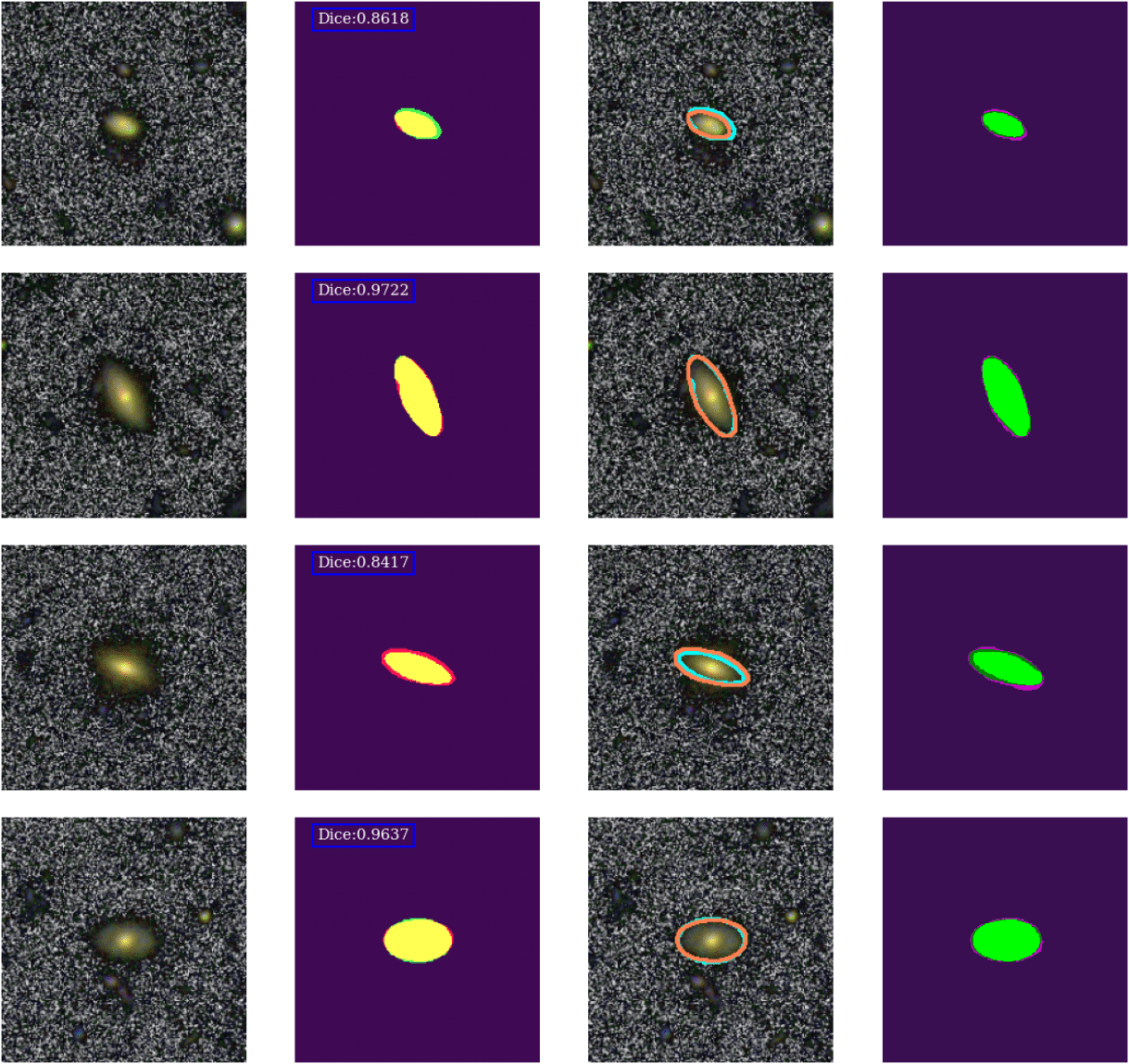}
         \caption{RGB images (first column), superimposed predicted and real masks (second column), superimposed predicted and real truncations (third column), and probabilistic ensemble outputs for galaxies 13 to 16 in the test set.}
        \label{Appendix-Figure-4}
\end{figure}

\begin{figure}[!ht]
        \centering
        \includegraphics[width=1\linewidth]{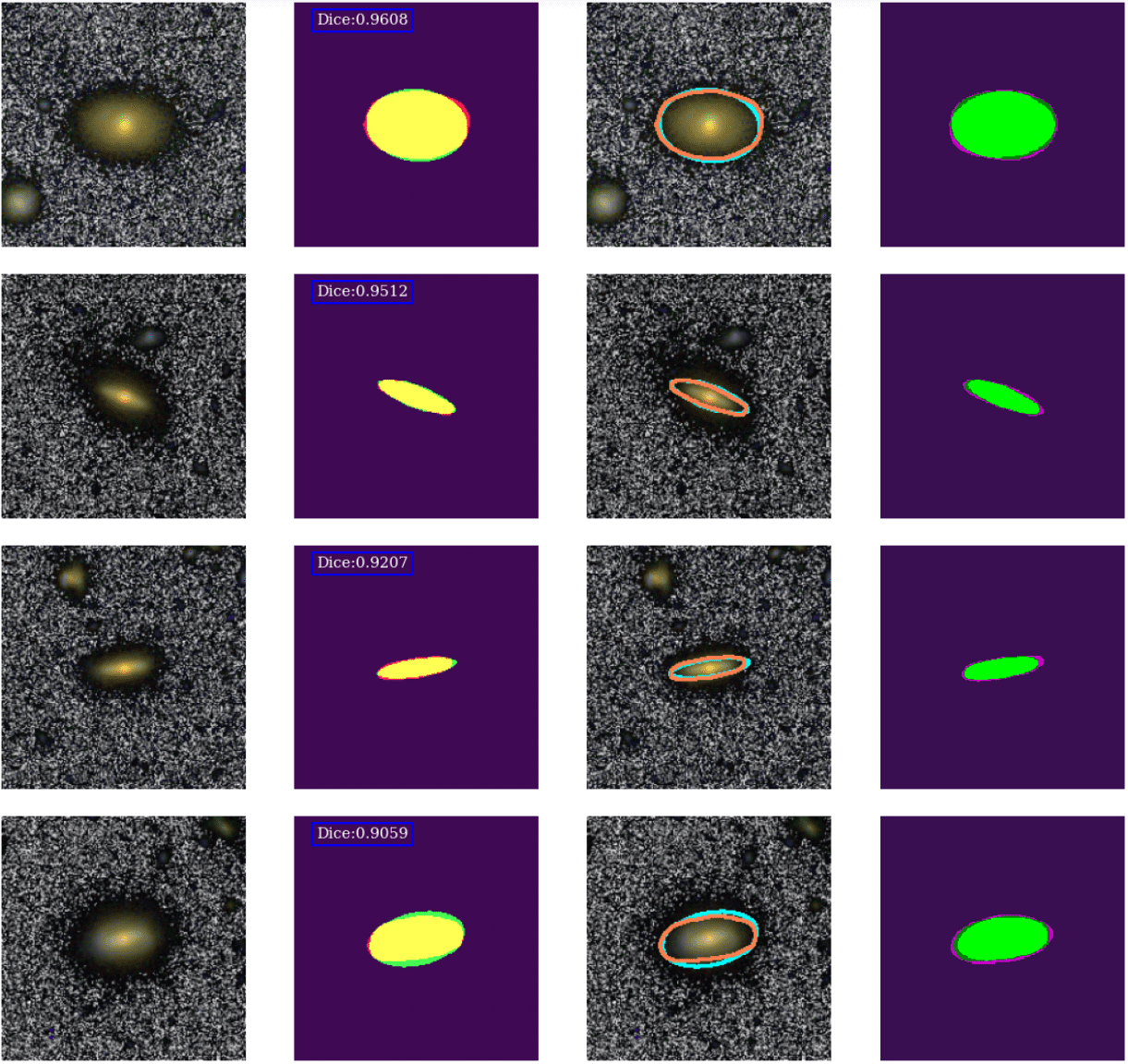}
         \caption{RGB images (first column), superimposed predicted and real masks (second column), superimposed predicted and real truncations (third column), and probabilistic ensemble outputs for galaxies 17 to 20 in the test set.}
        \label{Appendix-Figure-5}
\end{figure}

\begin{figure}[!ht]
        \centering
        \includegraphics[width=1\linewidth]{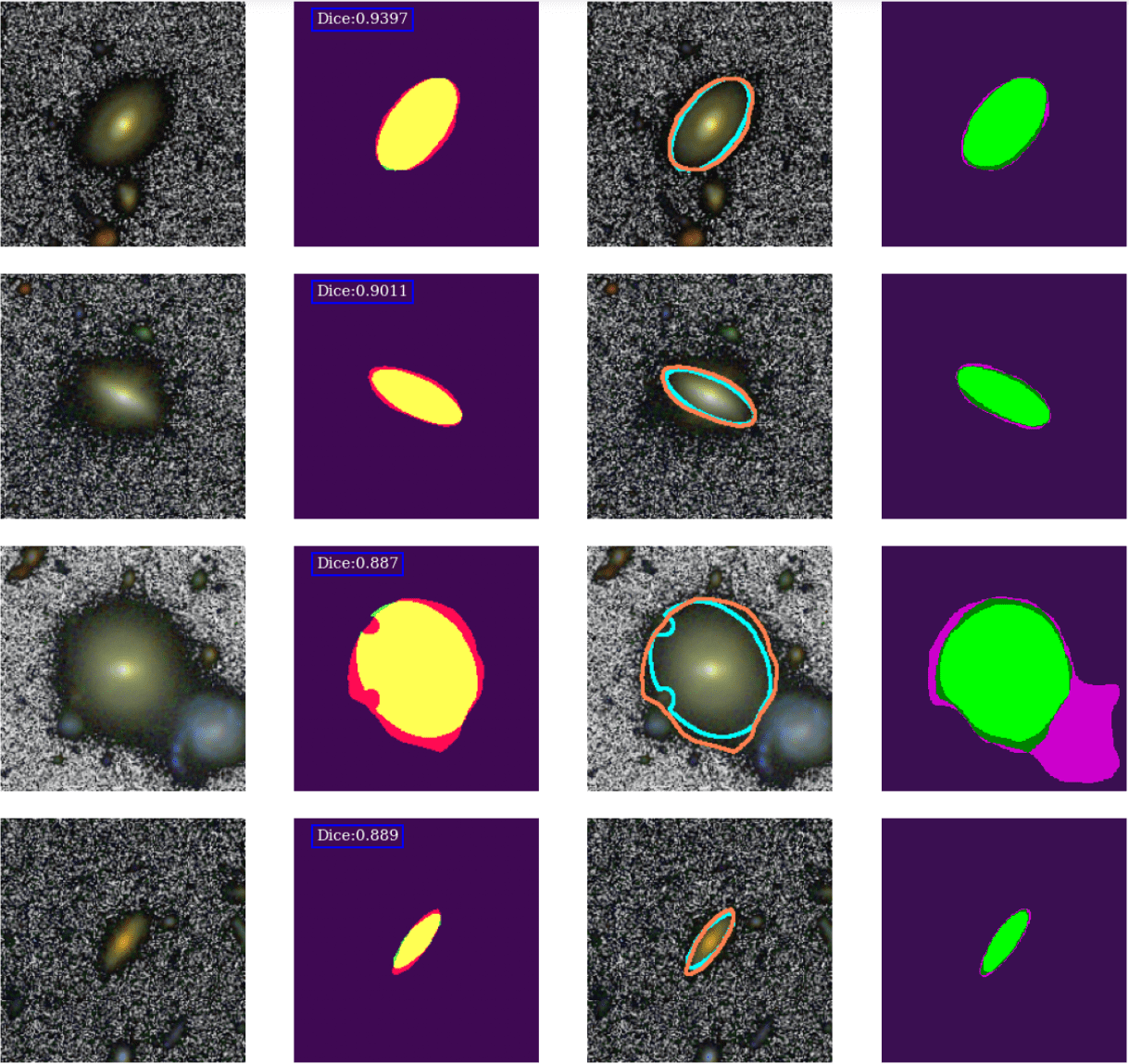}
        \caption{RGB images (first column), superimposed predicted and real masks (second column), superimposed predicted and real truncations (third column), and probabilistic ensemble outputs for galaxies 21 to 24 in the test set.}
        \label{Appendix-Figure-6}
\end{figure}

\begin{figure}[!ht]
        \centering
        \includegraphics[width=1\linewidth]{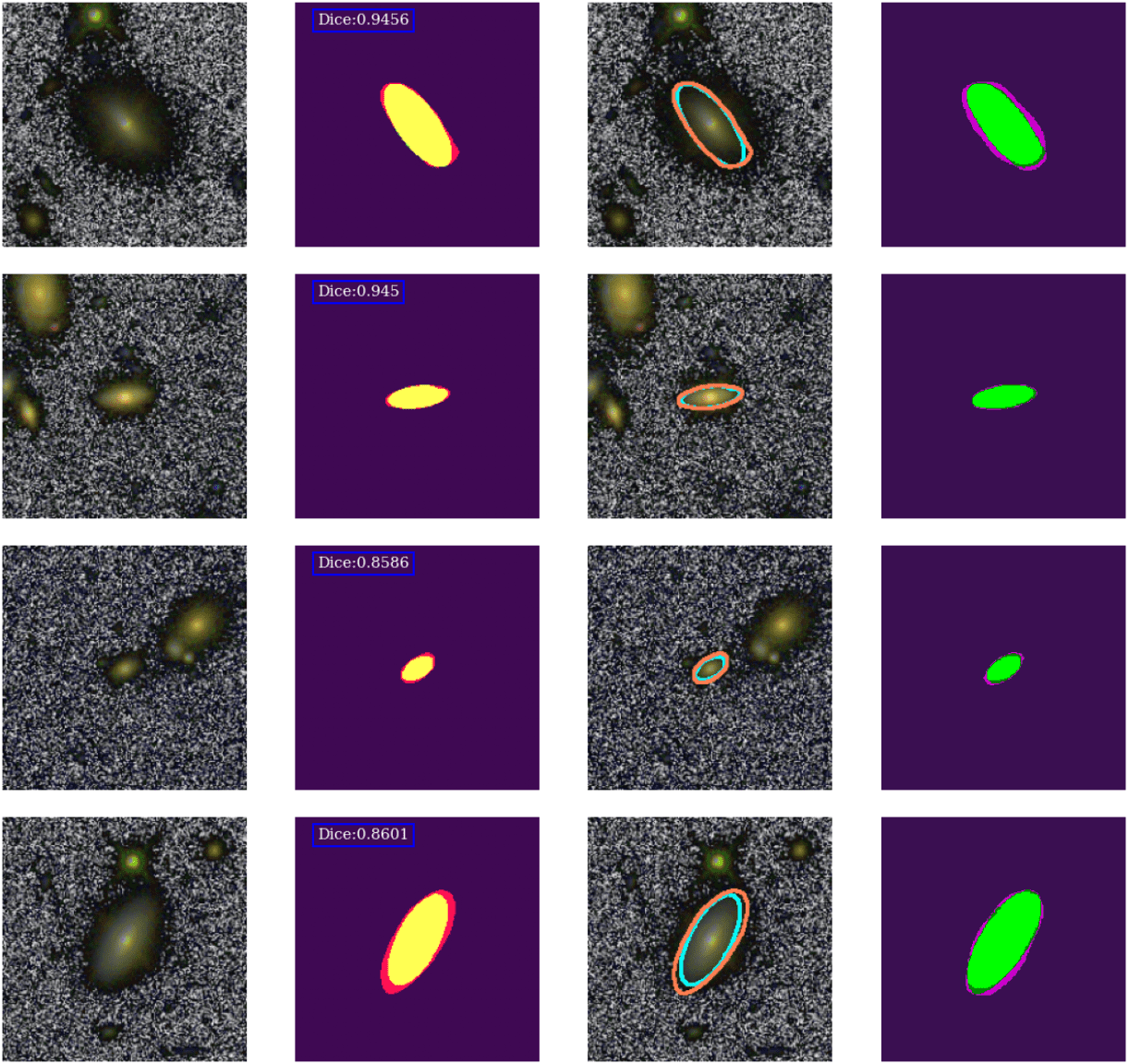}
        \caption{RGB images (first column), superimposed predicted and real masks (second column), superimposed predicted and real truncations (third column), and probabilistic ensemble outputs for galaxies 25 to 28 in the test set.}
        \label{Appendix-Figure-7}
\end{figure}

\begin{figure}[!ht]
        \centering
        \includegraphics[width=1\linewidth]{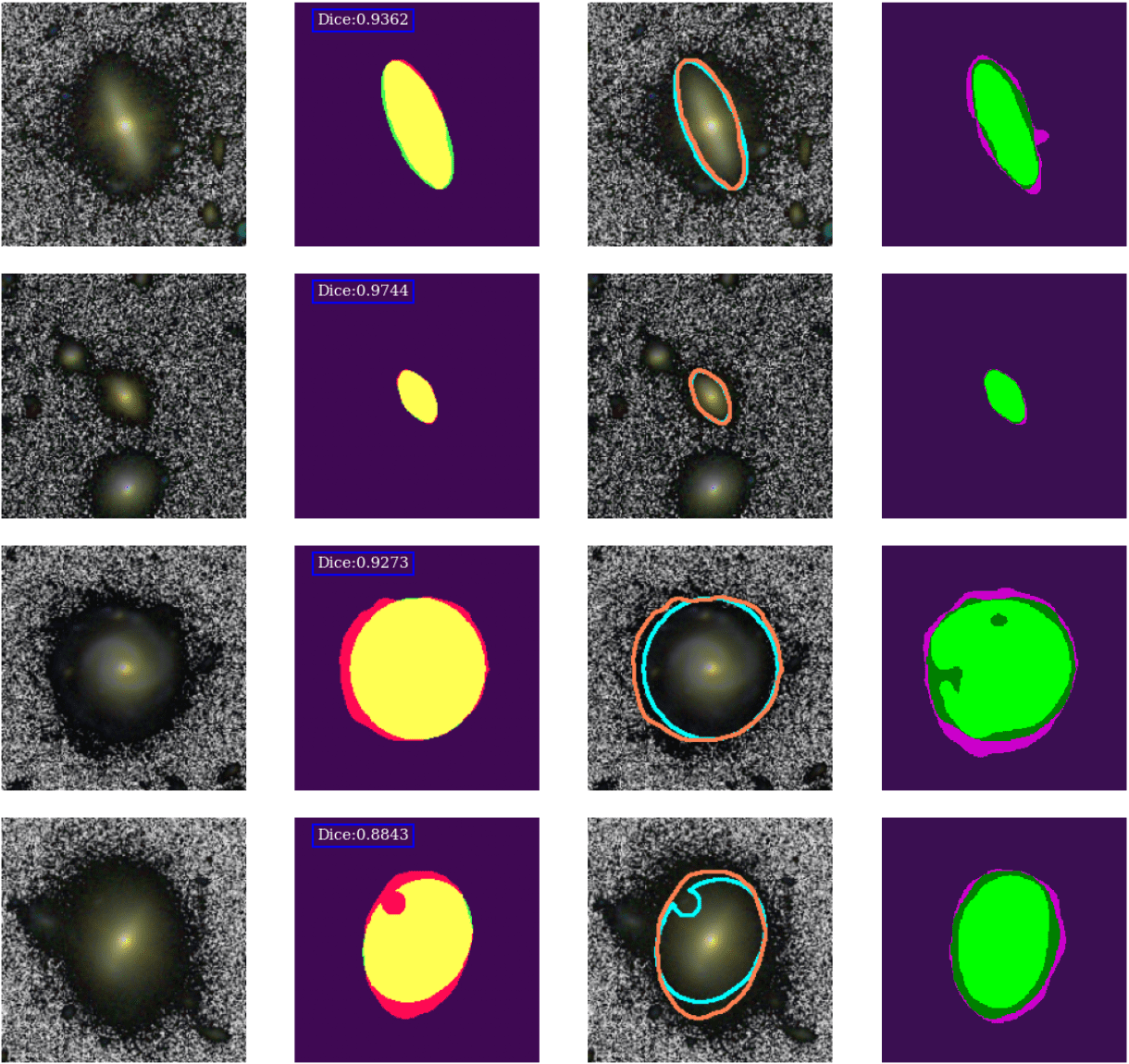}
        \caption{RGB images (first column), superimposed predicted and real masks (second column), superimposed predicted and real truncations (third column), and probabilistic ensemble outputs for galaxies 29 to 32 in the test set.}
        \label{Appendix-Figure-8}
\end{figure}

\begin{figure}[!ht]
        \centering
        \includegraphics[width=1\linewidth]{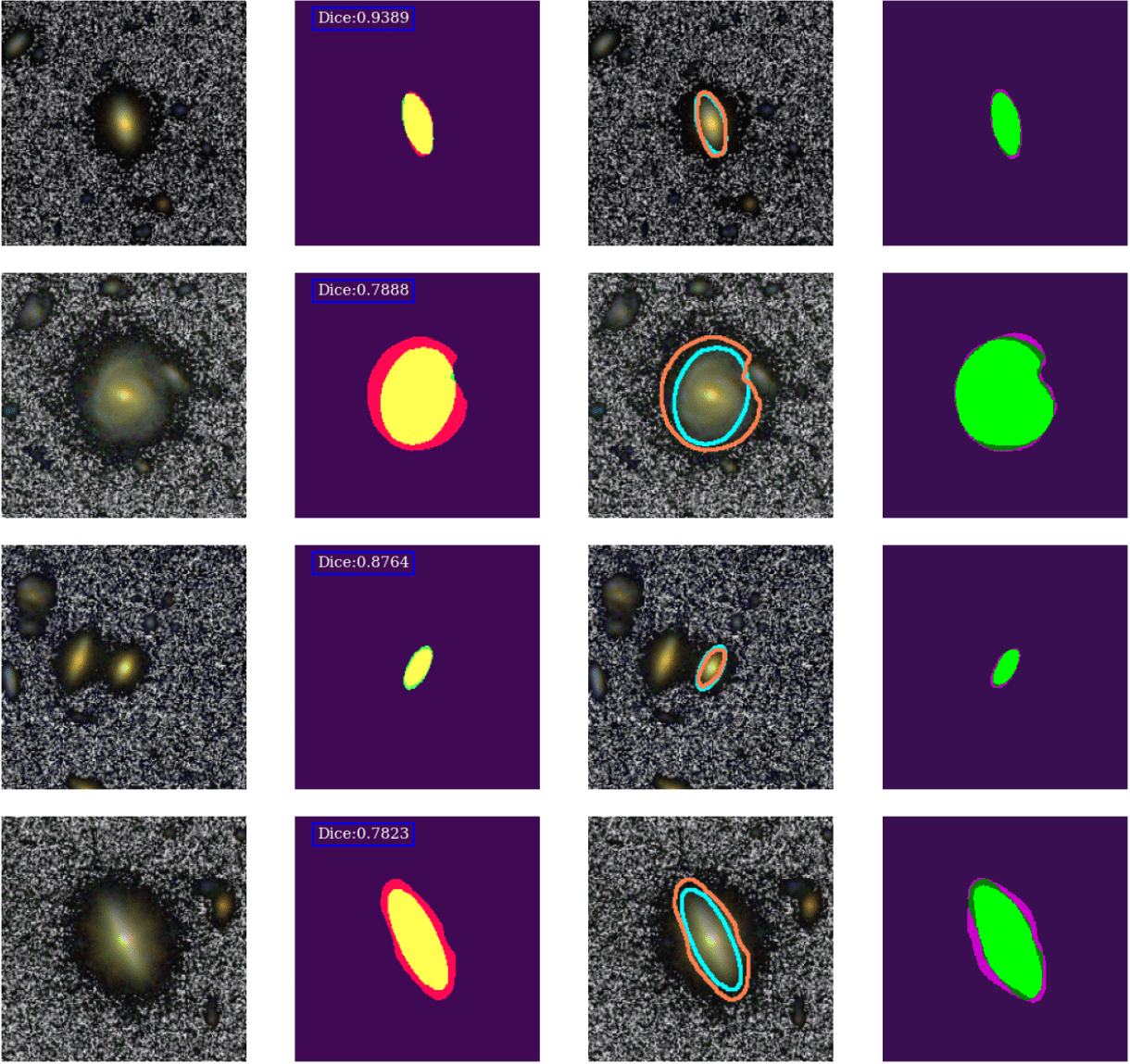}
        \caption{RGB images (first column), superimposed predicted and real masks (second column), superimposed predicted and real truncations (third column), and probabilistic ensemble outputs for galaxies 33 to 36 in the test set.}
        \label{Appendix-Figure-9}
\end{figure}

\begin{figure}[!ht]
        \centering
        \includegraphics[width=1\linewidth]{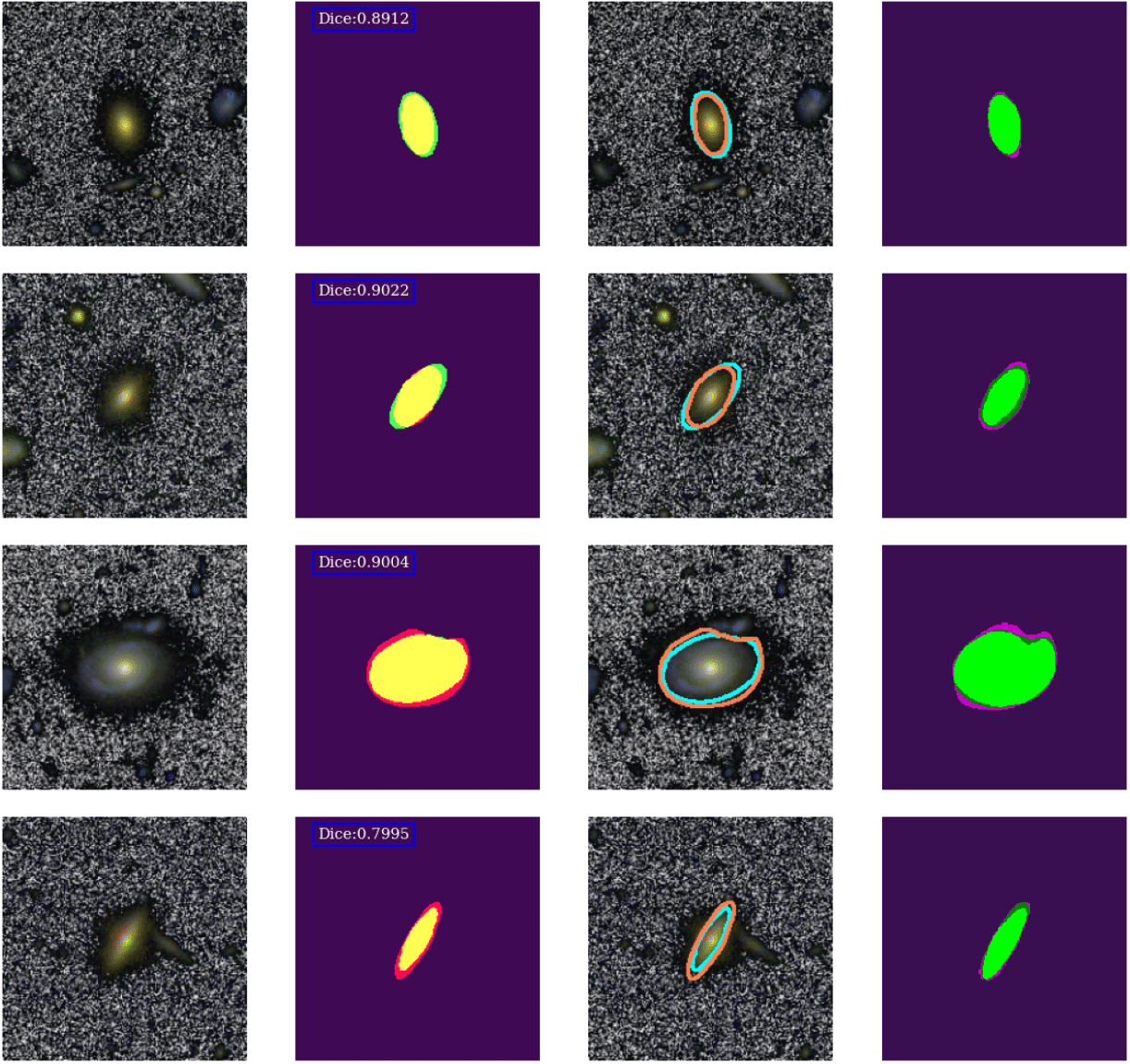}
        \caption{RGB images (first column), superimposed predicted and real masks (second column), superimposed predicted and real truncations (third column), and probabilistic ensemble outputs for galaxies 37 to 40 in the test set.}
        \label{Appendix-Figure-10}
\end{figure}

\begin{figure}[!ht]
        \centering
        \includegraphics[width=1\linewidth]{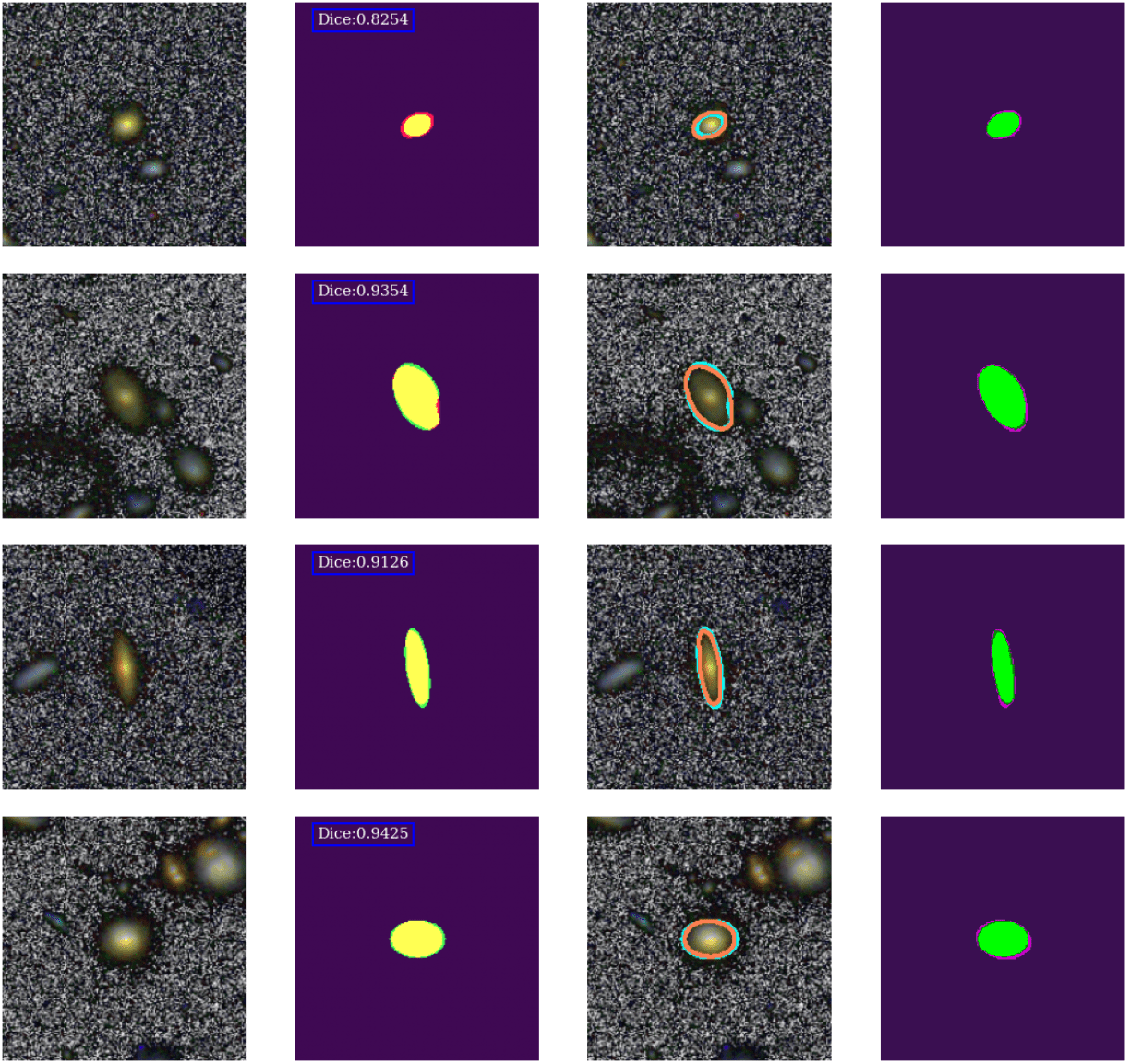}
        \caption{RGB images (first column), superimposed predicted and real masks (second column), superimposed predicted and real truncations (third column), and probabilistic ensemble outputs for galaxies 41 to 44 in the test set.}
        \label{Appendix-Figure-11}
\end{figure}

\begin{figure}[!ht]
        \centering
        \includegraphics[width=1\linewidth]{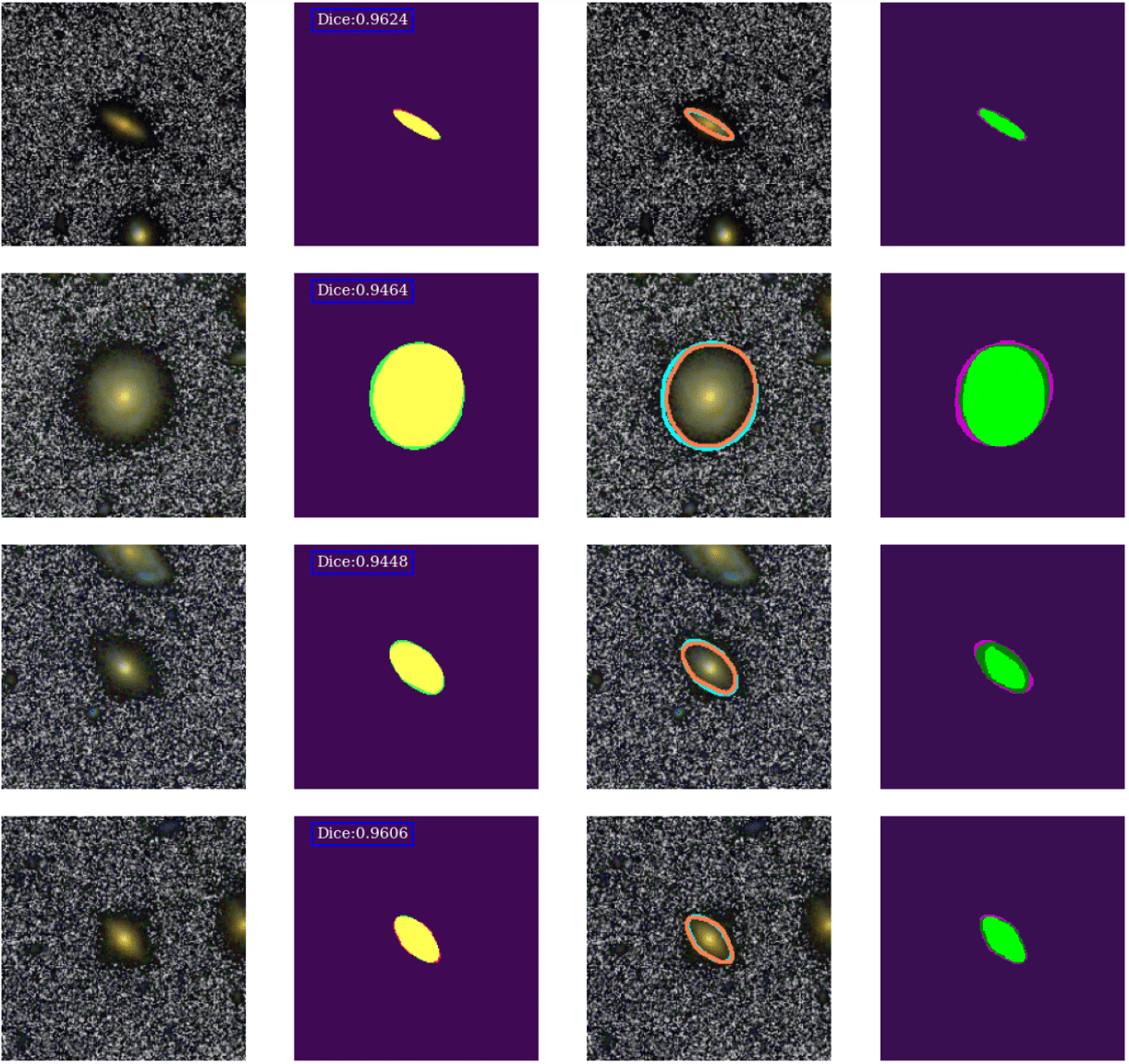}
        \caption{RGB images (first column), superimposed predicted and real masks (second column), superimposed predicted and real truncations (third column), and probabilistic ensemble outputs for galaxies 45 to 48 in the test set.}
        \label{Appendix-Figure-12}
\end{figure}

\begin{figure}[!ht]
        \centering
        \includegraphics[width=1\linewidth]{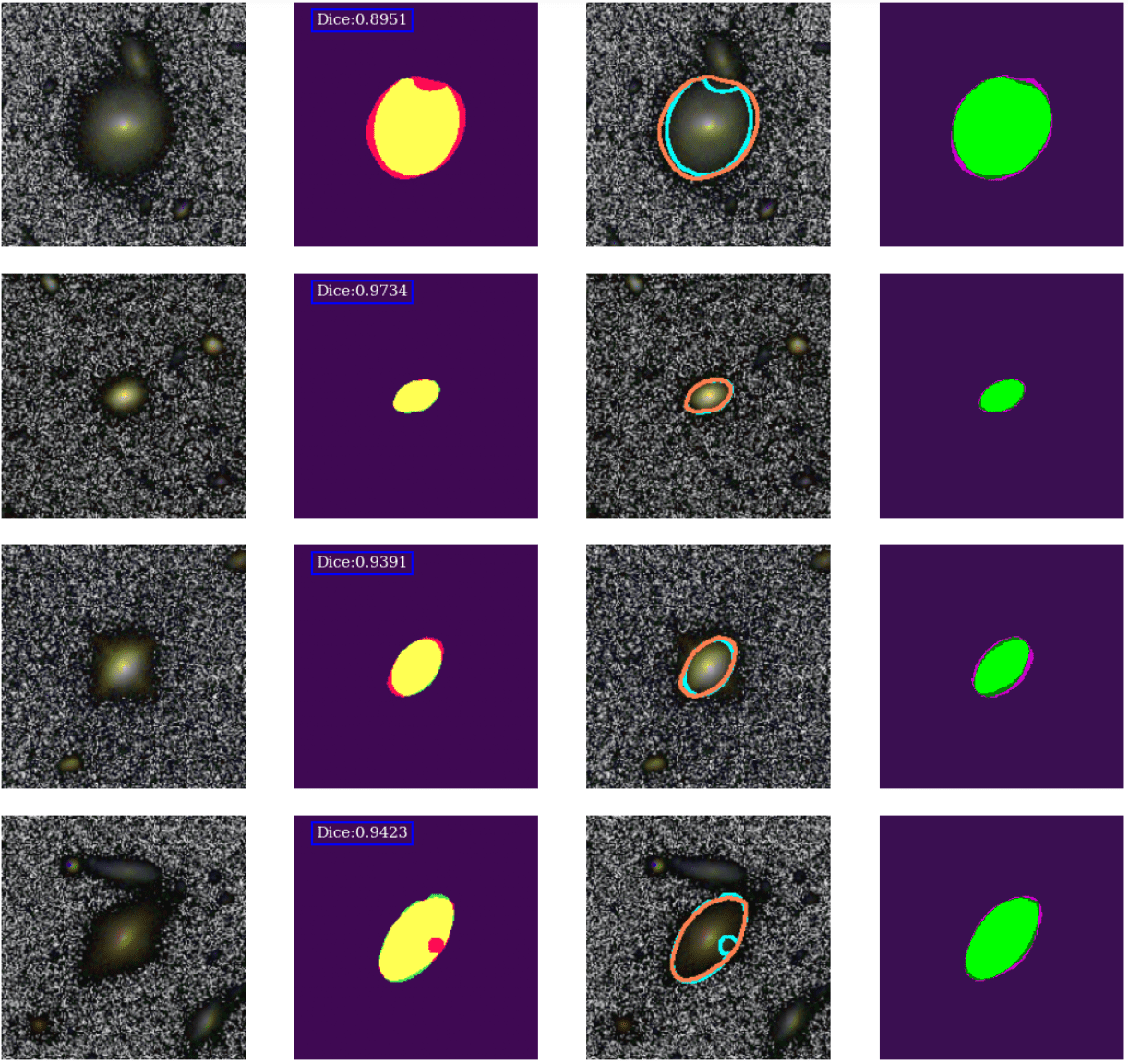}
        \caption{RGB images (first column), superimposed predicted and real masks (second column), superimposed predicted and real truncations (third column), and probabilistic ensemble outputs for galaxies 49 to 52 in the test set.}
        \label{Appendix-Figure-13}
\end{figure}

\begin{figure}[!ht]
        \centering
        \includegraphics[width=1\linewidth]{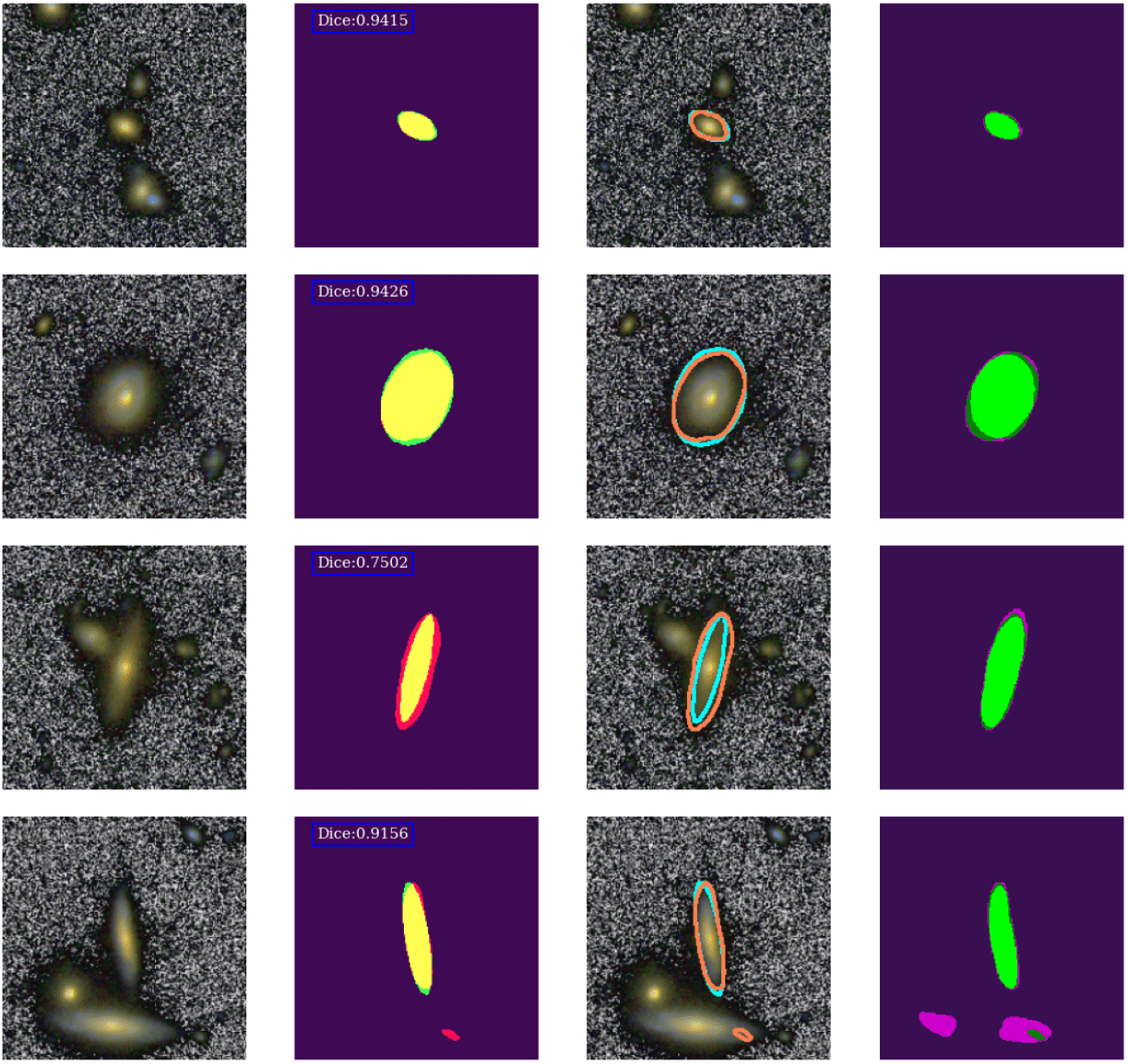}
        \caption{RGB images (first column), superimposed predicted and real masks (second column), superimposed predicted and real truncations (third column), and probabilistic ensemble outputs for galaxies 53 to 56 in the test set.}
        \label{Appendix-Figure-14}
\end{figure}

\begin{figure}[!ht]
        \centering
        \includegraphics[width=1\linewidth]{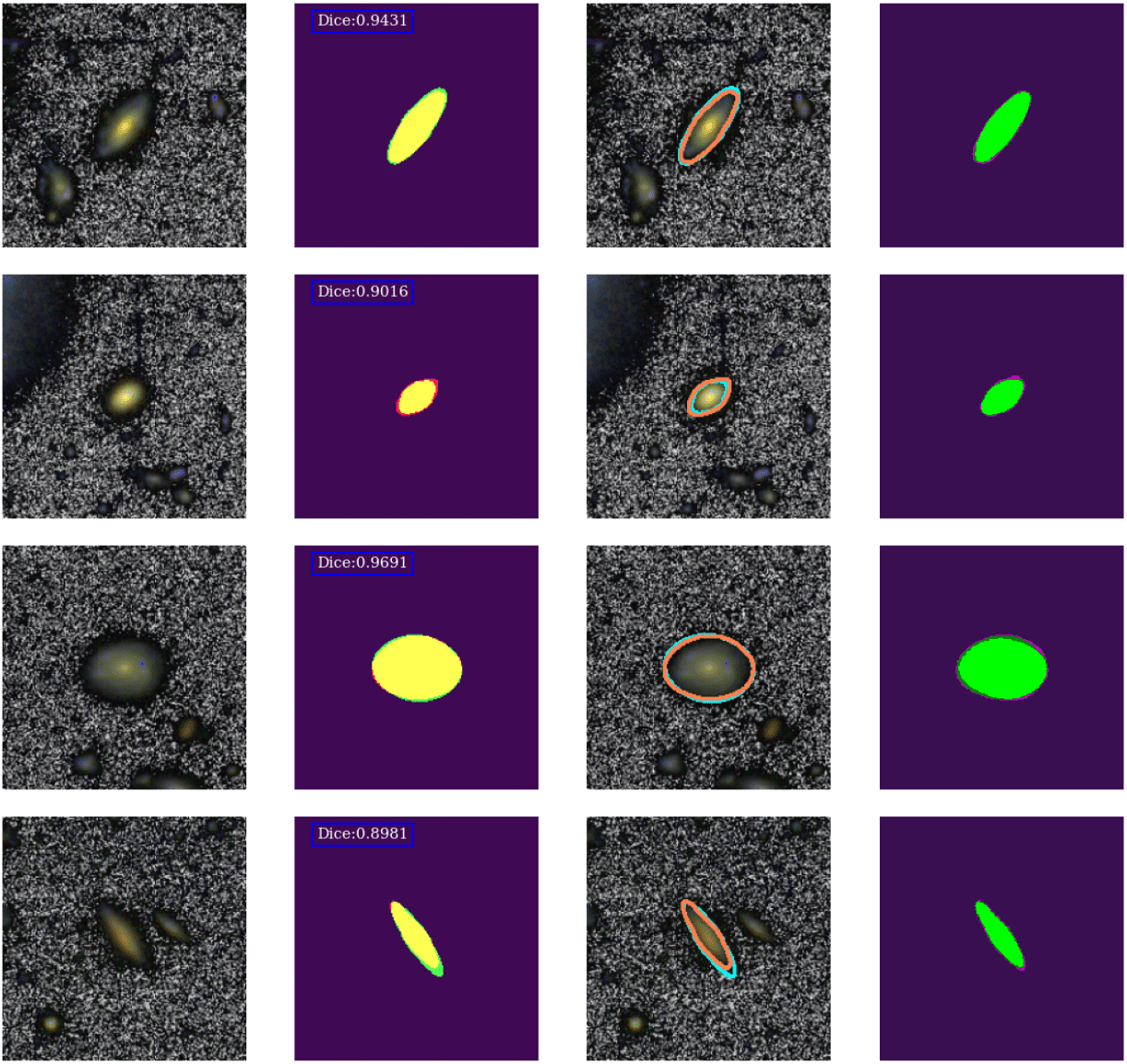}
        \caption{RGB images (first column), superimposed predicted and real masks (second column), superimposed predicted and real truncations (third column), and probabilistic ensemble outputs for galaxies 57 to 60 in the test set.}
        \label{Appendix-Figure-15}
\end{figure}

\begin{figure}[!ht]
        \centering
        \includegraphics[width=1\linewidth]{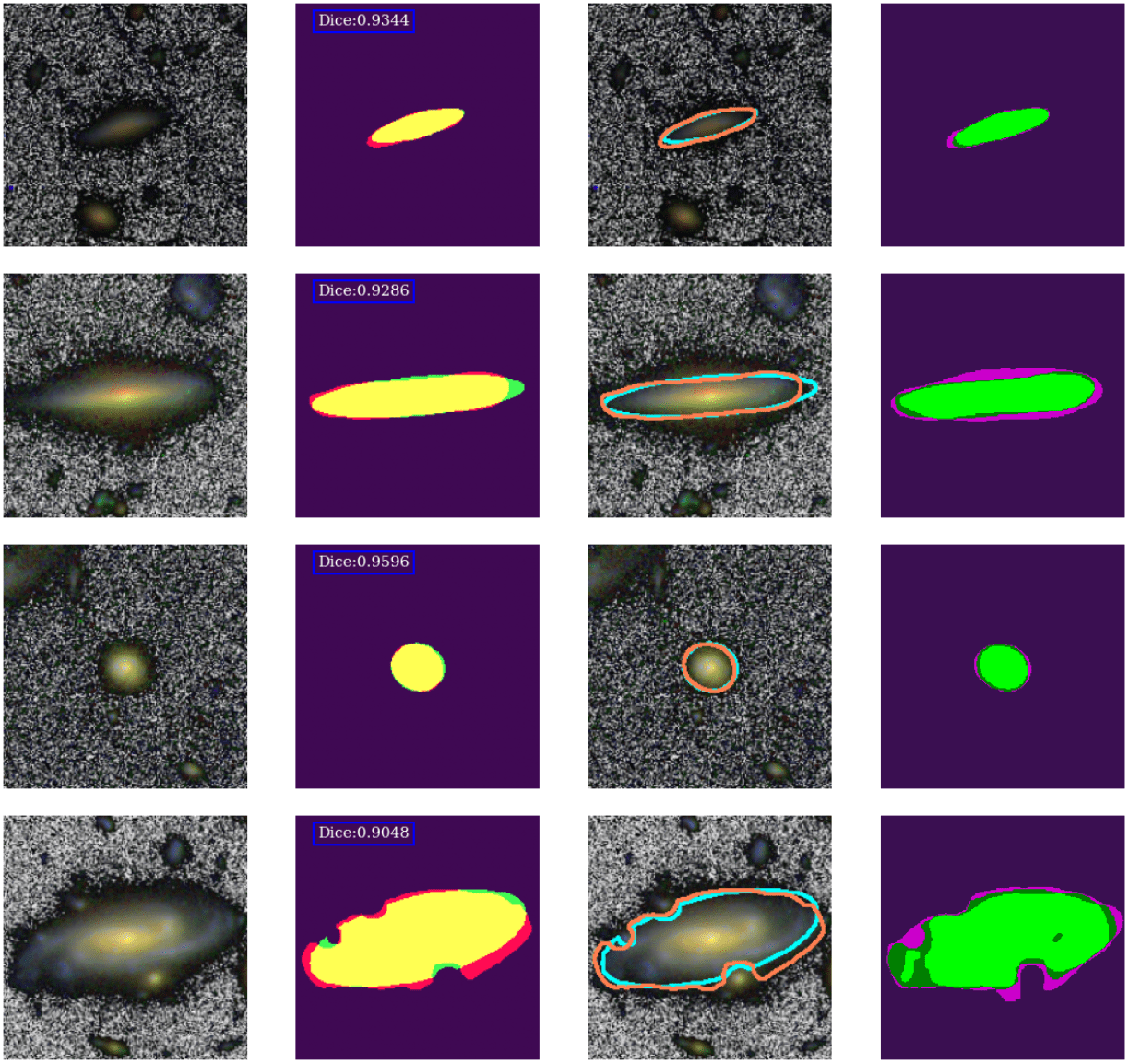}
        \caption{RGB images (first column), superimposed predicted and real masks (second column), superimposed predicted and real truncations (third column), and probabilistic ensemble outputs for galaxies 61 to 64 in the test set.}
        \label{Appendix-Figure-16}
\end{figure}

\begin{figure}[!ht]
        \centering
        \includegraphics[width=1\linewidth]{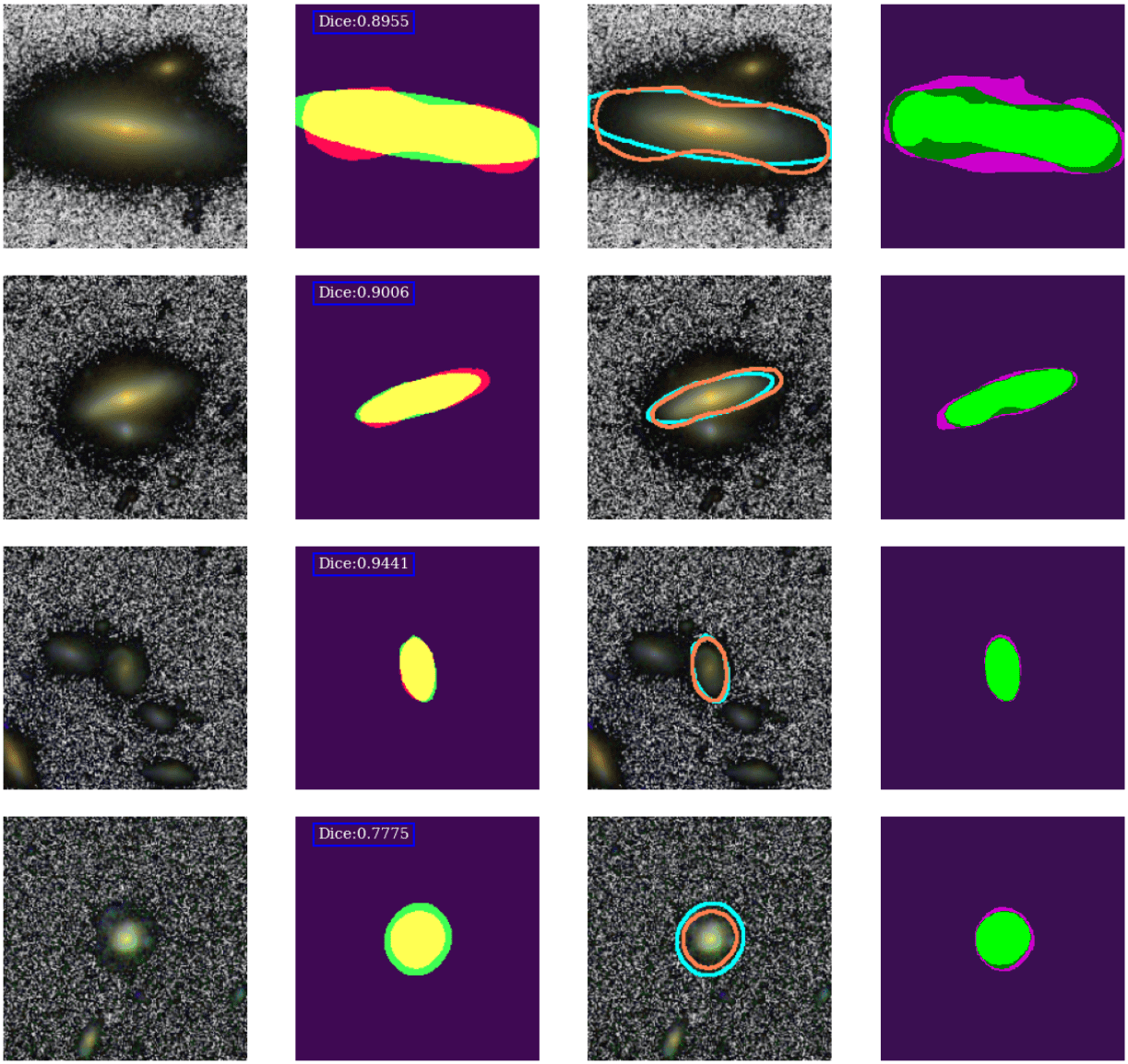}
        \caption{RGB images (first column), superimposed predicted and real masks (second column), superimposed predicted and real truncations (third column), and probabilistic ensemble outputs for galaxies 65 to 68 in the test set.}
        \label{Appendix-Figure-17}
\end{figure}

\begin{figure}[!ht]
        \centering
        \includegraphics[width=1\linewidth]{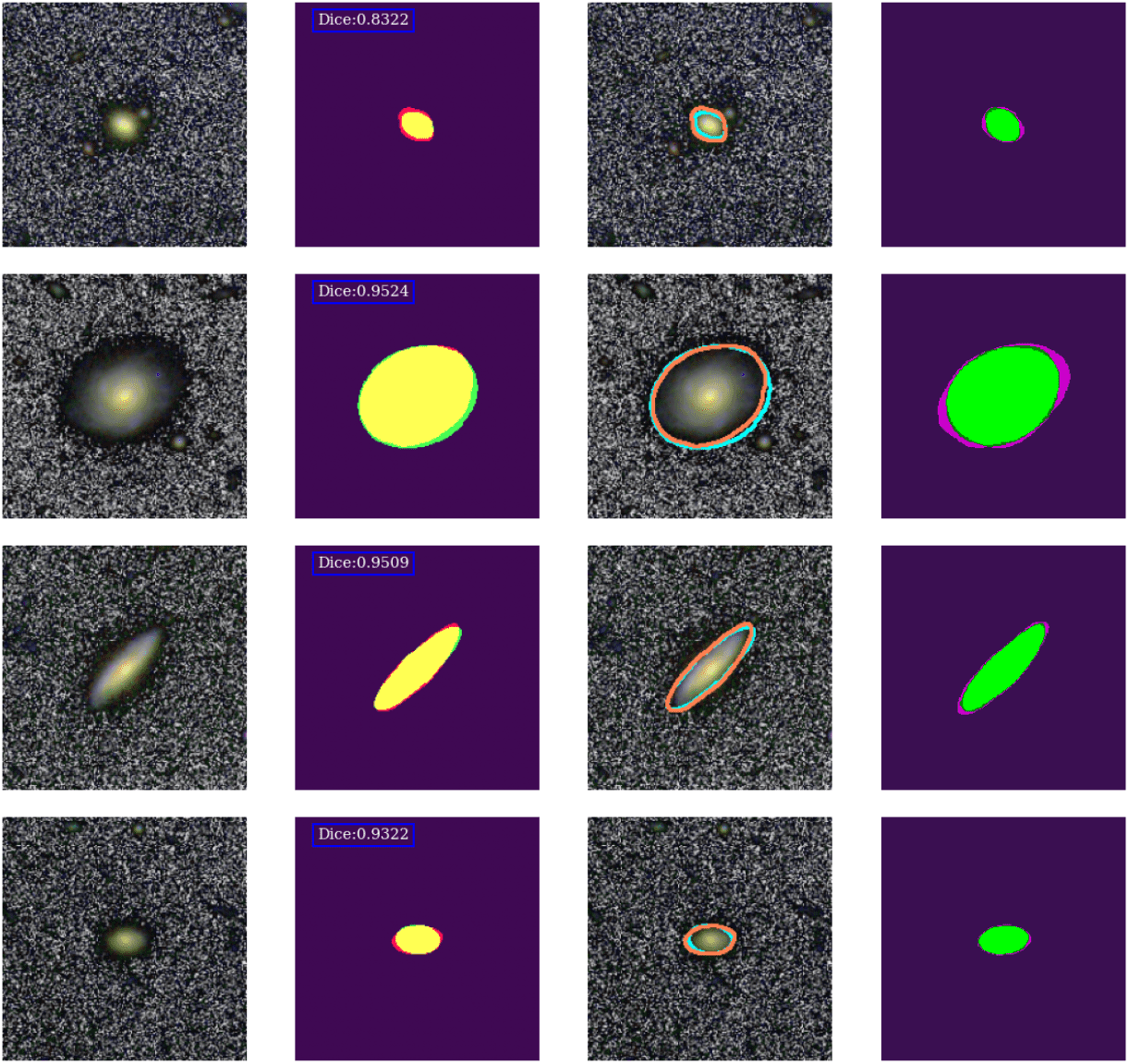}
        \caption{RGB images (first column), superimposed predicted and real masks (second column), superimposed predicted and real truncations (third column), and probabilistic ensemble outputs for galaxies 69 to 72 in the test set.}
        \label{Appendix-Figure-18}
\end{figure}

\begin{figure}[!ht]
        \centering
        \includegraphics[width=1\linewidth]{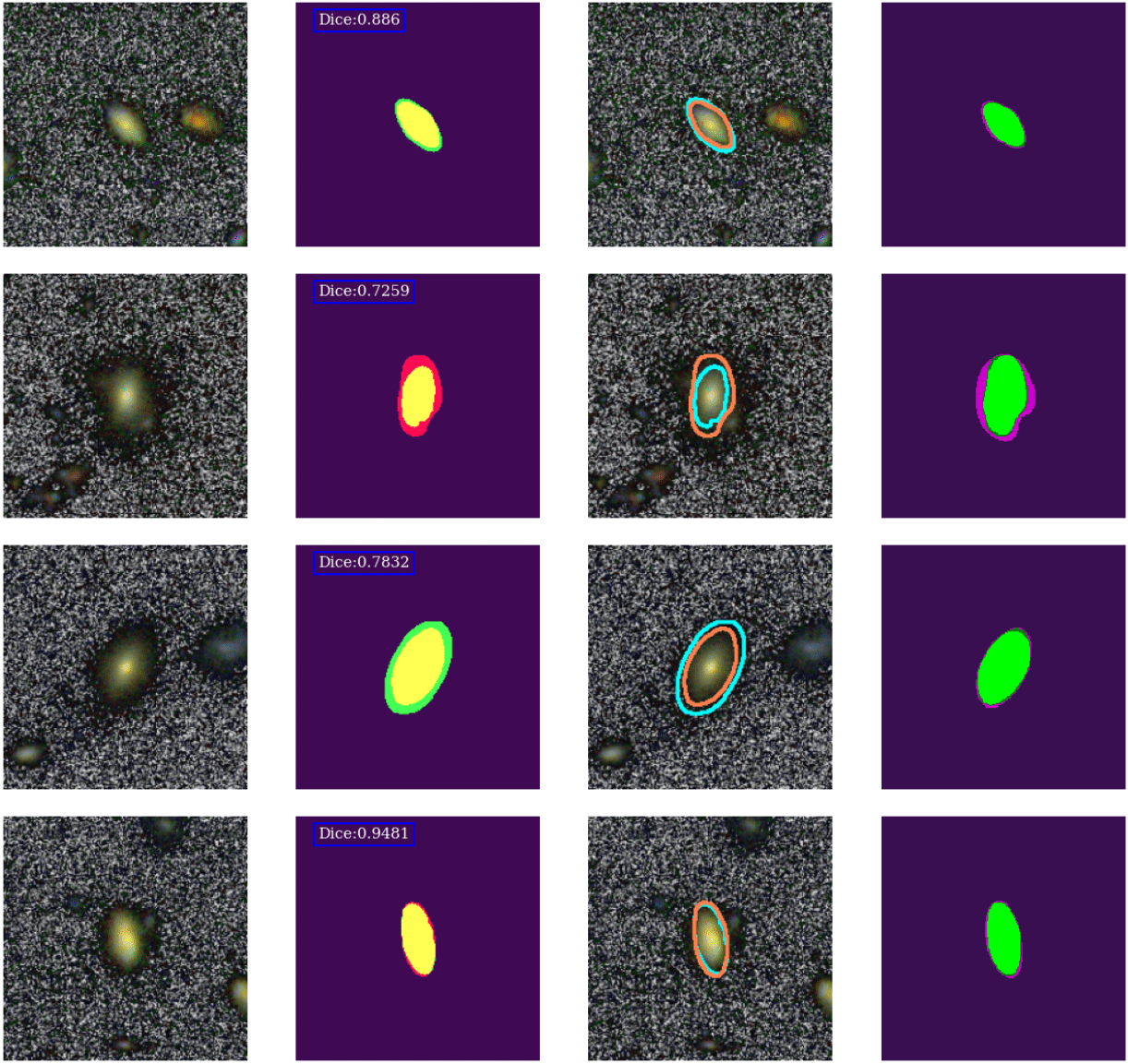}
        \caption{RGB images (first column), superimposed predicted and real masks (second column), superimposed predicted and real truncations (third column), and probabilistic ensemble outputs for galaxies 73 to 76 in the test set.}
        \label{Appendix-Figure-19}
\end{figure}

\begin{figure}[!ht]
        \centering
        \includegraphics[width=1\linewidth]{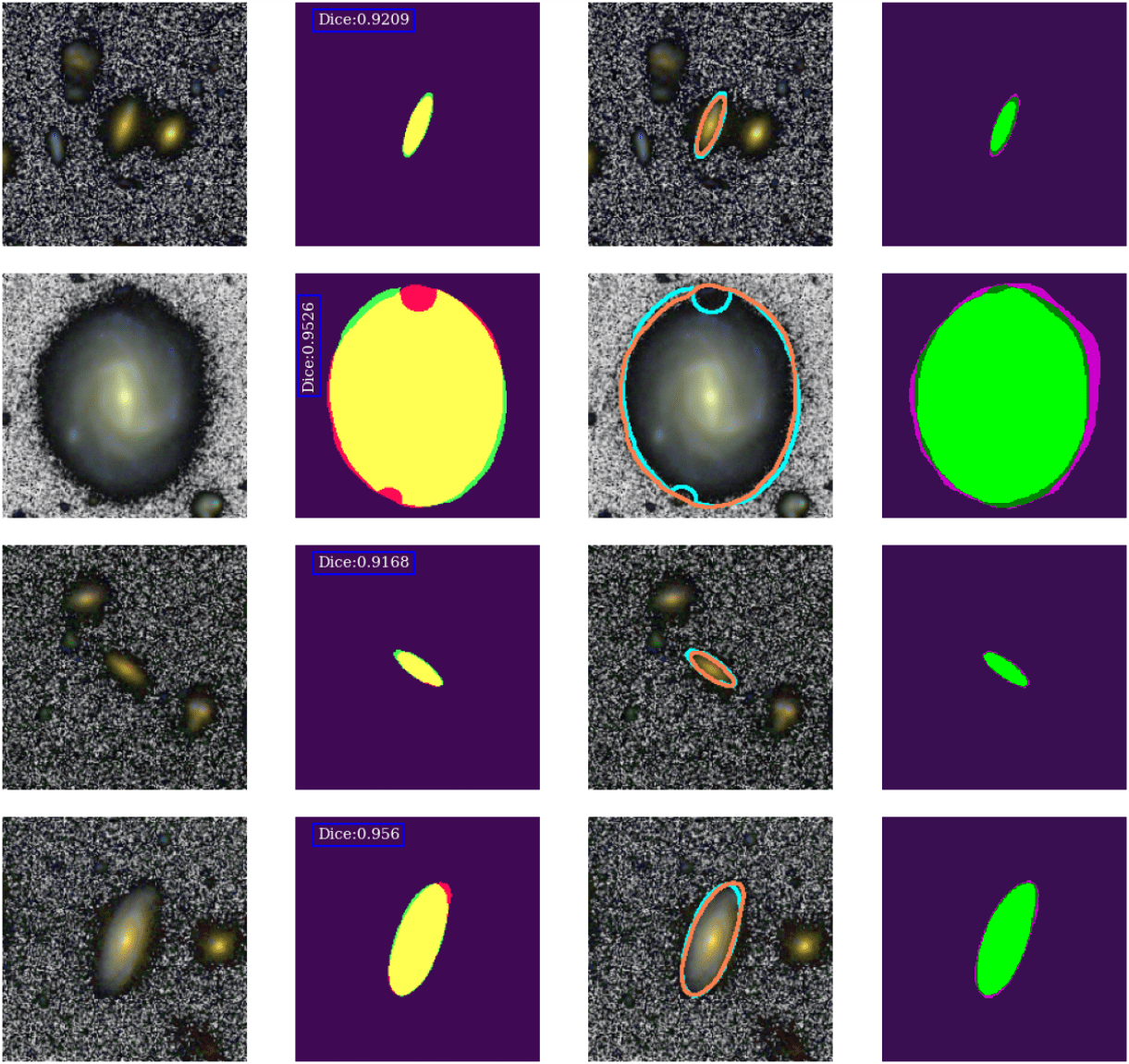}
        \caption{RGB images (first column), superimposed predicted and real masks (second column), superimposed predicted and real truncations (third column), and probabilistic ensemble outputs for galaxies 77 to 80 in the test set.}
        \label{Appendix-Figure-20}
\end{figure}

\begin{figure}[!ht]
        \centering
        \includegraphics[width=1\linewidth]{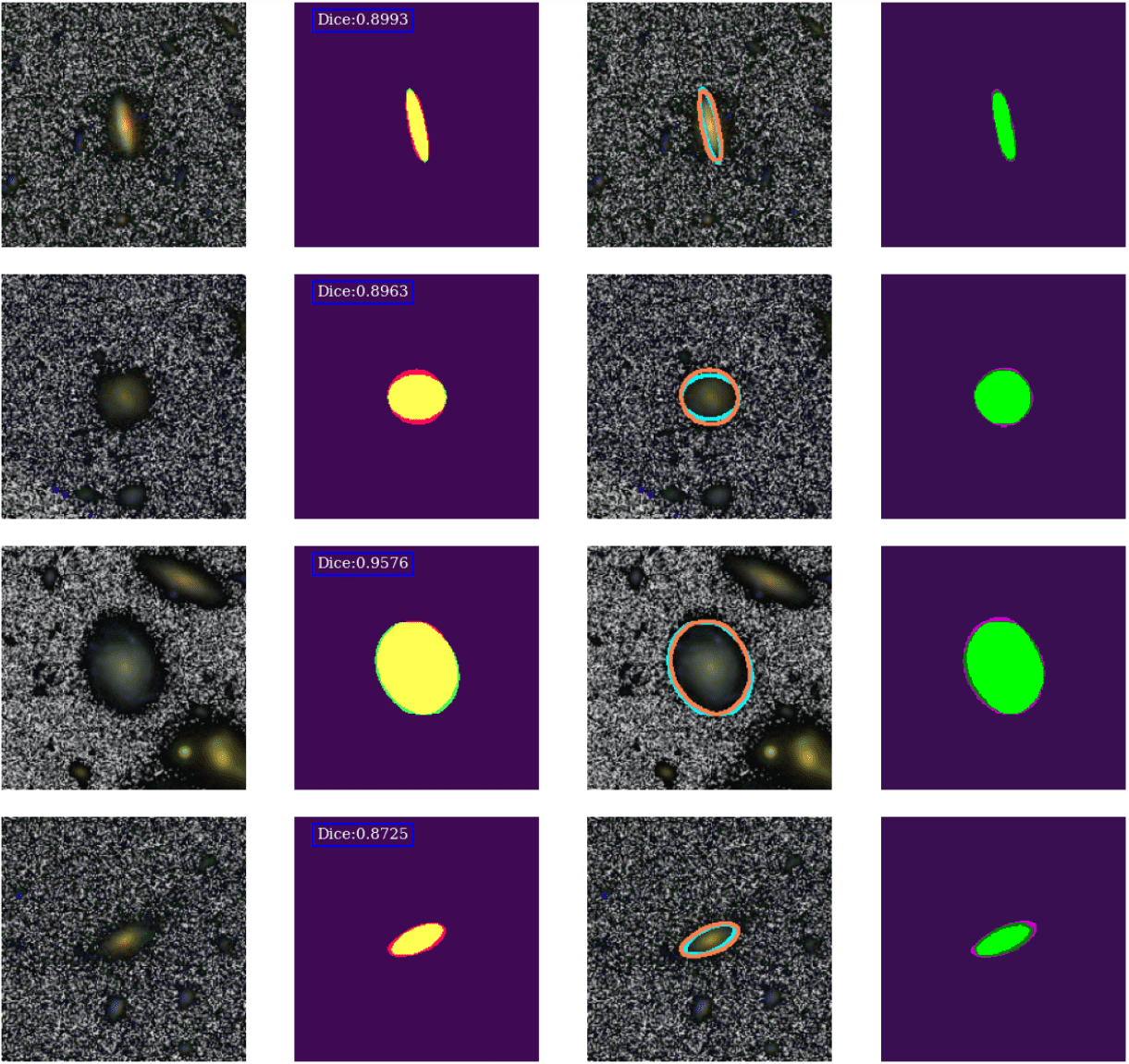}
        \caption{RGB images (first column), superimposed predicted and real masks (second column), superimposed predicted and real truncations (third column), and probabilistic ensemble outputs for galaxies 81 to 84 in the test set.}
        \label{Appendix-Figure-21}
\end{figure}

\begin{figure}[!ht]
        \centering
        \includegraphics[width=1\linewidth]{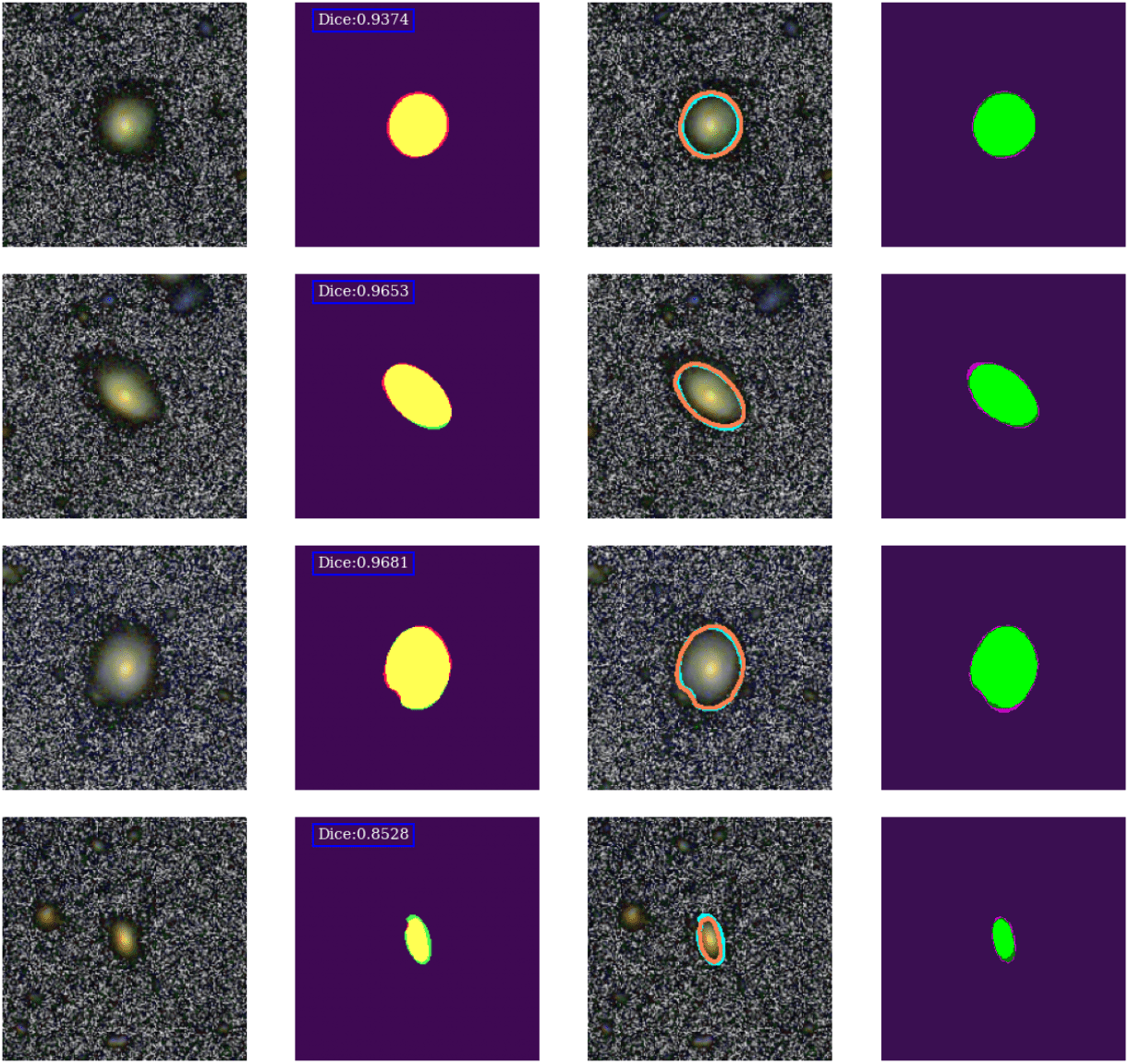}
        \caption{RGB images (first column), superimposed predicted and real masks (second column), superimposed predicted and real truncations (third column), and probabilistic ensemble outputs for galaxies 85 to 88 in the test set.}
        \label{Appendix-Figure-22}
\end{figure}

\begin{figure}[!ht]
        \centering
        \includegraphics[width=1\linewidth]{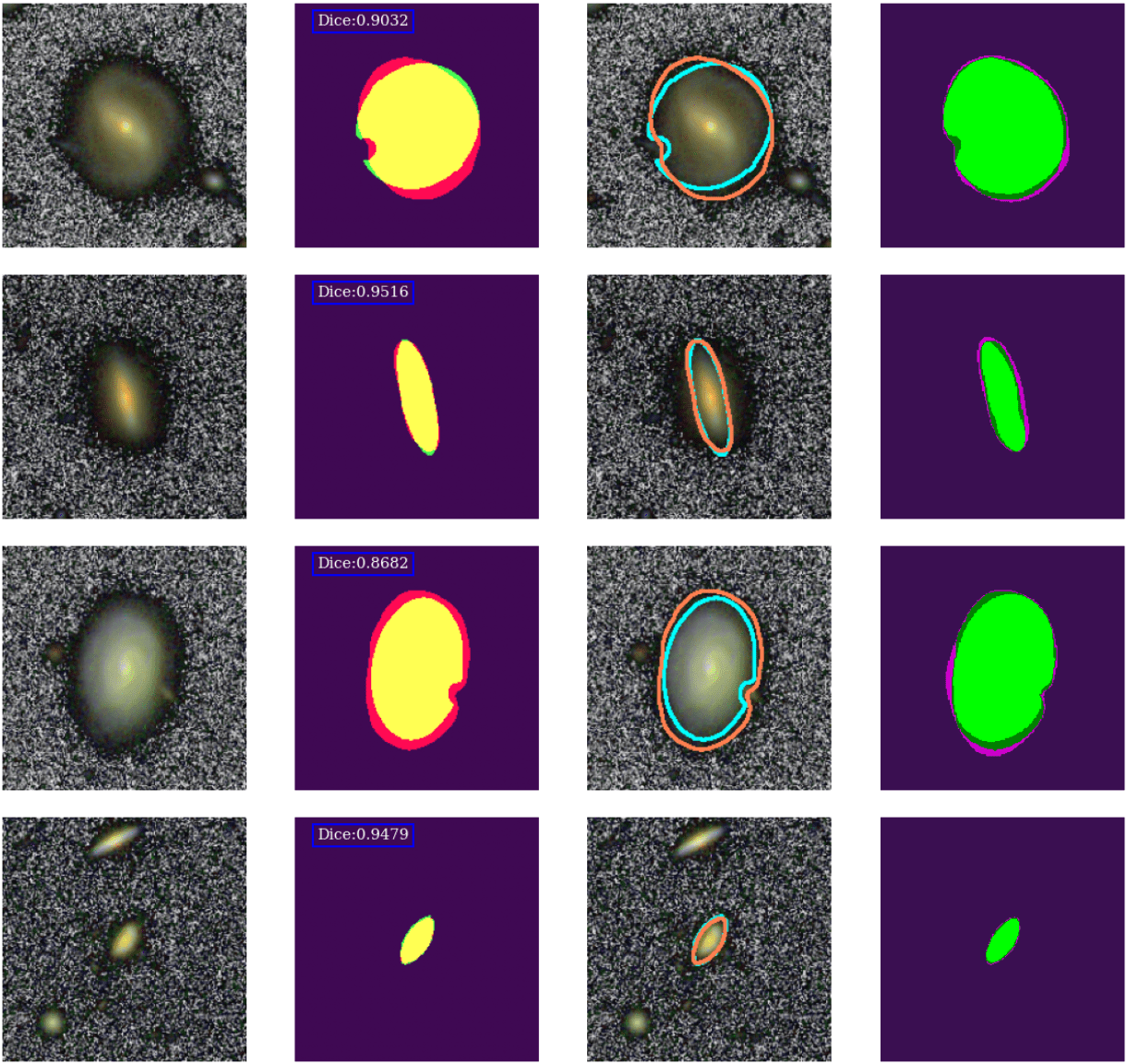}
        \caption{RGB images (first column), superimposed predicted and real masks (second column), superimposed predicted and real truncations (third column), and probabilistic ensemble outputs for galaxies 89 to 92 in the test set.}
        \label{Appendix-Figure-23}
\end{figure}

\begin{figure}[!ht]
        \centering
        \includegraphics[width=1\linewidth]{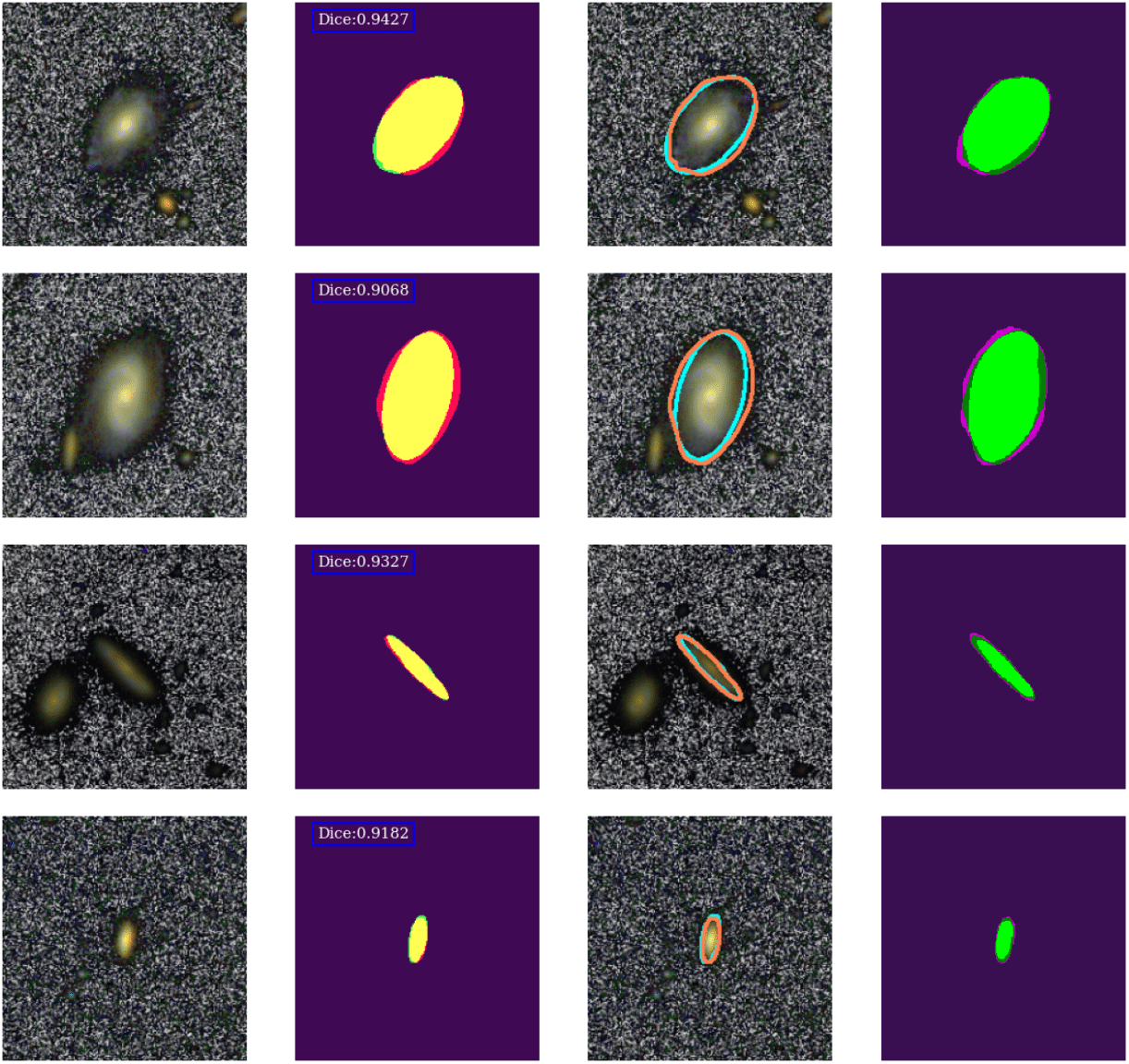}
        \caption{RGB images (first column), superimposed predicted and real masks (second column), superimposed predicted and real truncations (third column), and probabilistic ensemble outputs for galaxies 93 to 96 in the test set.}
        \label{Appendix-Figure-24}
\end{figure}

\begin{figure}[!ht]
        \centering
        \includegraphics[width=1\linewidth]{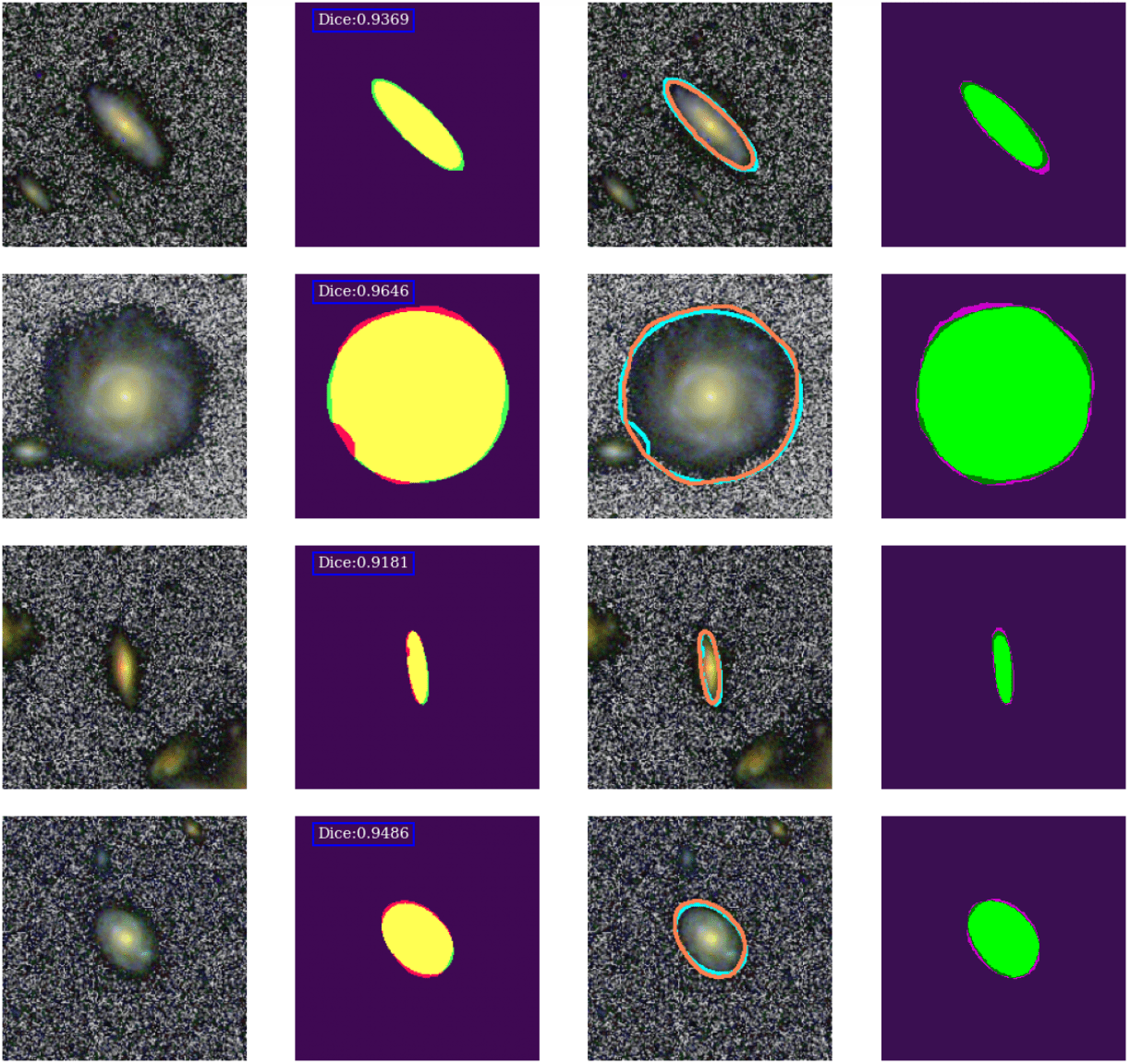}
        \caption{RGB images (first column), superimposed predicted and real masks (second column), superimposed predicted and real truncations (third column), and probabilistic ensemble outputs for galaxies 97 to 100 in the test set.}
        \label{Appendix-Figure-25}
\end{figure}

\end{appendix}

\end{document}